\documentclass[sn-chicago]{sn-jnl}

\usepackage{graphicx}%
\usepackage{multirow}%
\usepackage{amsmath,amssymb,amsfonts}%
\usepackage{amsthm}%
\usepackage{mathrsfs}%
\usepackage[title]{appendix}%

\usepackage{textcomp}%
\usepackage{manyfoot}%
\usepackage{booktabs}%
\usepackage{algorithm}%
\usepackage{algorithmicx}%
\usepackage{algpseudocode}%
\usepackage{listings}%
\usepackage{natbib}
\usepackage{url} 
\usepackage{hyperref}
\usepackage[table,xcdraw]{xcolor}
\usepackage[subrefformat=parens,labelformat=parens]{subfig}
\usepackage{float}
\usepackage{url}
\usepackage{xcolor}

\newcommand{\ybold}    {\mbox{\bf y}}

\newcounter{xxx}
\begin{document}

\title[Article Title]{Robust Bayesian Functional Principal Component Analysis}


\author[1]{\fnm{Jiarui} \sur{Zhang}}\email{jiaruiz@sfu.ca}

\author[1]{\fnm{Jiguo} \sur{Cao}}\email{jiguo\_cao@sfu.ca}

\author*[1]{\fnm{Liangliang} \sur{Wang}}\email{lwa68@sfu.ca}

\affil[1]{\orgdiv{Department of Statistics and Actuarial Science}, \orgname{Simon Fraser University}, \orgaddress{\street{8888 University Dr W}, \city{Burnaby}, \postcode{V5A 1S6}, \state{BC}, \country{Canada}}}

\abstract{We develop a robust Bayesian functional principal component analysis (RB-FPCA) method that utilizes the skew elliptical class of distributions to model functional data, which are observed over a continuous domain. This approach effectively captures the primary sources of variation among curves, even in the presence of outliers, and provides a more robust and accurate estimation of the covariance function and principal components. The proposed method can also handle sparse functional data, where only a few observations per curve are available. We employ annealed sequential Monte Carlo for posterior inference, which offers several advantages over conventional Markov chain Monte Carlo algorithms. To evaluate the performance of our proposed model, we conduct simulation studies, comparing it with well-known frequentist and conventional Bayesian methods. The results show that our method outperforms existing approaches in the presence of outliers and performs competitively in outlier-free datasets. Finally, we demonstrate the effectiveness of our method by applying it to environmental and biological data to identify outlying functional observations.
The implementation of our proposed method and applications are available at \url{https://github.com/SFU-Stat-ML/RBFPCA}.}

\keywords{Functional data analysis, Robust estimation, Sparse functional data, Multivariate skew elliptical distribution, Model comparison, Outlier detection}



\maketitle

\section{Introduction}
\label{sec:intro}

\begin{figure}[htbp]
  \centering
  \subfloat[]{\includegraphics[scale = 0.35]{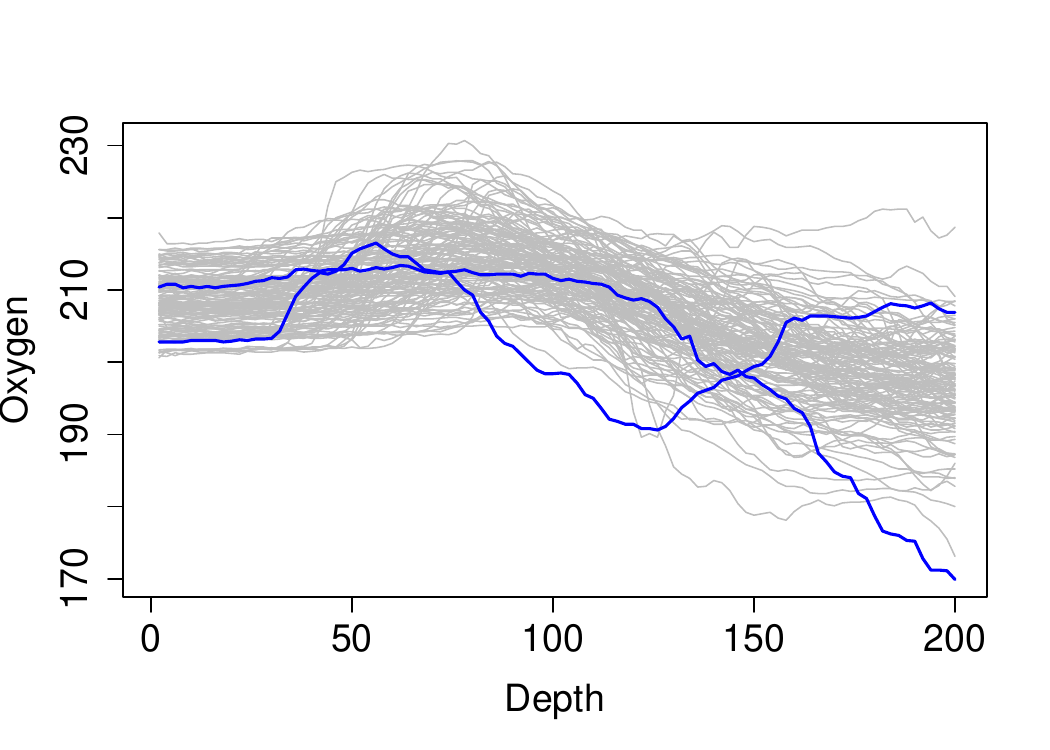}\label{fig:intro_data1}}
  \subfloat[]{\includegraphics[scale = 0.35]{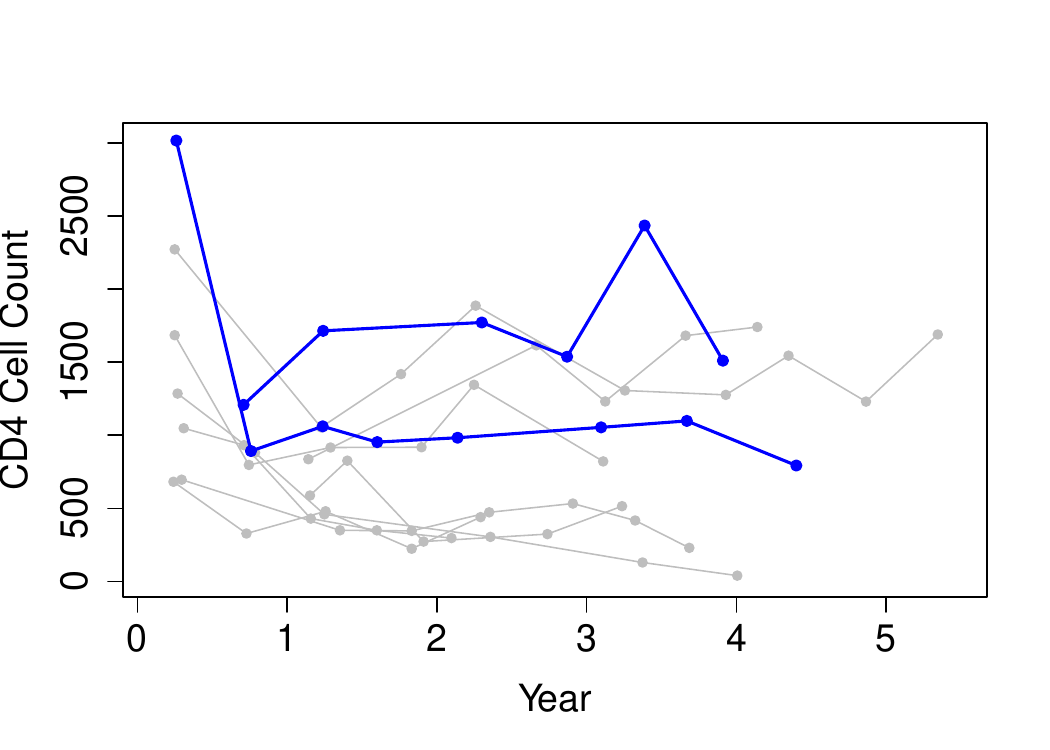}\label{fig:intro_data2}}
  \caption{(a) Hawaii Ocean Oxygen data contains 133 densely observed trajectories for oxygen concentrations measuring at different depths below the sea surface. The two blue trajectories exhibit atypical behaviors in comparison with the remaining ones. (b) A subset of CD4 data with 10 sparsely observed trajectories for CD4 cell counts are shown to demonstrate the sparse nature. The two trajectories colored in blue display abnormal patterns.} \label{fig:intro_data_outlier}
\end{figure}

The development of modern technology has resulted in the continuous recording of data during a given period in many scientific fields, such as neuroscience, biology, and environmental science. These data can be categorized as functional data \citep{Ramsay05,Ferraty06,Kokoszka12,Eubank15}, which are usually observed over time, space or any other continuous domain. For instance, Figure \ref{fig:intro_data_outlier} displays two examples of functional data. Figure \ref{fig:intro_data1} shows a collection of densely observed trajectories from the Hawaii Ocean Oxygen dataset \citep{HOT-DOGS}, wherein certain outlying trajectories have been highlighted. Figure \ref{fig:intro_data2} presents one illustration of sparse function data from the CD4 dataset \citep{zeger1994semiparametric}, with only a few observations per curve. Both datasets have a few curves demonstrating unusual patterns in contrast to the rest. 

Functional principal component analysis (FPCA) plays a crucial role in functional data analysis (FDA) as a dimension reduction technique. FPCA can be viewed as an exploratory analysis tool to discover the hidden structure of the data by capturing the optimal low-dimensional representation and the major source of variations among curves. FPCA can also be used to help reconstruct partially observed functions or used as a preprocessing step for regression analysis \citep{reiss2017methods} or clustering tasks \citep{margaritella2021parameter}. The frequentist analysis of FPCA is a mature field. Numerous works have been proposed to explore fully or densely observed function data \citep{dauxois1982asymptotic, rice1991estimating, cardot2000nonparametric, hall2006properties}. Studies of the FPCA approach for sparsely observed data, a more challenging situation, can also be seen in \cite{james2000principal}, \cite{yao2005functional}, and \cite{paul2009consistency}. \cite{wang2016functional} provided a detailed review of methods, open questions and applications of FPCA.

In contrast, the Bayesian perspective on FPCA has received less attention. The lack of Bayesian methodologies in FDA and FPCA may stem from the challenges in specifying a full probability model and the limited availability of suitable inferential tools for implementation. Nonetheless, the Bayesian framework offers significant advantages from at least three key aspects. First, Bayesian methods provide a natural way to quantify uncertainty, offering credible intervals and full posterior distributions that give a more comprehensive understanding of parameter estimates. Second, the flexibility of Bayesian inference allows for the seamless incorporation of domain knowledge through the specification of informative priors, making it especially useful in complex or expert-driven applications.  \textcolor{black}{Lastly, the Bayesian framework enables a principled approach to robust model comparison through Bayes factors, using marginal likelihoods.}

One example of applying Bayesian methodologies in FPCA is presented by \cite{behseta2005hierarchical}, who utilized a hierarchical Gaussian process model to assess variability among functions. Later, \cite{van2008variational} proposed analyzing the modes of variation of curves via variational inference. The prevalent strategy has been to work within the hierarchical representation of the FPCA model and specify the prior distributions for all model parameters \citep{crainiceanu2010bayesian, margaritella2021parameter}. Alternatively, \cite{suarez2017bayesian} proposed a Bayesian FPCA method using approximate spectral decomposition and modelled the number of principal components with truncated Poisson distributions. For partially observed functional data, \cite{jiang2020bayestime} considered a Bayesian model for sparse FPCA using a reduced rank mixed-effects framework. An alternative method closely related to FPCA is functional factor models. In a Bayesian context, \cite{montagna2012bayesian} introduced a Bayesian latent factor model for functional data, integrating a sparse latent factor regression model over the basis coefficients to effectively accomplish dimension reduction. \cite{kowal2017bayesian} employed a multivariate functional dynamic linear model to model functional, time-dependent, and multivariate data, interpreting it as a form of dynamic factor analysis. \cite{shamshoian2022bayesian} focused on the longitudinal functional data and captured low-dimensional interpretable features through latent factor models from a Bayesian perspective. Both FPCA and functional factor models are dimension reduction techniques. In functional factor models, the factors are not subject to orthogonality constraints. Consequently, a common practice involves rotating the estimated factors to enhance their interpretability. While in FPCA, orthonormal eigenfunctions span the subspaces that provide the best approximation. These orthogonality constraints contribute to the interpretability of the components within the decomposition. To the best of our knowledge, no existing literature has explored outlier detection within the context of a robust FPCA or functional factor modelling framework from a Bayesian standpoint, that is capable of handling both dense and sparse functional data.

Most previous works on the Bayesian FPCA model employ multivariate normal distribution to model the discretized, noise-contaminated observations. Despite the fact that the multivariate normal distribution has several desirable properties for modelling data, it is common that observations are not normally distributed in general. Various alternative distributions are available to accommodate higher moments. For instance, the multivariate Student-$t$ distribution works well for fat-tailed data but does not consider asymmetry. The log-normal distribution has been used to model skewed data, but its skewness is reflected as a function of the mean and variance. This is somewhat inflexible because skewness is not a separate parameter. \cite{branco2001general} proposed a general class of skew elliptical distributions by extending the previous work on a multivariate skew normal distribution as described by \cite{azzalini1996multivariate}. The general class developed by \cite{branco2001general} includes the multivariate normal, Student-$t$, and exponential power, but with an additional parameter to control skewness. This family of distributions was improved upon by \cite{sahu2003new} for capturing multivariate asymmetry and adding more flexibility in adjusting the correlation structure, which often leads to better-fitting models. Moreover, the family of distributions proposed by \cite{sahu2003new} offers a convenient implementation within a Bayesian framework.

\textcolor{black}{
Almost all existing Bayesian FPCA methods utilize Markov Chain Monte Carlo (MCMC) approaches for implementation.  Sequential Monte Carlo (SMC) methods, however, provide a versatile alternative framework for sampling from target distributions through a series of intermediate distributions, which can help overcome challenges encountered with traditional MCMC approaches. Originally developed for state-space models, SMC methods have since evolved and are now applied to a broader range of Bayesian inference problems and shown their ability to handle complex posterior distributions more effectively than MCMC techniques \citep{DelMoral2006,wang2020annealed,wang2022adaptive}.  The annealed sequential Monte Carlo (ASMC) algorithm, in particular, generates a sequence of artificial intermediate distributions by annealing the likelihood---raising it to the power of annealing parameters---and combining it with the prior, providing an efficient strategy for exploring challenging posterior landscapes. Moreover, SMC methods offer the advantage of yielding unbiased estimates of the marginal likelihood with minimal additional computational effort, making them especially useful for Bayesian model selection \citep{zhou2016,wang2020annealed,Dai2022}.}

\textcolor{black}{Building upon the numerous advantages of Bayesian inference and the advancements in the skew  elliptical class of distributions, we propose a general method for fitting observations using multivariate skew elliptical distributions within the Bayesian FPCA framework. Our proposed method presents several significant contributions. First, our method produces more robust estimations of the covariance function and the corresponding principal components by integrating the class of skew elliptical distributions. This integration circumvents the need for ad hoc transformations often required for skewed data, thus providing a more straightforward and effective modeling approach.  Our methodology incorporates a range of distributions, including the multivariate skew normal distribution, multivariate skew $t$ distribution, and multivariate finite mixture models. This enhancement allows the model to better accommodate diverse data characteristics, thereby broadening its applicability and improving robustness across various scenarios. Second, we extend our method to fit sparse and irregular functional data. This extension is particularly crucial for analyzing complex datasets where traditional methods may falter. Third, we implement an annealed sequential Monte Carlo method for posterior inference, offering several advantages over conventional MCMC methods. This method effectively navigates the multimodal posterior surface and provides an efficient way to obtain an unbiased estimator of the marginal likelihood, which is valuable for model comparison. Fourth, we present credible intervals within our Bayesian framework for FPCA to facilitate the interpretation of uncertainty measures, empowering researchers to better evaluate the reliability and robustness of their models and make informed decisions.  Fifth, we use the marginal likelihood estimates from ASMC for model comparison, allowing us to choose between models with skew normal, skew $t$, and a mixture of distributions. Additionally, this approach allows us to compare variations of the same model with different numbers of components.  Finally, we provide R code for the implementation of our proposed method, enabling practitioners to readily apply our method in their analyses.}

The rest of the article is organized as follows: We introduce the setup of Bayesian FPCA in Section \ref{sec:background_BFPCA}. The proposed model is presented in Section \ref{sec:model_prior}, along with computational details in Section \ref{sec:posterior}, \textcolor{black}{an extension to multivariate finite mixture model in Section \ref{sec:mixture_models}}, a description of outlier detection method in Section \ref{sec:outlier_detection}, and an extension to sparse data in Section \ref{sec:model_sparse}. Simulation studies are presented in Section \ref{sec:simulation}, in which we compare the proposed model to other Bayesian and frequentist FPCA approaches in the presence of different data generation processes and noise levels. Three examples including dense and sparse data are discussed in Section \ref{sec:data_application}. A summary and some directions for future work are given in Section \ref{sec:discussion}.

\section{Robust Bayesian FPCA}
\label{sec:model}

In this section, we will present the proposed robust Bayesian FPCA model and describe the features that contribute to its improvement over the existing FPCA methods.

\subsection{Functional Principal Component Analysis}
\label{sec:background_BFPCA}

Functional principal component analysis (FPCA) extends the idea of the principal component analysis of multivariate data \cite[see][for a comprehensive introduction to classical multivariate PCA]{jolliffe2002principal} to functional data. FPCA is the most prevalent tool in FDA and serves as a standard first step when processing functional data in most cases \citep{ramsay2005principal}. This is due to FPCA's capability to convert inherently infinite-dimensional functional data into a finite-dimensional vector of random scores. The general goals of FPCA include finding patterns in data of high dimensions, capturing the major sources of variation and reducing the dimensionality.

Denote $\{ X(t): t \in [0, T] \subset \mathcal{R} \}$ as a square integrable stochastic process. Consider $n$ independent and identically distributed realizations, $X_1(t), X_2(t),\ldots, X_n(t)$, of a mean-zero stochastic process with covariance function $\text{Cov}(s, t)$ at a sequence of random points on $\mathcal{T} = [0, T]$. The covariance function $\text{Cov}(s, t)$ specifies the covariance between curve values $X_i(s)$ and $X_i(t)$ at times $s$ and $t$, respectively. We assume the observations are noise-corrupted. That is, the observed data $Y_i(t)$, $i = 1,\ldots,n$,  is
\begin{equation} \label{eq:fda_model}
    Y_i(t) = \mu(t) + X_i(t) + \epsilon_i(t),
\end{equation}
where $Y_i(t)$ is the noise-contaminated observed data for the $i$th curve at time $t$, for $i = 1,\ldots,n$; $X_i(\cdot)$ is the true underlying detrended curve from a Gaussian process with mean zero and covariance function $\text{Cov}(s, t)$ for $s, t \in \mathcal{T}$, independently for $i = 1,\ldots,n$; and $\epsilon_{i}(t)$ is the error term. A conventional assumption is  $\mathbb{E}(\epsilon_{i}(t)) = 0$ and $\text{Var}(\epsilon_{i}(t)) = \sigma^2$.

Mercer's theorem \citep{AshRobertB1965It/b} states that, under mild assumptions, the spectral decomposition of the covariance function can be approximated as
\begin{equation}
    \text{Cov}(s, t) \approx Q(s,t) =  
    \sum^{K}_{k=1} \lambda_k \phi_{k}(s) \phi_{k}(t), \;  \text{for} \: s, t \in \mathcal{T},
\label{eq:cov_eigen}
\end{equation}
where $Q(s, t)$ converges uniformly to $\text{Cov}(s, t)$ as $K \rightarrow \infty$, $\{\lambda_1, \lambda_2, \dots\}$ are the non-negative eigenvalues in non-increasing order of the covariance operator, and $\{\phi_{1}, \phi_{2}, \dots\}$ are the corresponding orthogonal eigenfunctions.
By the Karhunen-Loève expansion \citep{fukunaga1970representation}, each curve $X_i(t)$ can be approximated by a linear combination of the eigenfunctions $\phi_{k}(t)$ and FPC scores $\xi_{ik}$:
\begin{equation}
    X_i(t) \approx \sum^{K}_{k=1} \xi_{ik} \phi_{k}(t).
\end{equation}
By truncating the infinity expansion at some level, $K \in \mathbb{N}$, the resulting finitely truncated series provides a good approximation to $X_i(t)$.

The eigenfunctions $\{\phi_{1}, \phi_{2}, \dots\}$ can be further represented with the following basis expansion:
\begin{equation}
    \phi_{k}(t) = \sum^{\infty}_{p=1} c_{kp} h_p(t) \approx \sum^{P}_{p=1} c_{kp} h_p(t) ,
\label{eq:phi_expansion}
\end{equation}
where $h_p(t)$'s are some given basis functions and $c_{kp}$'s are the corresponding basis coefficients. The expansion in Equation (\ref{eq:phi_expansion}) is truncated at some level, $P \in \mathbb{N}$. 
Let $\boldsymbol{H}_P(t)$ be the $P \times 1$ vector with elements consisting of the given basis functions evaluated at time $t$. 
Let $\boldsymbol{C}_{KP}$ denote the coefficient matrix of size $K \times P$ and $\boldsymbol{\Lambda}_{K}$ be a $K \times K$ diagonal matrix whose diagonal is composed of the non-negative eigenvalues $\{\lambda_1, \lambda_2, \dots, \lambda_K\}$ in non-increasing order. Then the covariance function can be represented by the following relation:
\begin{equation}
    Q(s,t) = \boldsymbol{H}'_{P}(s) \boldsymbol{C}'_{KP} \boldsymbol{\Lambda}_{K} \boldsymbol{C}_{KP} \boldsymbol{H}_{P}(t), \;  \text{for} \: s, t \in \mathcal{T}.
\label{eq:cov_expansion}
\end{equation}
Instead of directly working on $\boldsymbol{C}_{KP}$ and $\boldsymbol{\Lambda}_{K}$, we reparameterize Equation (\ref{eq:cov_expansion}) by defining $\boldsymbol{\Omega}  = \boldsymbol{C}'_{KP} \boldsymbol{\Lambda}_{K} \boldsymbol{C}_{KP}$; i.e. $Q(s,t) = \boldsymbol{H}'_{P}(s) \boldsymbol{\Omega} \boldsymbol{H}_{P}(t)$.

\subsection{Models and Prior Specification}
\label{sec:model_prior}

In this section, we give details on the structure of the proposed robust Bayesian FPCA (RB-FPCA) method. The main purpose of the inference is to estimate the covariance function $Q(s,t)$ and the principal components from the data. We introduce a Bayesian hierarchical model to perform robust estimations of the covariance function and the functional principal components. We propose to fit multivariate skew elliptical distributions to model the observations, which produces robust estimates in the Bayesian FPCA framework. To construct the skew elliptical class of distributions, we first present the settings for multivariate symmetric elliptical distributions, followed by an introduction to the skew elliptical class of distributions, where skewness is induced via transformation and conditioning.

We first consider the case when the data are dense and equally spaced, i.e., the number of measurements $n_i$ for each curve is the same and the sequence $t_{i1}, \ldots, t_{in_i}$ is equally spaced for every curve $i = 1, \ldots, n$. We simplify the notations by letting $n_i = m$, where all curves have the same number of measurements, and $\boldsymbol{Y}_{i} = (Y_i(t_{i1}), Y_i(t_{i2}), \ldots, Y_i(t_{im}))'$ for $i = 1,\ldots,n$.  For simplicity of notation, we assume $\mu(t)=0$, for $t\in \mathcal{T}$; in case $\mu(t)\neq 0$, 
$\mu(t)$ can be easily estimated, e.g. using local linear
smoothers, and subtracted from $Y_i(t)$.

We propose to fit $\boldsymbol{Y}_i$ with the multivariate skew elliptical distribution, which is constructed by implementing the transformation presented in \cite{sahu2003new}:
\begin{equation}\label{skew_transformation}
    \boldsymbol{Y}_{i} = \boldsymbol{D} \boldsymbol{z}_{i} + \boldsymbol{\eta}_i,
\end{equation}
where $\boldsymbol{Y}_{i}$ is the transformed variable which followed the skew elliptical distribution, $\boldsymbol{D}$ is a diagonal matrix with diagonals $\boldsymbol{d} = (d_1,\ldots,d_m)'$, which accommodates skewness, $\boldsymbol{z}_{i}$ and $\boldsymbol{\eta}_i$ are two $m$-dimensional random vectors. The skew elliptical class of distributions is formulated by introducing transformations and conditioning on the class of elliptically symmetric distributions. We use the following notation to represent the class of elliptically symmetric distributions:
\begin{equation}
    \left(\begin{array}{c}\boldsymbol{\eta}_i\\ \boldsymbol{z}_{i}\end{array}\right) \sim 
    \text{El}\left( \left(\begin{array}{c}\boldsymbol{X}_i\\ \boldsymbol{0}\end{array}\right), \begin{bmatrix}\boldsymbol{\Sigma} & \boldsymbol{0} \\\boldsymbol{0} & \boldsymbol{I} \end{bmatrix}; \: g_{2m} \right),
\end{equation}
where $\boldsymbol{X}_i = (X_i(t_{i1}), X_i(t_{i2}), \ldots, X_i(t_{im}))'$ is a $m$-dimensional vector, $\boldsymbol{\Sigma}$ is a positive definite matrix of size $m \times m$, $\boldsymbol{0}$ and $\boldsymbol{I}$ are the null matrix and identity matrix, respectively, and $g_{2m}$ is the generator of the probability density function (PDF). We emphasize that the density generator $g_{2m}$ depends on the dimension $2m$ of the random vector $( \boldsymbol{\eta}_i', \boldsymbol{z}_{i}' ) '$. The density generator $g_{2m}$ is a function from $\mathbb{R}^+$ to $\mathbb{R}^+$ which satisfies
\begin{equation}
    \int_0^{\infty} w^{m-1} g_{2m}(w) \,dw = \pi^{-m} \Gamma\left(m\right),
\end{equation}
where $\Gamma(m) = (m-1)!$ is the gamma function. The choice of the density generator function will determine the distribution of the random vector $( \boldsymbol{\eta}_i', \boldsymbol{z}_{i}' ) '$. This flexible nature of the elliptical class of distributions allows for including several widely recognized symmetric distributions as proper members, for example, the multivariate normal distribution. Assuming a random vector $\boldsymbol{W}$ of dimension $u$ follows a multivariate normal distribution with mean vector $\boldsymbol{\zeta}$ and covariance matrix $\boldsymbol{\check{\Sigma}}$, i.e., $\boldsymbol{W} \sim \text{MVN}(\boldsymbol{\zeta}, \boldsymbol{\check{\Sigma}})$, the probability density function can be formulated via the following generalized expression in terms of the generator function $g_{u}$:
\begin{center}
    $f(\boldsymbol{W} \: | \: \boldsymbol{\zeta}, \boldsymbol{\check{\Sigma}}; g_{u}) = |\boldsymbol{\check{\Sigma}}|^{-1/2} g_{u} ( (\boldsymbol{W} - \boldsymbol{\zeta})'\boldsymbol{\check{\Sigma}}^{-1}(\boldsymbol{W} - \boldsymbol{\zeta}) )$,
\end{center}
where the density generator function has the form of $g_{u}(v) = e^{-v/2} / (2\pi)^{u/2}$. The conventional form of the multivariate normal distribution’s PDF can be retrieved with this density generator function.

By establishing the class of elliptically symmetric distributions, the skew elliptical class is derived through the consideration of the random variable $\boldsymbol{Y}_i \, |\, \boldsymbol{z}_{i} \! > \! 0$, where $\boldsymbol{z}_{i} = (z_{i1},\ldots,z_{im})' > 0$ implies every element of $\boldsymbol{z}_{i}$ is positive. The construction in Equation (\ref{skew_transformation}) along with the conditioning introduces skewness. Specifically, positive values of elements of $\boldsymbol{d}$ result in positively skewed distributions, while negative values lead to negatively skewed distributions. Note that the elliptically symmetric distribution is retrieved if the diagonals of $\boldsymbol{D}$ are zeros, i.e., $d_1,\ldots,d_m = 0$. With the transformation and conditioning, the random variable $\boldsymbol{Y}_{i}$ follows the skew elliptical distribution and we denote it with the notation $\boldsymbol{Y}_i \sim \text{SE}(\boldsymbol{X}_i, \boldsymbol{\Sigma}, \boldsymbol{D}; g_{m})$, $i = 1,\ldots,n$.

In the Bayesian framework, we specify a prior distribution for the covariance function $Q(s,t) = \boldsymbol{H}'_{P}(s) \boldsymbol{\Omega} \boldsymbol{H}_{P}(t)$ through a prior on $\boldsymbol{\Omega}^{-1} = (\boldsymbol{C}'_{KP} \boldsymbol{\Lambda}_{K} \boldsymbol{C}_{KP})^{-1}$, which is assumed to be a Wishart distribution. When the number of principal components $(K)$ is equal to the number of basis functions $(P)$ used for approximation, we can incorporate the following prior distribution $\boldsymbol{\Omega}^{-1} \sim \text{Wishart}(\nu, \: \boldsymbol{\Psi}^{-1})$, where $\nu$ and $\boldsymbol{\Psi}$ are hyperparameters. However, it is commonly assumed that more basis functions are required than the number of principal components for the reconstruction of the covariance function. In the case when $K \leq P$, the Wishart matrix becomes singular \citep{uhlig1994singular}. Because Wishart distribution gives zero density to singular matrices, to allow $K \leq P$, it is necessary to consider the inclusion of singular Wishart matrices of rank $K$ into the specification of the prior. First, we decompose $\boldsymbol{\Psi} = \boldsymbol{U} \boldsymbol{L} \boldsymbol{U}'$, where $\boldsymbol{U}$ is a $P \times P$ orthogonal matrix and $\boldsymbol{L}$ is a $P \times P$ diagonal matrix with ordered eigenvalues along the diagonal. Now choose $\boldsymbol{\Psi}_K = \boldsymbol{U}_K \boldsymbol{L}_K \boldsymbol{U}_K'$, where $\boldsymbol{U}_K$ is the first $K$ columns of $\boldsymbol{U}$ and $\boldsymbol{L}_K$ is the first $K$ rows and $K$ columns of $\boldsymbol{L}$. It follows that $\boldsymbol{\Psi}_K$ has rank $K$ even though its dimension is $P \times P$. By defining the singular center matrix $\boldsymbol{\Psi}_K$ of rank $K$, we allow for singular Wishart matrices in specifying the prior. We consider incorporating the following prior distribution $\boldsymbol{U}_K \boldsymbol{\Omega}^{-1} \boldsymbol{U}_K' \sim \text{Wishart}(\nu, \: \boldsymbol{\Psi}_K^{+})$, where $\boldsymbol{\Psi}_K^{+}$ is the Moore-Penrose inverse of the matrix $\boldsymbol{\Psi}_K$, and $\boldsymbol{\Omega}^{-1} \sim \text{Wishart}(\nu, \boldsymbol{L}_k^{-1})$. This forms the desired singularity in the prior specification (see \cite{suarez2017bayesian}). With the aforementioned specification, the covariance function $Q(s,t) = \boldsymbol{H}'_{P}(s)  \boldsymbol{U}_K \boldsymbol{\Omega} \boldsymbol{U}_K'\boldsymbol{H}_{P}(t)$. This is implemented by modelling $X_i(t) = \boldsymbol{H}'_P(t) \boldsymbol{U}_K \boldsymbol{\beta}_{i},$ where the variance–covariance matrix of $\boldsymbol{\beta}_{i}$ is $\boldsymbol{\Omega}$.

We can now integrate the family of skew elliptical distributions within the Bayesian FPCA framework and present the full RB-FPCA model. We assume $\boldsymbol{Y}_i \sim \text{SE}(\boldsymbol{X}_i, \boldsymbol{\Sigma}, \boldsymbol{D}; g_{m})$, \textcolor{black}{exemplified by the multivariate skew $t$ distribution and multivariate skew normal distribution}, and further relax the assumption of homoscedasticity on noises. Let $\boldsymbol{H}_P$ be the $m \times P$ matrix with columns consisting of the given basis functions evaluated at all time points. \textcolor{black}{The hierarchical representation of the proposed RB-FPCA model with multivariate skew $t$ distribution for $i = 1, \ldots ,n,$ is
\begin{align} 
\boldsymbol{Y}_i  &\sim \text{MVN}_{m}(\boldsymbol{H}_P \boldsymbol{U}_K \boldsymbol{\beta}_{i} + \boldsymbol{D}\boldsymbol{z}_i, \: \boldsymbol{\Sigma}/w_i), \label{eqn:likelihood_skew_t} \\ 
\boldsymbol{\beta}_{i} &\stackrel{i.i.d}{\sim}  \text{MVN}_{K}(\boldsymbol{0}, \: \boldsymbol{\Omega}), \label{eqn:beta_skew_t} 
 \\
\boldsymbol{\Omega}^{-1} &\sim \text{Wishart}_K(\nu, \: \boldsymbol{L}_k^{-1}), \\
\boldsymbol{z}_i &\sim \text{MVN}_{m}(\boldsymbol{0}, \boldsymbol{I})I(\boldsymbol{z}_i > 0), \\
\text{vec}(\boldsymbol{D}) &\sim \text{MVN}_{m}(\boldsymbol{0}, \boldsymbol{\Gamma}), \\
\boldsymbol{\Sigma}^{-1} &\sim \text{Wishart}_{m}(2r, 2\boldsymbol{\kappa}), \\
w_i &\sim \text{Gamma}(\nu_{w_i}/2, \nu_{w_i}/2), \\
\nu_{w_i} &\sim \text{Gamma}(1, 0.1)I (\nu_{w_i} > 2), \label{eqn:nv_in_skew_t}
\end{align}
}
where 
$\boldsymbol{\beta}_{i}$ are the coefficients, $\text{vec}(\boldsymbol{D})$ denotes a vector of diagonals of $\boldsymbol{D}$, \textcolor{black}{$w_i$'s are $n$ i.i.d. random variables introduced to obtain the multivariate skew $t$ distribution}, and $\nu, \boldsymbol{\Gamma}, 2r$ and $\boldsymbol{\kappa}$ are hyperparameters. \textcolor{black}{We denote our proposed model with the multivariate skew $t$ distribution as  RB-FPCA-ST. For the multivariate skew normal distribution, each of $w_i$ will be set to 1 and the last two specifications for $w_i$ and $\nu_{w_i}$ are omitted. We denote this model as  RB-FPCA-SN.}

\subsection{Posterior Inference}
\label{sec:posterior}
\textcolor{black}{In Bayesian framework, the target distribution is the posterior distribution $\pi(\boldsymbol{\theta})$, which is proportional to the product of the likelihood function $\pi(\ybold|\boldsymbol{\theta})$ in Equation \ref{eqn:likelihood_skew_t} and prior distributions, denoted by $\pi_0(\boldsymbol{\theta})$,  in Equations \ref{eqn:beta_skew_t} to \ref{eqn:nv_in_skew_t}, where  $\ybold$ presents all the data and  $\boldsymbol{\theta}$ denotes all the unknown parameters in the model. A Gibbs algorithm can be used to sample}
 the parameters from the posterior distribution by deriving their full conditional distributions. All parameters' conditional distributions can be derived analytically in closed forms and are given as follows. In each conditional distribution, we use $| \cdots$ to denote conditioning on the data and all other parameters.

\begin{itemize}
    \item For $i = 1,\ldots,n$, the full conditional distribution for $\boldsymbol{\beta}_{i}$ is
    \begin{equation}\label{eq:cond_dist_beta}
        \begin{split}
            \boldsymbol{\beta}_{i} | \cdots \sim \text{MVN}_{K}((\boldsymbol{U}_K' \boldsymbol{H}_P' & \boldsymbol{\Sigma}^{-1} \boldsymbol{H}_P \boldsymbol{U}_K + \boldsymbol{\Omega}^{-1})^{-1}(\boldsymbol{U}_K' \boldsymbol{H}_P' \boldsymbol{\Sigma}^{-1} \boldsymbol{Y}_i), \\
            & (\boldsymbol{U}_K' \boldsymbol{H}_P' \boldsymbol{\Sigma}^{-1} \boldsymbol{H}_P \boldsymbol{U}_K + \boldsymbol{\Omega}^{-1})^{-1}),
        \end{split}
    \end{equation}
    where $\boldsymbol{H}_P$ is picked to be the Legendre polynomial basis functions and the choice of $\boldsymbol{U}_K$ is discussed in the conditional distribution for $\boldsymbol{\Omega}^{-1}$.
    
    \item For $\boldsymbol{\Omega}^{-1}$, the full conditional distribution is
    \begin{equation} \label{eq:cond_dist_Omega}
         \boldsymbol{\Omega}^{-1} | \cdots \sim \text{Wishart}_K(\nu + n + 1, (\boldsymbol{L}_k + \sum^n_{i = 1} \boldsymbol{\beta}_{i} \boldsymbol{\beta}_{i}')^{-1}),
    \end{equation}
    where $\nu$ is the number of degrees of freedom. 
    \textcolor{black}{Here, $\boldsymbol{L}_K$ in Equation \ref{eq:cond_dist_Omega} and $\boldsymbol{U}_K$ in Equation \ref{eq:cond_dist_beta} are derived by first decomposing $\boldsymbol{\Psi}= \boldsymbol{U} \boldsymbol{L} \boldsymbol{U}'$, where $\boldsymbol{\Psi}$ is the inverse of the scale matrix in Wishart prior for $\boldsymbol{\Omega}^{-1}$ when $K=P$. The matrices $\boldsymbol{L}$ and $\boldsymbol{U}$ are then subsetted, as outlined in Section \ref{sec:model_prior}, to obtain the desired $\boldsymbol{L}_K$ and $\boldsymbol{U}_K$ for $K \leq P$. A sensible choice of $\boldsymbol{\Psi}$ is given as $\boldsymbol{\Psi} = (\boldsymbol{H}_P'\boldsymbol{H}_P)^{-1} \boldsymbol{H}_P' \boldsymbol{\Omega}^{*} \boldsymbol{H}_P (\boldsymbol{H}_P'\boldsymbol{H}_P)^{-1}$, where $\boldsymbol{\Omega}^{*}$ is the prior covariance function corresponding to the time grid being used. 
    }
    
    \item For $\boldsymbol{z}_i$, the full conditional distribution is
    \begin{equation} \label{eq:cond_dist_z}
        \boldsymbol{z}_i | \cdots \sim \text{MVN}_{m}(\boldsymbol{A}_i^{-1} \boldsymbol{a}_i, \boldsymbol{A}_i^{-1})I(\boldsymbol{z}_i > 0),
    \end{equation}
    \textcolor{black}{where $\boldsymbol{A}_i = \boldsymbol{I} + w_i \boldsymbol{D} \boldsymbol{\Sigma}^{-1} \boldsymbol{D}$ and $\boldsymbol{a}_i = w_i \boldsymbol{D} \boldsymbol{\Sigma}^{-1}(\boldsymbol{Y}_i - \boldsymbol{H}_P \boldsymbol{U}_K \boldsymbol{\beta}_{i})$. For multivariate skew normal distribution, each of $w_i$ is set to 1.}
    
    \item For $\text{vec}(\boldsymbol{D})$, the full conditional distribution is
    \begin{equation} \label{eq:cond_dist_vecD}
        \text{vec}(\boldsymbol{D}) | \cdots \sim \text{MVN}_{m}(\boldsymbol{B}^{-1}\boldsymbol{b}, \boldsymbol{B}^{-1}),
    \end{equation}
    with $\boldsymbol{B} = \boldsymbol{\Gamma}^{-1} + \sum^{n}_{i = 1} \text{diag}(\boldsymbol{z}_i) \boldsymbol{\Sigma}^{-1} \text{diag}(\boldsymbol{z}_i)$, $\boldsymbol{b} = \sum^{n}_{i = 1} \text{diag}(\boldsymbol{z}_i) \boldsymbol{\Sigma}^{-1} (\boldsymbol{Y}_i - \boldsymbol{H}_P \boldsymbol{U}_K \boldsymbol{\beta}_{i})$.
    
    \item For $\boldsymbol{\Sigma}^{-1}$, the full conditional distribution is
    \begin{equation} \label{eq:cond_dist_Sigma}
    \begin{split}
        \boldsymbol{\Sigma}^{-1} | \cdots \sim & \text{Wishart}_{m}(2r + n, ((2\boldsymbol{\kappa})^{-1} + \\
        & \sum^n_{i = 1} (\boldsymbol{Y}_i - \boldsymbol{H}_P \boldsymbol{U}_K \boldsymbol{\beta}_{i} - \boldsymbol{D}\boldsymbol{z}_i) (\boldsymbol{Y}_i - \boldsymbol{H}_P \boldsymbol{U}_K \boldsymbol{\beta}_{i} - \boldsymbol{D}\boldsymbol{z}_i)')^{-1}.
    \end{split}
    \end{equation}

    \textcolor{black}{\item For $w_i$, the full conditional distribution is
    \begin{equation} \label{eq:cond_dist_wi}
        w_i | \cdots \sim \text{Gamma}(\nu_{w_i}/2 + m/2, \nu_{w_i}/2 + m \times s_i / 2),
    \end{equation}
    where $m$ is the number of measurements in each curve and $s_i$ is the sample variance of the measurements in the $i$th curve.}
    
\end{itemize}

\color{black}

Sequential Monte Carlo (SMC) methods \cite[see][for an introduction to SMC]{doucet2009tutorial} have demonstrated several advantages over the Markov chain Monte Carlo algorithms, including Gibbs samplers and Metropolis-Hastings (MH) algorithms. 
We propose an annealed sequential Monte Carlo (ASMC) algorithm to perform Bayesian inference for the model parameters. The core idea is to utilize a sequence of intermediate distributions to facilitate the exploration of the parameter space. The SMC method approximates the intermediate distributions through an iterative process of reweighting, propagating, and resampling a set of random samples called particles. In ASMC, the sequence of intermediate distributions is constructed by raising the likelihood to fractional powers, effectively flattening the posterior distribution and facilitating the movement of particles between modes. The power applied to the likelihood, called the annealing parameter, ranges from zero to one. When the annealing parameter is small, the intermediate distribution is relatively flat, allowing particles to move more freely between modes. As the annealing parameter increases, the particles gradually move closer to the target posterior distribution. We implement an adaptive scheme to select the annealing parameters by controlling the increase in particle degeneracy, which in turn determines the sequence of intermediate distributions. Each iteration involves one step of the Gibbs sampler, based on Equations \ref{eq:cond_dist_beta} to \ref{eq:cond_dist_wi}, followed by an MH step to propagate the particles and approximate the intermediate distribution.  Given the annealing parameters, the ASMC can produce an unbiased estimator of the marginal likelihood as a byproduct of the sampling process, which can be used for model comparison.  Details of the ASMC implementation are provided in the supplementary materials. 

\color{black}

For posterior inference, the covariance function is estimated with the posterior mean of $\boldsymbol{H}'_{P}(s) \boldsymbol{U}_K' \boldsymbol{\Omega}^{(r)} \boldsymbol{U}_K \boldsymbol{H}_{P}(t)$, with the resampled particles $\boldsymbol{\Omega}^{(r)}$ from the final iteration in ASMC. The principal components $\hat{\phi}_{k}(t)$ are estimated by decomposing the estimated posterior mean of the covariance function. The corresponding FPC scores are estimated with $\hat{\xi}_{ik} = $ \(\int_{\mathcal{T}} \hat{\phi}_{k}(t) Y_i(t) \,dt\).

\color{black}
\subsection{Extension to Finite Mixture Models}
\label{sec:mixture_models}

In this section, we extend the proposed RB-FPCA model to include mixture models, which is a powerful technique for integrating multiple data generation processes within a single framework. When observations arise from two or more underlying groups, a mixture model can model the different groups in the data with different components. We adopt a finite mixture model of the form
\begin{equation}
\boldsymbol{Y}_i \sim \sum_{f=1}^{F} \pi_{f} \mathcal{M}_{f} (\boldsymbol{Y}_i \: | \: \boldsymbol{\theta}_{f}),  \quad  i = 1, \ldots ,n,
\end{equation}
where $\{\mathcal{M}_{f} ( \cdot \: | \: \boldsymbol{\theta}_{f})\}_{f=1}^{F}$ is a set of parametric distributions, $ \boldsymbol{\theta}_{f}$ is the set of parameters specific to component $f = 1, \ldots ,F$, and $\pi_{f}$ is the associated weight for each component with $\sum_{f=1}^{F} \pi_{f} = 1$. Essentially, a mixture model is a combination of multiple statistical distributions, each representing a distinct underlying population \cite[see][for a comprehensive review of mixture models]{fruhwirth2019handbook}. To facilitate the inference, an auxiliary variable $\boldsymbol{\tau} = (\tau_1, \ldots , \tau_n )$ is introduced to assign each observation to a component. Each variable $\tau_i$, $i = 1, \ldots ,n$, takes values in the set $\{ 1, \ldots ,F \}$. Thus, the mixture model can also be represented as follows: 
\begin{equation}
\boldsymbol{Y}_i \: | \: \tau_i \sim \mathcal{M}_{\tau_i} (\boldsymbol{Y}_i \: | \: \boldsymbol{\theta}_{\tau_i}),  \quad  i = 1, \ldots ,n.
\end{equation}

In our context, $\mathcal{M}_{\tau_i} (\boldsymbol{Y}_i \: | \: \boldsymbol{\theta}_{\tau_i})$ can take various forms, including multivariate normal distribution, multivariate skew $t$ distribution, and multivariate skew normal distribution. Here, we focus on the case with two components, where each component represents a distinct data generation process. For illustration purposes, we consider a mixture of multivariate skew normal distributions and multivariate skew $t$ distributions. In this special case, the mixture model can be represented as follows:
\begin{equation}
\boldsymbol{Y}_i \sim \pi_{1} \mathcal{M}_{1} (\boldsymbol{Y}_i \: | \: \boldsymbol{\theta}_{1}) + (1 - \pi_{1}) \mathcal{M}_{2} (\boldsymbol{Y}_i \: | \: \boldsymbol{\theta}_{2}),  \quad  i = 1, \ldots ,n,
\end{equation}
where $\mathcal{M}_{1} (\boldsymbol{Y}_i \: | \: \boldsymbol{\theta}_{1})$ is  $\text{MVN}_{m}(\boldsymbol{H}_P \boldsymbol{U}_K \boldsymbol{\beta}_{i,1} + \boldsymbol{D}_1\boldsymbol{z}_{i,1}, \: \boldsymbol{\Sigma}_1)$,  $\mathcal{M}_{2} (\boldsymbol{Y}_i \: | \: \boldsymbol{\theta}_{2})$ is $\text{MVT}_{m}(\boldsymbol{H}_P \boldsymbol{U}_K \boldsymbol{\beta}_{i,2} + \boldsymbol{D}_2\boldsymbol{z}_{i,2}, \: \boldsymbol{\Sigma}_2/w_{i,2})$, and $0 \leq \pi_{1} \leq 1$. We construct the mixture so that $\boldsymbol{\theta}_{1}  = \{ \boldsymbol{\beta}_{i,1}, \boldsymbol{D}_1, \boldsymbol{z}_{i,1}, \boldsymbol{\Sigma}_1 \}$ and $\boldsymbol{\theta}_{2}  = \{ \boldsymbol{\beta}_{i,2}, \boldsymbol{D}_2, \boldsymbol{z}_{i,2}, \boldsymbol{\Sigma}_2, w_{i,2} \}$ are parameters in the multivariate skew normal and multivariate skew $t$ distributions, respectively. We denote this model as  RB-FPCA-MM.

We adopt a Beta prior on $\pi_{1}$, i.e. $\pi_{1} \sim \text{Beta}(1,1)$, which represents a flat, uninformative prior over the weight parameter. This leads to conditional independence between $\boldsymbol{\theta}_{1}$, $\boldsymbol{\theta}_{2}$ and $\pi_{1}$, conditional on $\boldsymbol{\tau}$. The conditional posterior distributions for parameters in this model are provided in the supplementary material. In Bayesian mixture models, a common challenge known as ``label switching" \citep{stephens2000dealing} arises due to the symmetry in the likelihood of the model parameters. To address this issue, we remap the mixture component labels to a unique canonical space at each sampling step \citep{10.1214/16-EJS1163}, thereby preventing label switching.
\color{black}

\subsection{Outlier Detection }
\label{sec:outlier_detection}

Outlier detection is among the subsequent tasks following the acquisition of a robust estimate of the covariance function through the posterior mean derived from all particles. The corresponding FPC scores can be used to detect any potential outlying trajectories. In this work, we employ the approach described in \cite{boente2021robust} to conduct a robust outlier detection for data applications. The first step is calculating the robust estimates for multivariate location and scatter of the estimated FPC scores. These estimates, named MM-estimates proposed by \cite{yohai1987high}, have a high breakdown point while remaining efficient. The breakdown point is a measure of robustness, defined as the maximum proportion of contamination or atypical points the data may contain while the estimator remains informative about the underlying parameter. Thus, estimators with higher breakdown points are desired. Then we compute the robust Mahalanobis distance, which is defined as
\begin{center}
    $d_{\text{Mahalanobis}}(\hat{\boldsymbol{\xi}}_{i\cdot}) = \sqrt{(\hat{\boldsymbol{\xi}}_{i\cdot} - \hat{\boldsymbol{\mu}}_{MM})' \hat{\boldsymbol{\Sigma}}_{MM}^{-1} (\hat{\boldsymbol{\xi}}_{i\cdot} - \hat{\boldsymbol{\mu}}_{MM})}$,
\end{center}
where $\hat{\boldsymbol{\xi}}_{i\cdot} = (\hat{\xi}_{i1}, \ldots, \hat{\xi}_{iK})'$, $\hat{\boldsymbol{\mu}}_{MM}$ and $\hat{\boldsymbol{\Sigma}}_{MM}$ denote the MM-estimators for multivariate location and scatter, respectively. The trajectory $i$ is flagged as an outlier if its FPC scores have a distance larger than a threshold quantile of a $\chi^2_K$ distribution. Different threshold values can be applied to detect mild and extreme outliers.

Alternatively, the utilization of the posterior distribution of the estimated covariance function provides a way for identifying outliers and quantifying uncertainties. One suggested approach involves employing all particles $\boldsymbol{\Omega}^{(r)}$ for estimation. For each particle, the covariance function $\boldsymbol{H}'_{P}(s) \boldsymbol{U}_K' \boldsymbol{\Omega}^{(r)} \boldsymbol{U}_K \boldsymbol{H}_{P}(t)$ and the corresponding FPC scores $\hat{\xi}_{ik}^{(r)}$ are estimated, and outlying trajectory $i$ are identified based on distances exceeding a threshold quantile of a $\chi^2_K$ distribution. Extending this process to all particles yields a point estimate of the likelihood of a trajectory being an outlier.

\subsection{Extend to Sparse Longitudinal Data}
\label{sec:model_sparse}

In this section, we extend the robust Bayesian FPCA model in \ref{sec:model_prior} to fit longitudinal data that are sparsely and irregularly observed. We assume the observations are contaminated by measurement errors and adopt the FDA model given in Equation \ref{eq:fda_model}. Now we consider the case when the number of measurements $n_{i}$ made per subject is random due to the sparse and irregular designs. We also assume that the random variables $n_{i}$ are i.i.d. The sparse data are centered before fitting the RB-FPCA model by subtracting the estimated mean function $\hat{\mu}(t)$ based on the pooled data from all individuals.

Write $\boldsymbol{Y}_{i} = (Y_i(t_{i1}),\ldots,Y_i(t_{in_{i}}))'$. \textcolor{black}{The hierarchical representation of the proposed RB-FPCA model with multivariate skew $t$ distribution for $i=1,\ldots,n$ with sparse longitudinal data is
\begin{align*} 
\boldsymbol{Y}_{i}  &\sim \text{MVN}_{n_i}(\boldsymbol{H}_P^{(i)} \boldsymbol{U}_K^{(i)} \boldsymbol{\beta}_{i} + \boldsymbol{D}^{(i)}\boldsymbol{z}^{(i)}, \: \boldsymbol{\Sigma}^{(i)} / w^{(i)}),\\ 
\boldsymbol{\beta}_{i} &\stackrel{i.i.d}{\sim}  \text{MVN}_{K}(\boldsymbol{0}, \: \boldsymbol{\Omega}), \\
\boldsymbol{\Omega}^{-1} &\sim \text{Wishart}_K(\nu, \: \boldsymbol{L}_k^{-1}), \\
\boldsymbol{z}^{(i)} &\sim \text{MVN}_{n_i}(\boldsymbol{0}, \boldsymbol{I})I(\boldsymbol{z}^{(i)} > 0), \\
\text{vec}(\boldsymbol{D}^{(i)}) &\sim \text{MVN}_{n_i}(\boldsymbol{0}, \boldsymbol{\Gamma}^{(i)}), \\
(\boldsymbol{\Sigma}^{(i)})^{-1} &\sim \text{Wishart}_{n_i}(2r, 2\boldsymbol{\kappa}^{(i)}), \\
w^{(i)} &\sim \text{Gamma}(\nu_{w^{(i)}}/2, \nu_{w^{(i)}}/2), \\
\nu_{w^{(i)}} &\sim \text{Gamma}(1, 0.1)I (\nu_{w^{(i)}} > 2), 
\end{align*}
}
where the superscript $(i)$ represents the parameters whose values and dimensions depend on the curve index $i$. The value of $\boldsymbol{L}_k^{-1}$ is set to be the element-wise average over matrices $(\boldsymbol{L}_k^{(i)})^{-1}$ for $i=1,\ldots,n$. \textcolor{black}{For the multivariate skew normal distribution, each of $w^{(i)}$ will be set to 1 and the last two specifications for $w^{(i)}$ and $\nu_{w^{(i)}}$ are omitted.}
The full conditional distributions for the sparse longitudinal data have similar forms. To ensure the matrix $\boldsymbol{\Psi}$ described in the full conditional distribution of $\boldsymbol{\Omega}^{-1}$ is valid, we assume the number of principal components is less than or equal to the number of basis functions, and the number of basis functions used for approximation is less than or equal to the minimum number of observations per subject, i.e., $K \leq P \leq \min(1,\ldots,n_{i})$.

The posterior covariance function is $\boldsymbol{H}'_{P}(s) \boldsymbol{U}_K' \boldsymbol{\Omega}^{(r)} \boldsymbol{U}_K \boldsymbol{H}_{P}(t)$, where $\boldsymbol{H}_{P}$ and $\boldsymbol{U}_K$ are constructed using a given number of support points in each direction of the covariance surface. The default value of support points is set as 51 (same as in \texttt{fdapace} package in \texttt{R}). The principal components are $\boldsymbol{C}_{KP}\boldsymbol{H}'_{P}(t)$. The traditional way to estimate the FPC scores with numerical integration is used and performs well when a sufficiently dense measurement grid for each subject is available. However, when dealing with sparse longitudinal data, the numerical integration approximated by sums could not produce reasonable approximations due to the sparseness of the data. Thus, the alternative PACE method described by \cite{yao2005functional} is implemented here to find FPC scores for sparse longitudinal data. Specifically, using the data from each subject, the prediction of the FPC scores for the $i$th subject is given by the conditional expectation:
\begin{equation}\label{eq:fpc_scores_sparse}
    \hat{\xi}_{ik} = \hat{\mathbb{E}}[\xi_{ik} | \boldsymbol{Y}_{i}] = \hat{\lambda}_k \hat{\phi}'_{ik}(t) \hat{Q}^{-1}_{\boldsymbol{Y}_{i}} \boldsymbol{Y}_{i},
\end{equation}
where $\hat{\lambda}_k$ is the $k$th estimated eigenvalues of $\hat{Q}(s,t)$, $\hat{\phi}_{ik}(t)$ is the estimates of the eigenfunctions $\phi_{ik}(t) = (\phi_{k}(t_{i1}),\ldots,\phi_{k}(t_{in_{i}}))'$, and the $(j,l)$th element of $\hat{Q}_{\boldsymbol{Y}_{i}}$ is $(\hat{Q}_{\boldsymbol{Y}_{i}})_{j,l} = \hat{Q}(t_{ij}, t_{il}) + \boldsymbol{\hat{\Sigma}}^{(i)}$. This conditioning method works with sparse data with the existence of measurement errors and gives the best prediction of the FPC scores under Gaussian assumptions. The estimate $\hat{\xi}_{ik}$ in Equation \ref{eq:fpc_scores_sparse} is the best linear prediction of $\xi_{ik}$ from the information in the $i$th subject, regardless of whether the Gaussian assumption holds (as previously observed in \cite{yao2005functional}).

\section{Simulation Studies}
\label{sec:simulation}

\color{black}
In the simulation studies, we evaluate the proposed model for dense functional data in the presence of two types of outliers. The first type arises from heavy tails in the noise process, resulting in a few observations with extreme values. The second type of outlier is characterized by unusual weights on the principal components, causing variations in the curve patterns. In the first simulation study, we investigate the method's performance in selecting the optimal model when the true underlying data generation model is known. In simulations II to IV, we assess the model's performance under conditions where the two types of outliers exist independently as well as in combination. The final simulation study examines sparse functional data containing outliers generated from various distributions, including symmetric and skewed distributions.

\subsection{\textcolor{black}{Simulation I: Comparison of Model Fit}}
\label{sec:sim1}

In the first simulation, we evaluate the performance of different models. Each simulated dataset consists of 20 noisy observations measured at 50 evenly spaced time points within the interval $[ -1, 1 ]$. We adopted $\mu(t) = \text{sin}(2 \pi t)$ as the underlying true mean function and chose the linear covariance function $\text{Cov}_2(s,t) = \text{min} \{s+1, t+1\}$ as the true underlying covariance function. We simulated three datasets: Setting 1, Setting 2, and Setting 3. In Setting 1, the true underlying model follows a multivariate skew normal distribution; in Setting 2, it follows a multivariate skew $t$ distribution; and in Setting 3, it combines a half multivariate skew normal distribution and half multivariate skew $t$ distribution. We assessed the performance of each model when different forms of independent noises were added during the sampling process. Specifically, we considered normal noises, $N(0, 0.3)$; skew normal noises with location 0, scale 1, and shape 5, denoted as $SN(0, 1, 5)$; and skew $t$ noises with location 0, scale 1, shape 5, and 5 degrees of freedom, denoted as $ST(0, 1, 5, 5)$. The three methods for comparison are listed below:
\begin{enumerate}
    \item our proposed robust Bayesian FPCA (RB-FPCA) method with three candidate models: RB-FPCA-SN, RB-FPCA-ST, and RB-FPCA-MM
    \item Fast Covariance Estimation (FACE) method of \cite{xiao2016fast}, implemented in \texttt{refund} package in \texttt{R}
    \item Principal Analysis by Conditional Estimation (PACE) method of \cite{wang2016functional}, implemented in \texttt{fdapace} package in \texttt{R}
\end{enumerate}

We established the values for the hyperparameters as outlined in \cite{sahu2003new}. Specifically, we set $\nu=2K$ for $\boldsymbol{\Omega}^{-1}$ in Equation \ref{eq:cond_dist_Omega}, $\boldsymbol{\Gamma} = \text{diag}(10,\dots,10)$ for $\text{vec}(\boldsymbol{D})$ in Equation \ref{eq:cond_dist_vecD}, $2r = m$ and $\boldsymbol{\kappa} = 100 \boldsymbol{R}^{-1}/(2r)$ for $\boldsymbol{\Sigma}^{-1}$ in Equation \ref{eq:cond_dist_Sigma}. Here, $\boldsymbol{R}$ is a diagonal matrix of dimensions $m \times m$, where the main diagonal elements are the squared ranges of the corresponding components in the data. In the ASMC, the number of particles is set to 200. The resampling threshold is 0.5, and the threshold for determining the sequence of annealing parameters is 0.9. These chosen hyperparameter and tuning parameter values yield satisfactory results in the simulation studies and are consistently employed in all examples unless explicitly stated otherwise. We predetermined the numbers of basis functions and eigenfunctions as $P = 15$ and $K = 5$. The prior covariance function was set to be the same as the true underlying covariance function. Each setting was repeated 30 times with different random seeds, and the results are reported as the average of the estimated posterior means across 30 runs.

The estimated marginal likelihood for each model is computed to facilitate model comparison. This measure represents the probability of the data given different models and their parameters, with higher values indicating a better fit. As shown in Table \ref{tab:sim1-logZ}, the correct models are identified in most cases. However, when the true model is ST with normal noises or skew $t$ noises, RB-FPCA tends to favor SN model, with ST being recognized as the second-best fit. In addition, to validate the estimation accuracy of each model, we compared the correlation estimates from the proposed Bayesian method with those obtained from two frequentist methods. The results, summarized in Table \ref{tab:sim1-corr_est}, indicate that the correct models consistently yield the most accurate estimations, as measured by the distances between the estimates and the true correlation function using the $L_2$ norm. It is also observed that RB-FPCA-SN model performs comparably to the best model in most cases, demonstrating both robustness and computational efficiency. Consequently, we propose implementing RB-FPCA-SN model for all subsequent simulation studies unless otherwise specified. For simplicity in notation, we will refer to the method utilizing the SN model as RB-FPCA throughout all subsequent simulations unless stated otherwise.

\begin{table}[htbp]
\centering
\small
\begin{tabular}{|c|c|c|c|c|}
\hline
\begin{tabular}[c]{@{}c@{}}True\\ Data\\ Model\end{tabular} & Noise & \begin{tabular}[c]{@{}c@{}}RB-FPCA-SN\end{tabular} & \begin{tabular}[c]{@{}c@{}}RB-FPCA-ST\end{tabular} & \begin{tabular}[c]{@{}c@{}}RB-FPCA-MM\end{tabular} \\ \hline
\multirow{3}{*}{SN} & $N(0,0.3)$ & \textbf{-636.004} & -685.938 & -654.820 \\ 
 & $SN(0,1,5)$ & \textbf{-713.534} & -753.233 & -716.121 \\  
 & $ST(0,1,5,5)$ & \textbf{-891.581} & -906.040 & -894.222 \\ \hline
\multirow{3}{*}{ST} & $N(0,0.3)$ & \textbf{-676.483} & -711.990 & -793.727 \\
 & $SN(0,1,5)$ & -922.145 & \textbf{-922.139} & -930.065 \\ 
 & $ST(0,1,5,5)$ & \textbf{-1064.57} & -1068.989 & -1088.487 \\ \hline
\multirow{3}{*}{MM} & $N(0,0.3)$ & -720.071 & -743.633 & \textbf{-694.378} \\
 & $SN(0,1,5)$ & -788.024 & -812.789 & \textbf{-754.130} \\
 & $ST(0,1,5,5)$ & -932.673 & -952.157 & \textbf{-919.148} \\ \hline
\end{tabular}
\caption{\textcolor{black}{Simulation I: Marginal likelihood (in log scale) estimates for different models.}  \label{tab:sim1-logZ}}
\end{table}

\begin{table}[htbp]
\centering
\small
\begin{tabular}{|c|c|c|c|c|c|c|}
\hline
\begin{tabular}[c]{@{}c@{}}True\\ Data\\ Model\end{tabular} & Noise & \begin{tabular}[c]{@{}c@{}}RB-FPCA\\-SN\end{tabular} & \begin{tabular}[c]{@{}c@{}}RB-FPCA\\-ST\end{tabular} & \begin{tabular}[c]{@{}c@{}}RB-FPCA\\-MM\end{tabular} & FACE & PACE \\ \hline
\multirow{3}{*}{SN} & $N(0,0.3)$ & \textbf{5.640} & 5.656 & 7.350 & 6.065 & 10.889 \\ 
 & $SN(0,1,5)$ & \textbf{5.614} & 5.621 & 7.101 & 6.343 & 11.851 \\ 
 & $ST(0,1,5,5)$ & \textbf{5.654} & 5.657 & 6.434 & 6.344 & 15.993 \\ \hline
\multirow{3}{*}{ST} & $N(0,0.3)$ & 5.683 & \textbf{5.665} & 7.030 & 7.929 & 11.425 \\ 
 & $SN(0,1,5)$ & 5.653 & \textbf{5.651} & 6.704 & 8.140 & 12.598 \\ 
 & $ST(0,1,5,5)$ & 5.721 & \textbf{5.695} & 6.154 & 7.811 & 14.957 \\ \hline
\multirow{3}{*}{MM} & $N(0,0.3)$ & 5.608 & 5.642 & \textbf{5.106} & 6.850 & 11.562 \\ 
 & $SN(0,1,5)$ & 5.636 & 5.661 & \textbf{5.101} & 6.849 & 11.054 \\ 
 & $ST(0,1,5,5)$ & 5.674 & 5.672 & \textbf{5.369} & 6.851 & 14.808 \\ \hline
\end{tabular}
\caption{\textcolor{black}{Simulation I: Estimations of the correlation function are evaluated by the distances between the estimates and the true correlation function using the $L_2$ norm.}  \label{tab:sim1-corr_est}}
\end{table}

\color{black}

\subsection{\textcolor{black}{Simulation II: Dense Functional Data with Outliers from Different Noise Distributions}}
\label{sec:sim2}

In the second simulation, we aim to compare the performance of the proposed RB-FPCA method with various frequentist and Bayesian FPCA methods. We followed a similar simulation setting as in \cite{suarez2017bayesian}. Each set of simulated data consists of \textcolor{black}{100} noisy observations over 50 time points. These time points are evenly spaced in the interval $[ -1, 1 ]$. The underlying true mean function is $\mu(t) = \text{sin}(2 \pi t)$. Depending on the experimental settings, the true underlying covariance function is of either $\text{Cov}_1(s,t) = \exp\{-3(t-s)^2\}$ or $\text{Cov}_2(s,t) = \text{min} \{s+1, t+1\}$. \textcolor{black}{Next, several forms of independent noises are considered during the sampling process. For symmetric noises, we included normal noises $N(0, 0.3)$ and $t$-distributed noises $t(5)$. For skewed noises, we incorporated skew normal noises $SN(0, 1, 5)$, and skew $t$ noises $ST(0, 1, 5, 5)$. 
}

\textcolor{black}{For the performance comparison, in addition to the two frequentist methods outlined in Section \ref{sec:sim1}, we also included the Bayesian FPCA (BFPCA) method of \cite{suarez2017bayesian}.} In RB-FPCA and BFPCA methods, we predetermined the numbers of basis functions and eigenfunctions as $P = 15$ and $K = 5$. \textcolor{black}{For BFPCA, a total of 50,000 Gibbs sampling iterations was run, and the first 30,000 iterations were discarded as the burn-in.} We compared the methods in the following aspects with different metrics: estimation of the covariance function with $L_2$ norms and estimation of the principal components with mean squared errors. Each simulation represents a choice between $\text{Cov}_1(s,t)$ and $\text{Cov}_2(s,t)$ as the true covariance function and the prior covariance function, which results in four experimental settings. Each setting was repeated 30 times with different random seeds and the results are the average of the estimated posterior means from 30 runs.

\begin{table}[htbp]
\small
\begin{tabular}{|c|c|c|c|c|c|c|c|}
\hline
Noise & Truth & \begin{tabular}[c]{@{}c@{}}RB-FPCA\\ Prior \\ Cov1\end{tabular} & \begin{tabular}[c]{@{}c@{}}RB-FPCA\\ Prior \\ Cov2\end{tabular} & \begin{tabular}[c]{@{}c@{}}BFPCA\\ Prior \\ Cov1\end{tabular} & \begin{tabular}[c]{@{}c@{}}BFPCA\\ Prior \\ Cov2\end{tabular} & FACE & PACE \\ \hline
\multirow{2}{*}{$N(0,0.3)$} & Cov1 & 5.440 & 8.832 & \textbf{5.139} & 8.204 & 5.423 & 5.571 \\ 
 & Cov2 & 12.732 & 9.412 & 12.672 & 9.262 & \textbf{9.159} & 10.737\\ \hline
\multirow{2}{*}{$SN(0,1,5)$} & Cov1 & \textbf{5.285} & 7.608 & 5.614 & 8.160 & 6.396 & 9.340 \\ 
 & Cov2 & 7.937 & \textbf{7.176} & 8.179 & 7.839 & 8.501 & 11.400 \\ \hline
\end{tabular}
\caption{\textcolor{black}{Simulation II: Estimations of the covariance function are evaluated by the distances between the estimates and the true covariance function using the $L_2$ norm.}  \label{tab:sim2-cov_est}}
\end{table}

\begin{figure}[htbp]
  \centering
  \subfloat[]{\includegraphics[scale = 0.6]{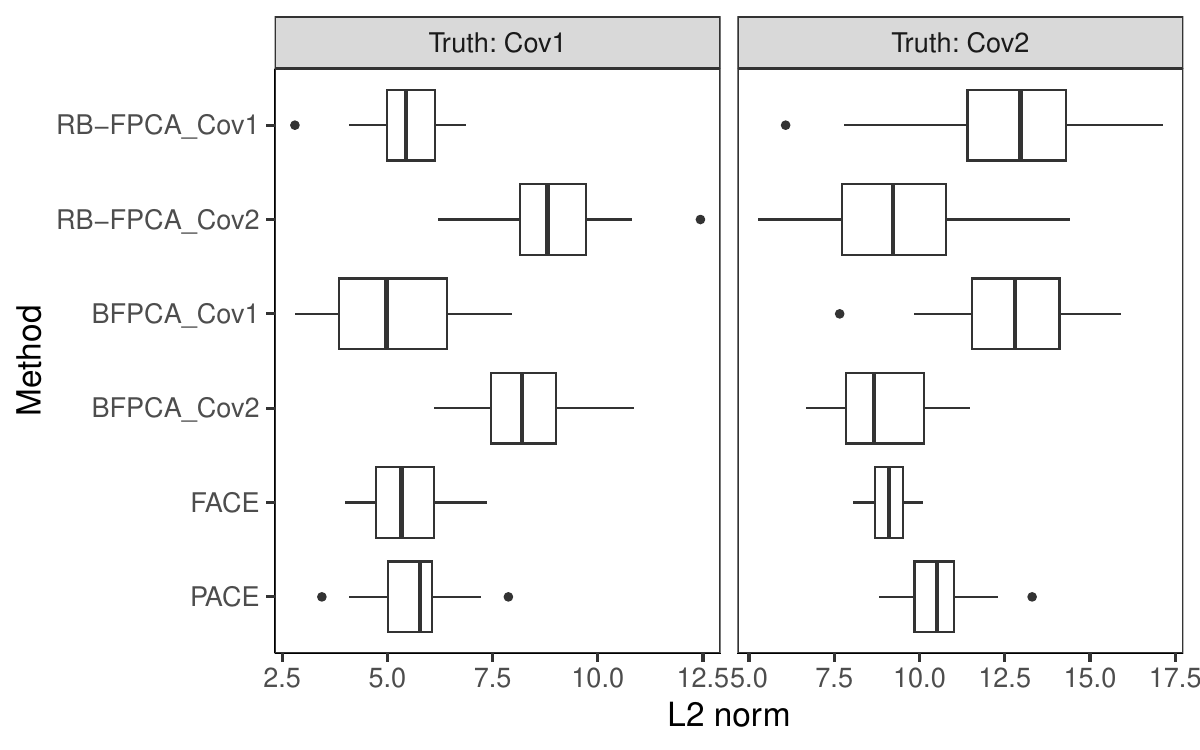}\label{fig:sim2-est_cov-n}} \\
  \subfloat[]{\includegraphics[scale = 0.6]{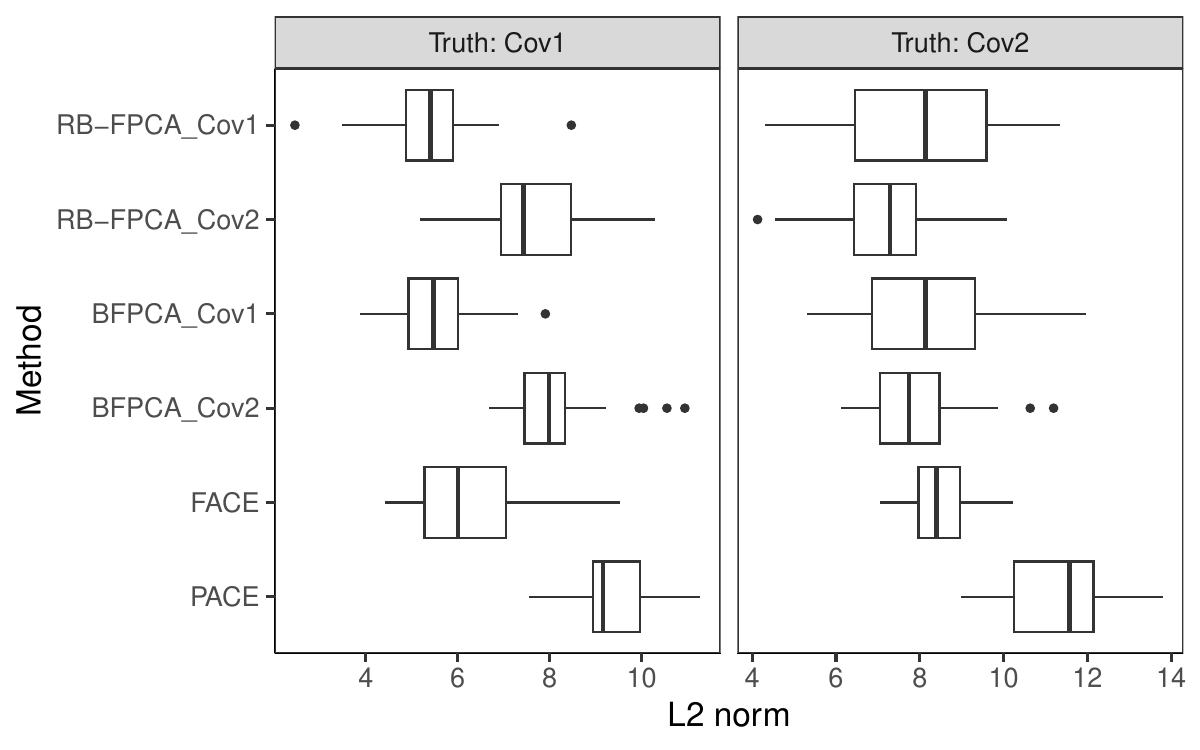}\label{fig:sim2-est_cov-sn}}
  \caption{\textcolor{black}{Simulation II: Boxplots of the distance in the $L_2$ norm between the estimates and the true covariance function. (a) $N(0,0.3)$ noises. (b) $SN(0,1,5)$ noises.}} \label{fig:sim2-est_cov}
\end{figure}

\begin{table}[htbp]
\small
\begin{tabular}{|c|c|c|c|c|c|c|c|c|}
\hline
Noise & PC & Truth & \begin{tabular}[c]{@{}c@{}}RB-\\ FPCA\\ Prior \\ Cov1\end{tabular} & \begin{tabular}[c]{@{}c@{}}RB-\\ FPCA\\ Prior \\ Cov2\end{tabular} & \begin{tabular}[c]{@{}c@{}}BFPCA\\ Prior \\ Cov1\end{tabular} & \begin{tabular}[c]{@{}c@{}}BFPCA\\ Prior \\ Cov2\end{tabular} & FACE & PACE \\ \hline
\multirow{6}{*}{$N(0,0.3)$} & \multirow{2}{*}{PC1} & Cov1 & 0.039 & 0.054 & \textbf{0.032} & 0.039 & 0.048 & 0.055 \\ 
 &  & Cov2 & 0.040 & \textbf{0.033} & 0.056 & 0.038 & 0.040 & 0.057 \\
 & \multirow{2}{*}{PC2} & Cov1 & 0.024 & 0.042 & \textbf{0.018} & 0.025 & 0.032 & 0.026 \\ 
 &  & Cov2 & 0.065 & 0.050 & 0.072 & 0.048 & \textbf{0.039} & 0.040 \\ 
 & \multirow{2}{*}{PC3} & Cov1 & \textbf{0.013} & 0.022 & 0.026 & 0.040 & 0.040 & 0.018 \\ 
 &  & Cov2 & 0.028 & 0.023 & 0.037 & \textbf{0.019} & 0.063 & 0.026 \\ \hline
\multirow{6}{*}{$SN(0,1,5)$} & \multirow{2}{*}{PC1} & Cov1 & 0.049 & 0.053 & \textbf{0.025} & 0.060 & 0.046 & 0.029 \\ 
 &  & Cov2 & 0.030 & \textbf{0.028} & 0.036 & 0.033 & 0.036 & 0.029 \\ 
 & \multirow{2}{*}{PC2} & Cov1 & \textbf{0.030} & 0.034 & 0.031 & 0.034 & 0.036 & 0.039 \\ 
 &  & Cov2 & 0.063 & 0.047 & 0.056 & 0.045 & \textbf{0.041} & 0.044 \\ 
 & \multirow{2}{*}{PC3} & Cov1 & \textbf{0.024} & 0.040 & 0.026 & 0.046 & 0.038 & 0.042 \\ 
 &  & Cov2 & 0.025 & \textbf{0.017} & 0.031 & 0.029 & 0.037 & 0.040 \\ \hline
\end{tabular}
\caption{\textcolor{black}{Simulation II: Estimations of the first 3 principal components are compared with the Mean Squared Errors (MSEs) between the estimates and true PCs.} \label{tab:sim2-PCs_est}}
\end{table}

Table \ref{tab:sim2-cov_est} and Figure \ref{fig:sim2-est_cov} demonstrate the efficiency of the proposed RB-FPCA method in estimating covariance functions \textcolor{black}{with $N(0,0.3)$ and $SN(0, 1, 5)$ noises.} RB-FPCA performs competitively against the other Bayesian method and the two frequentist methods when the true covariance function is selected as the prior covariance function. When the true covariance function, $\text{Cov}_1(s,t)$, is selected as the prior function, RB-FPCA outperforms other methods in covariance estimation \textcolor{black}{23\% of the time with $N(0,0.3)$ noises, and 47\% with $SN(0, 1, 5)$ noises.} If RB-FPCA with $\text{Cov}_1(s,t)$ as prior is excluded from the comparison, RB-FPCA with $\text{Cov}_2(s,t)$ as prior ranks as the best performer \textcolor{black}{7\% of the time with $N(0,0.3)$ noises, and 10\% with $SN(0, 1, 5)$ noises.} When the true covariance function, $\text{Cov}_2(s,t)$, is selected as the prior, RB-FPCA exhibits the best performance \textcolor{black}{43\% of the time with $N(0,0.3)$ noises, and 33\% with $SN(0, 1, 5)$ noises.} Conversely, when employing the alternative covariance function as the prior, RB-FPCA surpasses other methods \textcolor{black}{10\% of the time with $N(0,0.3)$ noises, and 33\% with $SN(0, 1, 5)$ noises}, excluding RB-FPCA with true covariance function as the prior from the comparison. 

\textcolor{black}{In general, when $N(0,0.3)$ noises are added to the data, the proposed RB-FPCA performs comparably to BFPCA in estimating covariance functions. Both Bayesian methods estimate the covariance functions well when the prior covariance function is correctly specified as the true covariance function. When skewed noises $SN(0,1,5)$ are added, our RB-FPCA method demonstrates notable advantages. It exhibits reduced variability in covariance estimation and provides more accurate estimations of the covariance function and principal components, even when the assumed prior covariance function differs from the true covariance function. The estimations of the principal components measured by MSEs in Table \ref{tab:sim2-PCs_est} also indicate that the proposed RB-FPCA method can provide comparable performance to the BFPCA and some well-known frequentist methods. Results with $t(5)$ and $ST(0, 1, 5, 5)$ noises are summarized in the supplementary material. Specifically, visualizations of data and noises are given, and estimations of the covariance functions and the first 3 principal components are provided. These analyses and results offer further insights into the robustness of the proposed model under varying conditions.
} Generally, constructing a prior covariance function similar to the true covariance function can enhance estimation. Some certain degrees of subject matter expertise in the structure of the covariance function can be incorporated into the analysis to improve the estimation further.

\subsection{\textcolor{black}{Simulation III: Dense Functional Data with Outliers from Different Principal Components' Weights}}
\label{sec:sim3}

\textcolor{black}{The third simulation considers the presence of outliers arising from different principal components' weights.} The goal is to investigate the finite-sample performance and robustness of the proposed model. Different percentages of atypical observations are considered. A total of 100 observations over 50 time points are generated in each case.

For clean observations, the data are generated from the following model:
\begin{equation} \label{formula:sim3_dgp}
    Y_i(t) = \mu(t) + \sum_{k=1}^K \sqrt{\lambda_{k}} Z_{i,k} \phi_{k}(t), \; i = 1,\ldots,n,
\end{equation}
with $Z_{i,k} \stackrel{i.i.d}{\sim} \text{N}(0,1)$ and eigenvalues $\lambda_1 \geq \lambda_2 \geq \cdots \geq \lambda_K > 0$. Time points are evenly spaced in the interval $[ 0, 1 ]$. The mean function is set to $\mu(t) = 10\sin(2\pi t)\exp(-3t)$. We set $K = 4$ and choose the eigenfunctions $\phi_{k}(t)$ as the first $K$ eigenfunctions of the Matérn Covariance function \citep{rasmussen2003gaussian}:
\begin{align*}
    \text{Cov}_{\text{Matérn}}(s,t) = \sigma^2 \frac{2^{1-\nu}}{\Gamma(\nu)} \left( \frac{\sqrt{2\nu}|s-t|}{\rho} \right) ^{\nu} K_{\nu} \left( \frac{\sqrt{2\nu}|s-t|}{\rho} \right),
\end{align*}
where $\Gamma(\cdot)$ is the Gamma function and $K_{\nu}$ is a modified Bessel function of the second kind. We set the parameters as $\sigma^2 = 1$, $\rho = 3$, and $\nu = 1/2$. The eigenvalues $\lambda_1 = 0.83$, $\lambda_2 = 0.08$, $\lambda_3 = 0.03$, and $\lambda_4 = 0.015$ were selected so that they have similar ratios to those of the first four eigenvalues of the Matérn covariance function. In this model, the first principal direction summarizes major sources of variation among the curves. At the same time, the third and fourth eigenfunctions will tend to add some complexity to the covariance function.

\begin{figure}[htbp]
  \centering
  \subfloat[]{\includegraphics[scale = 0.35]{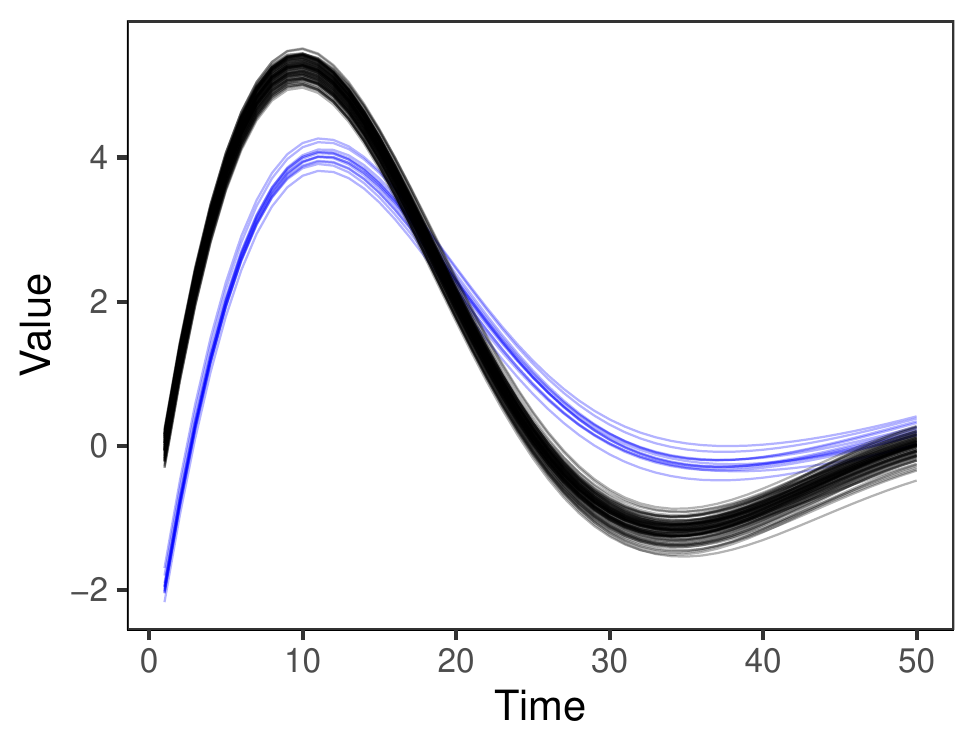}\label{fig:sim3-raw_data}} 
  \subfloat[]{\includegraphics[scale = 0.35]{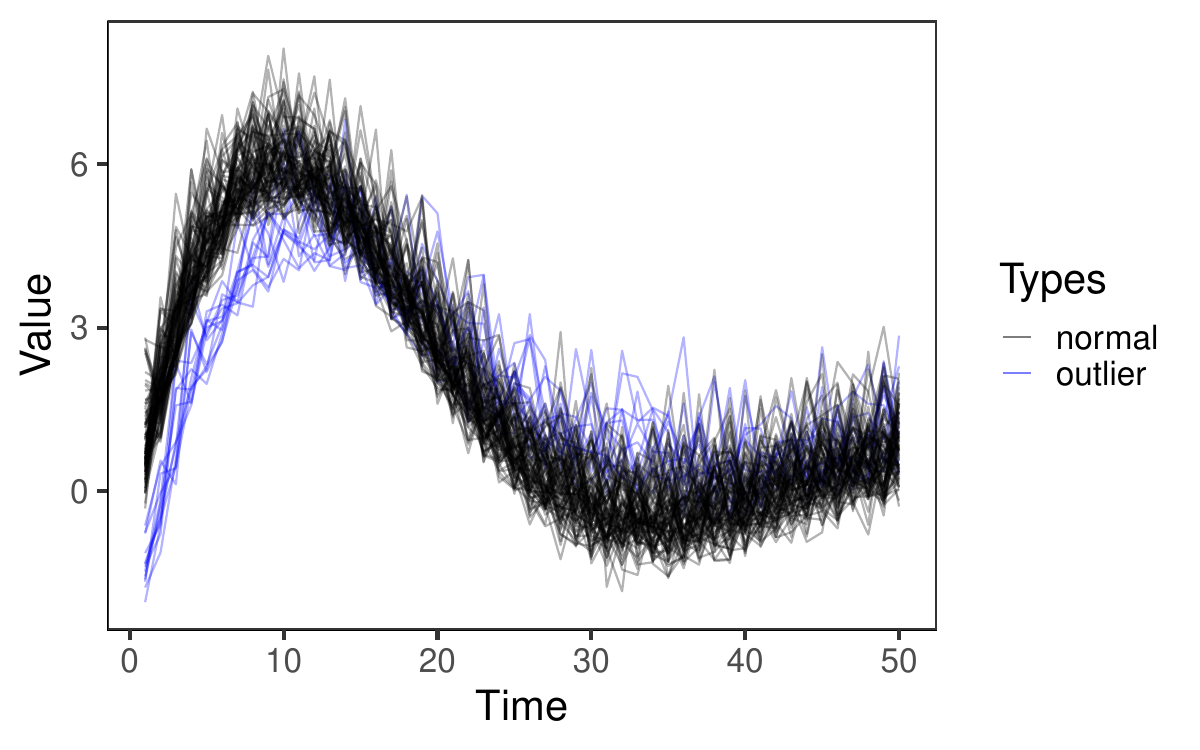}\label{fig:sim4-raw_data}}
  \caption{(a) Simulation III: Clean samples with contaminated samples from the data generation process described in \ref{sec:sim3}. The outlier percentage is 10\%. \textcolor{black}{(b) Simulation IV: Clean samples with contaminated samples from the data generation process described in \ref{sec:sim4}. $SN(0,1,5)$ noises are added. The outlier percentage is 10\%.}} 
\end{figure}

To add outliers to the clean observations, we introduced atypical observations to change the order of the principal directions to affect the estimated eigenvalues and eigenfunctions. We added outliers with a Bernoulli random variable $B_i \sim \text{Bernoulli}(1, p)$, where $p$ corresponds to the percentage of outliers. We contaminated the clean observations as follows:
\begin{enumerate}
    \item For curve $i$, generate the Bernoulli random variable $B_i$. If $B_i$ = 0, then generate the curve $Y_i(t)$ as described in Equation \ref{formula:sim3_dgp}.
    If $B_i$ = 1, then introduce outliers by altering the order of the principal directions. Sample the scores for the second and third principal directions with
        \begin{align*}
            \left(\begin{array}{c} Z_{i,2}\\ Z_{i,3}\end{array}\right) \sim \text{MVN}_2\left( \left(\begin{array}{c} 20 \\ 25 \end{array}\right), \begin{bmatrix}1/16 & 0 \\ 0 & 1/16 \end{bmatrix} \right).
        \end{align*}.        
    \item Repeat step $i$ for all curves, $i = 1,\ldots,n$.
\end{enumerate}
Figure \ref{fig:sim3-raw_data} shows one example of the 100 observations generated over 50 time points with an outlier percentage of $p=0.1$. This figure illustrates how contamination alters the pattern of the clean data. Black lines represent the clean samples, and blue lines correspond to 10 outlying samples.

We examined the behaviour of the estimated covariance function and the eigenfunctions for both clean and contaminated samples with $p = 0.05, 0.10, 0.15$. Each value of $p$ was repeated 30 times with different random seeds, and the final results are the average of the estimated posterior means from 30 runs. The prior covariance function used in this simulation is $\text{Cov}(s,t) = \exp\{-3(t-s)^2\}$. The numbers of basis functions and eigenfunctions are set to $P = 15$ and $K = 4$.

\begin{table}[htbp]
\small
\begin{tabular}{|c|c|c|c|c|c|c|}
\hline
                                                                                & $p$           & RB-FPCA                         & BFPCA   & FACE    & PACE    & \begin{tabular}[c]{@{}c@{}}RB-FPCA\\         outperforms others\end{tabular} \\ \hline
                                                                                & 0.05        & {\textbf{40.660}} & 44.057 & 44.082 & 44.074 & 93$\%$                    \\ 
                                                                                & 0.10        & {\textbf{40.450}} & 48.131 & 44.141 & 44.135 & 100$\%$                    \\ 
\multirow{-3}{*}{\begin{tabular}[c]{@{}c@{}}Cov\\ Function\end{tabular}} & 0.15        & {\textbf{38.249}} & 64.065 & 44.271 & 44.269 & 100$\%$                    \\ \hline
\end{tabular}
\caption{\textcolor{black}{Simulation III: Estimations of the covariance function are evaluated by the distances between the estimates and the true covariance function with the $L_2$ norms.}  \label{tab:sim3-cov_est}}
\end{table}

\begin{figure}[htbp]
    \centering
    \subfloat[]{\includegraphics[scale = 0.6]{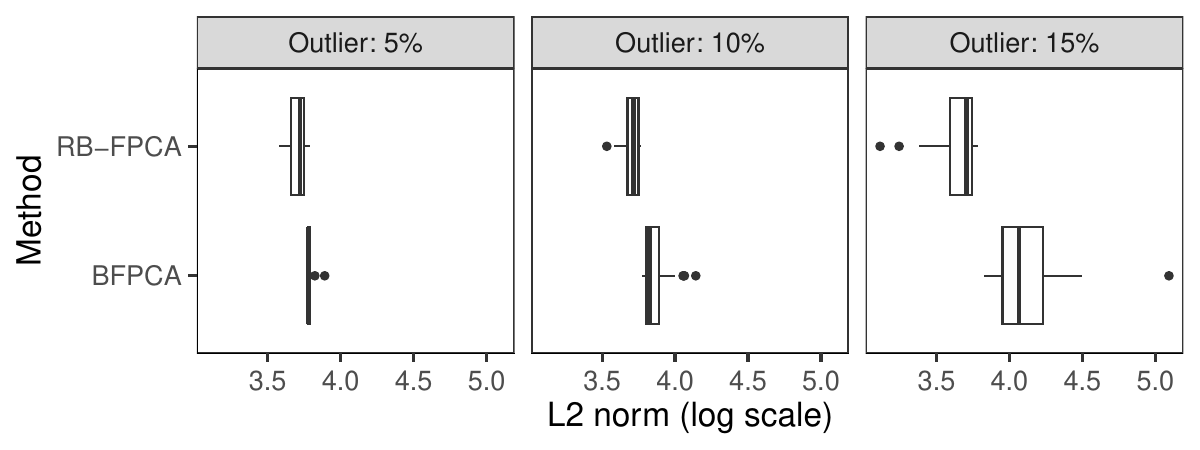}\label{fig:sim3-est_cov_1}} \\
    \subfloat[]{\includegraphics[scale = 0.6]{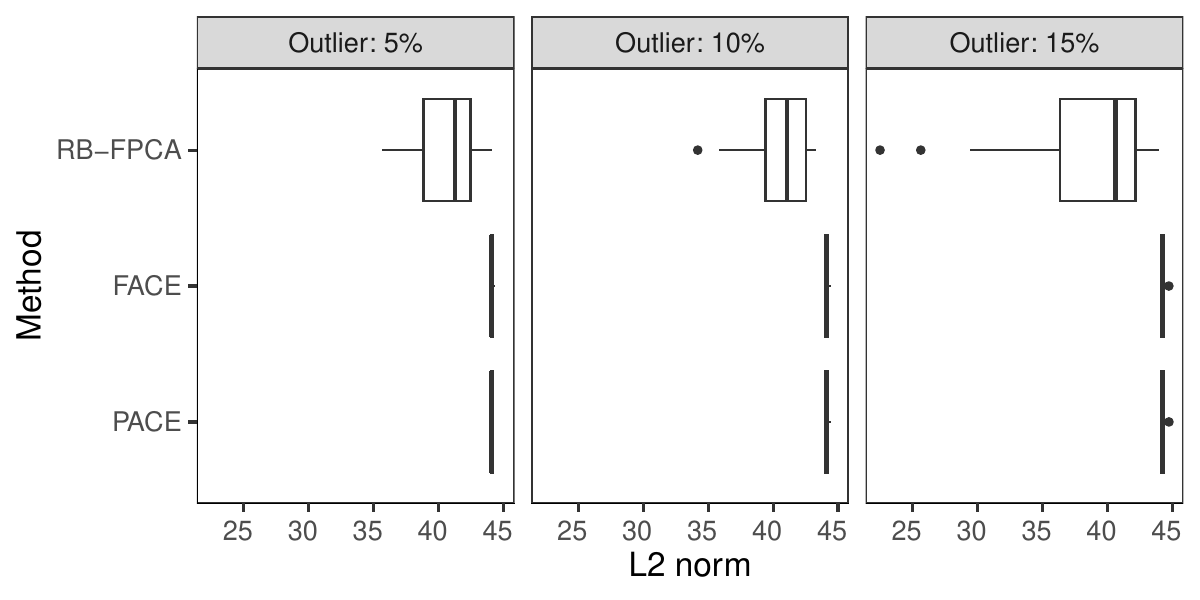}\label{fig:sim3-est_cov_2}}
    \caption{\textcolor{black}{Simulation III: (a) Boxplots of distances between the estimates and the true covariance function with the $L_2$ norms (in log scale). Comparison is between the proposed RB-FPCA method and the other Bayesian methods. (b) Boxplots of distances between the estimates and the true covariance function with the $L_2$ norms. Comparison is between the proposed RB-FPCA method and the two frequentist methods.} \label{fig:sim3-est_cov}}
\end{figure}

\begin{table}[htbp]
\small
\begin{tabular}{|c|c|c|c|c|c|c|}
\hline
                      & $p$                      & PCs & RB-FPCA                        & BFPCA                         & FACE                          & PACE                          \\ \hline
                      &                        & PC1 & {\textbf{1.376}} & 1.613                        & 1.601                        & 1.541                        \\ 
                      &                        & PC2 & {\textbf{1.280}} & 1.464                        & 1.590                        & 1.554                        \\ 
                      & \multirow{-3}{*}{0.05} & PC3 & \textbf{1.293}                        & 2.041                        & 2.495                        & 1.382 \\ 
                      &                        & PC1 & 1.552                        & 1.574                        & 1.593                        & {\textbf{1.549}} \\ 
                      &                        & PC2 & {\textbf{1.317}} & 1.503                        & 1.568                        & 1.552                        \\ 
                      & \multirow{-3}{*}{0.10} & PC3 & \textbf{1.163}                        & 2.234                        & 2.489                        & 1.507 \\ 
                      &                        & PC1 & {\textbf{0.998}} & 1.572                        & 1.584                        & 1.558                        \\ 
                      &                        & PC2 & {\textbf{1.260}} & 1.497                        & 1.569                        & 1.565                        \\ 
\multirow{-9}{*}{PCs} & \multirow{-3}{*}{0.15} & PC3 & \textbf{1.001}                        & 2.346                        & 2.487                        & 1.568 \\ \hline
\end{tabular}
\caption{\textcolor{black}{Simulation III: Estimations of the first 3 principal components are compared with angles (in radians) between the truth and estimates.} \label{tab:sim3-PCs_est}}
\end{table}

Table \ref{tab:sim3-cov_est} compares the estimations of the covariance function in terms of the distance between the estimated and the true covariance function, measured using the $L_2$ norm, with varying outlier percentages. These values are averaged across 30 replications. It can be observed that RB-FPCA shows significant improvement over BFPCA, providing a smaller error and a moderate improvement over the two modern frequentist methods. Figures \ref{fig:sim3-est_cov_1} and \ref{fig:sim3-est_cov_2} display the comparison of the boxplots of the distance between the estimated and true covariance functions. These boxplots confirm that RB-FPCA outperforms BFPCA in terms of median and dispersion and has a competitive performance compared with FACE and PACE. We also observe that the better performance of RB-FPCA is more pronounced when the outlier percentage is high. Table \ref{tab:sim3-PCs_est} summarizes the PC estimation results for various outlier percentage values. We can see that RB-FPCA outperforms other methods in estimating all the first 3 PCs.

\subsection{\textcolor{black}{Simulation IV: Dense Data with Outliers from Different Noise Distributions and Principal Components' Weights}}
\label{sec:sim4}

\textcolor{black}{The fourth simulation evaluates the model's performance when both types of outliers exist in the data. Different weights of the principal components were introduced to the observations, as detailed in Section \ref{sec:sim3}. Additionally, several forms of independent noises, such as $t(5)$, $SN(0,1,5)$ and $ST(0,1,5,5)$, were added to the observations. Figure \ref{fig:sim4-raw_data} provides a visualization of the clean and contaminated samples with $SN(0,1,5)$ noises. This figure illustrates the effects of the two types of outliers, which not only alter the pattern of the clean data but also add wiggles to the data due to the $SN(0,1,5)$ noises. The curves in \ref{fig:sim3-raw_data} appear smoother, with clear distinctions between normal samples and outliers. In contrast, the curves in \ref{fig:sim4-raw_data} are more jagged, showing higher variability and noise, thereby making it more challenging to distinguish between individual curves.}

\textcolor{black}{Analyses were conducted with an outlier percentage of  $p = 0.05, 0.10, 0.15$. Each value of $p$ was repeated 30 times using different random seeds, and the final results are obtained by averaging the estimated posterior means across these 30 runs. The prior covariance function used in this simulation is $\text{Cov}(s,t) = \exp\{-3(t-s)^2\}$. The numbers of basis functions and eigenfunctions are set to $P = 15$ and $K = 4$.}

\begin{table}[htbp]
\small
\begin{tabular}{|c|c|c|c|c|c|c|}
\hline
    & $p$           & RB-FPCA                         & BFPCA   & FACE    & PACE    & \begin{tabular}[c]{@{}c@{}}RB-FPCA \\         outperforms others\end{tabular} \\ \hline
    & 0.05        & \textbf{39.983}  & 49.944 & 43.661 & 44.122 & 100$\%$ \\ 
    & 0.10        & \textbf{39.873}  & 49.313 & 43.696 & 44.187 & 100$\%$ \\ 
    \multirow{-3}{*}{\begin{tabular}[c]{@{}c@{}}Cov\\ Function\end{tabular}} 
    & 0.15        & \textbf{39.254} & 51.338 & 43.797 & 44.329 & 100$\%$ \\ \hline
\end{tabular}
\caption{\textcolor{black}{Simulation IV: Estimations of the covariance function are evaluated by the distances between the estimates and the true covariance function with the $L_2$ norms. $SN(0, 1, 5)$ noises are added.}  \label{tab:sim4-cov_est}}
\end{table}

\begin{figure}[htbp]
    \centering
    \subfloat[]{\includegraphics[scale = 0.6]{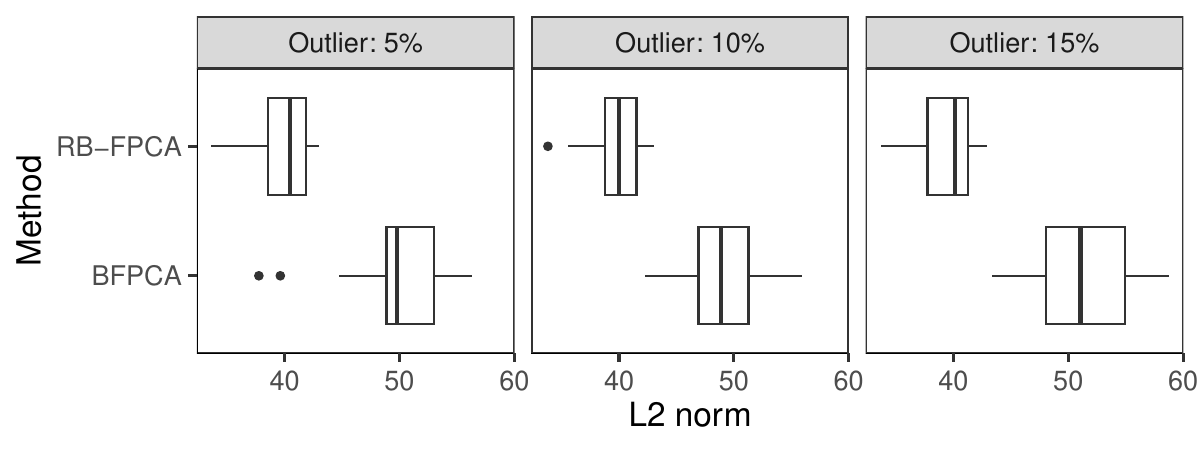}\label{fig:sim4-est_cov_1}} \\
    \subfloat[]{\includegraphics[scale = 0.6]{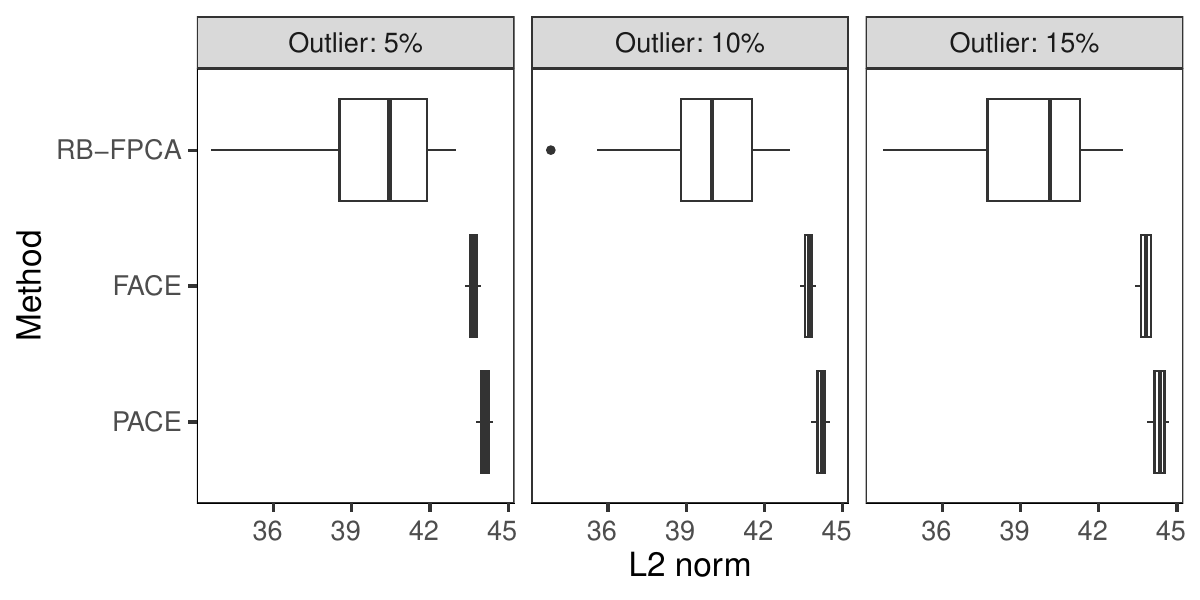}\label{fig:sim4-est_cov_2}}
    \caption{\textcolor{black}{Simulation IV: (a) Boxplots of distances between the estimates and the true covariance function with the $L_2$ norms. Comparison is between the proposed RB-FPCA method and the other Bayesian methods. (b) Boxplots of distances between the estimates and the true covariance function with the $L_2$ norms. Comparison is between the proposed RB-FPCA method and the two frequentist methods. $SN(0, 1, 5)$ noises are added.} \label{fig:sim4-est_cov}}
\end{figure}

\textcolor{black}{Table \ref{tab:sim4-cov_est} shows that RB-FPCA significantly improves covariance function estimation over BFPCA and the two modern frequentist methods in the presence of both types of outliers. Figure \ref{fig:sim4-est_cov} displays the boxplots of distances between the estimates and the true covariance function with the $L_2$ norms when $SN(0, 1, 5)$ noises are added. RB-FPCA outperforms BFPCA in terms of median and dispersion, as shown in Figure \ref{fig:sim4-est_cov_1}. We also observe that RB-FPCA outperforms the two frequentist methods, consistently yielding smaller errors across all 30 runs. Although the frequentist methods demonstrate a generally smaller dispersion, RB-FPCA's superior accuracy is evident. More results of the covariance function and the first 3 principal components with $t(5)$ and $ST(0, 1, 5, 5)$ noises are summarized in the supplementary material.
}

\subsection{Simulation V: Sparse Functional Data with Outliers} 
\label{sec:sim5}

The fifth simulation examines the performance of the Bayesian FPCA on the sparse longitudinal data. Each time we first generated a dense dataset with 100 noisy observations over 50 time points with the true mean function $\mu(t) = \text{sin}(2 \pi t)$. The true underlying covariance function is of either $\text{Cov}_1(s,t) = \exp\{-3(t-s)^2\}$ or $\text{Cov}_2(s,t) = \text{min} \{s+1, t+1\}$ depending on the experimental settings. Outlier percentage is chosen from $5\%, 10\% \text{ or } 15\%$ with equal probability. \textcolor{black}{To assess the impact of outliers, different types of noises were introduced into the sampling process. For symmetric noise distributions, we tested $N(0,3)$ and $t(5)$. For skewed noise distributions, we employed $SN(0,1,5)$ and $ST(0,1,5,5)$.} For clean samples, a small amount of noises of $N(0,0.3)$ was added. To get sparsely observed samples, a discrete uniform distribution $U(5, 10)$ was used to select a random number of times $n_i$ each curve was observed. The observed times $t_{ij}$ satisfy $t_{ij} \sim U(-1,1), \text{ i.i.d. for } i = 1,\ldots, 100, j = 1,\ldots, n_{i}$. This data generation process produces sparse longitudinal data with some noisy samples.

\textcolor{black}{The values of the hyperparameters in the full conditional distributions are set similarly to the dense case, except for $\boldsymbol{\kappa} = 100  (\boldsymbol{R}^{(i)})^{-1}/(2r)$ for $\boldsymbol{\Sigma}^{-1}$ in Equation \ref{eq:cond_dist_Sigma}. The hyperparameter $\boldsymbol{R}^{(i)}$ is defined as an $n_{i} \times n_{i}$ diagonal matrix where elements of the main diagonal are the squared ranges of the corresponding components in the data. For the sparse longitudinal data, the squared ranges of the components cannot be directly calculated since data are observed at different time points. We estimate the squared ranges with a search bandwidth denoted as $h_R$, with a default value $h_R = 0.05n$ used in simulations and examples. Specifically, a set of $h_R$ observations that are closest to the target observation is used to estimate the squared ranges for the target individual at a specific time point.
}

\begin{figure}[htbp]
    \centering
    \subfloat[]{\includegraphics[width=0.4\textwidth]{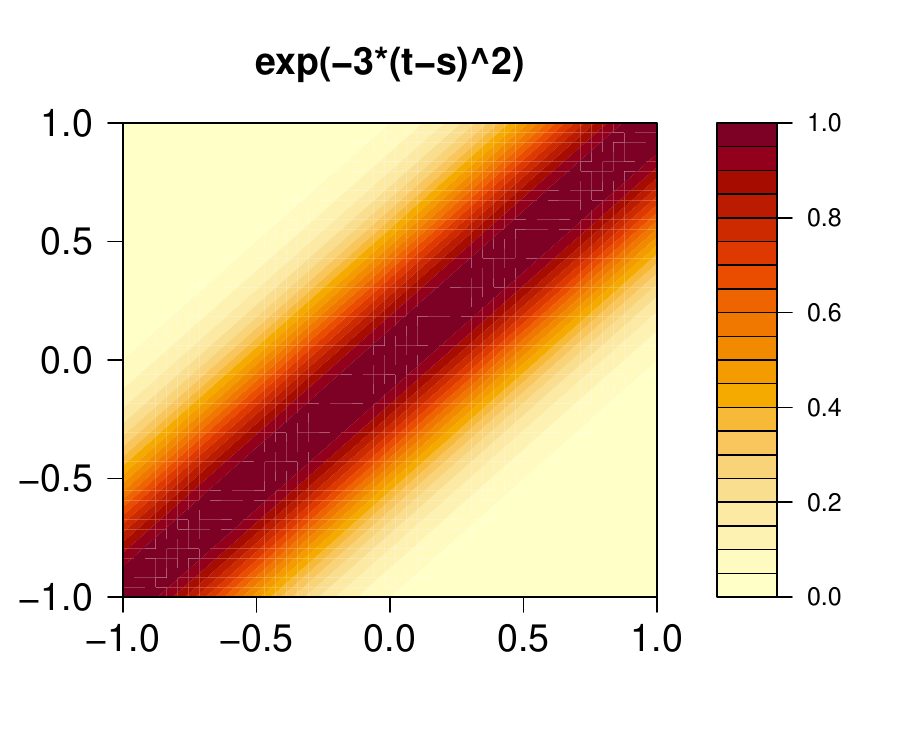}}
    \subfloat[]{\includegraphics[width=0.4\textwidth]{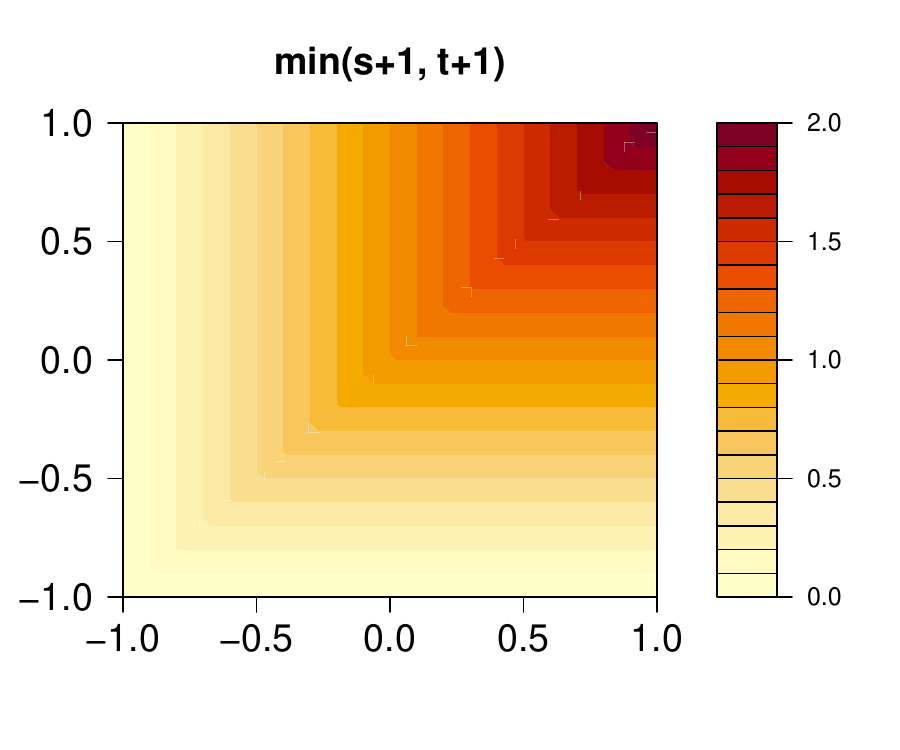}} \\
    \subfloat[]{\includegraphics[width=0.4\textwidth]{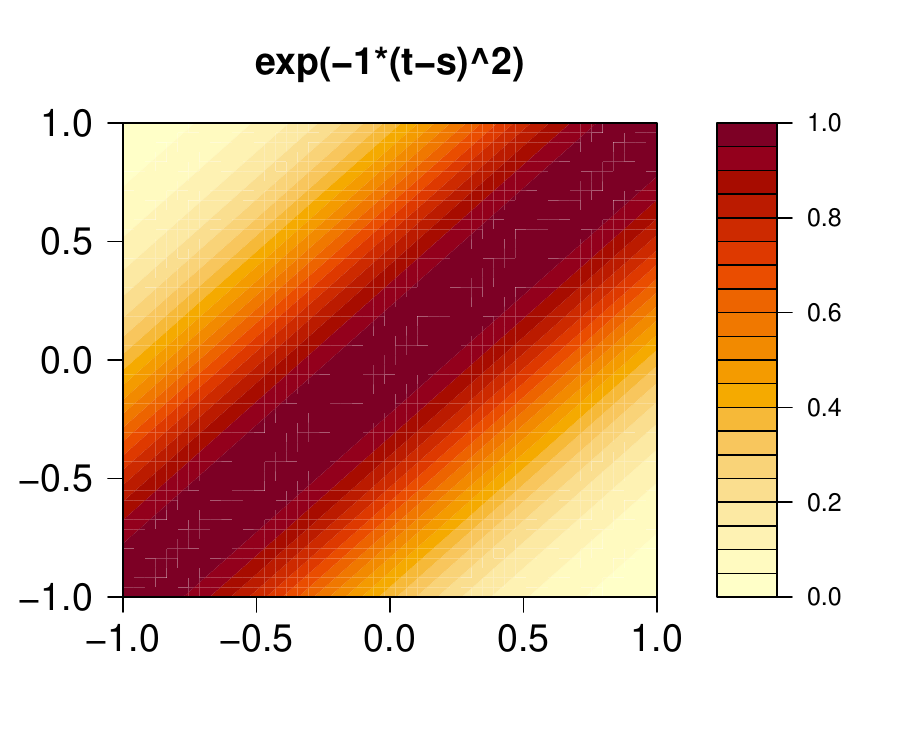}} 
    \subfloat[]{\includegraphics[width=0.4\textwidth]{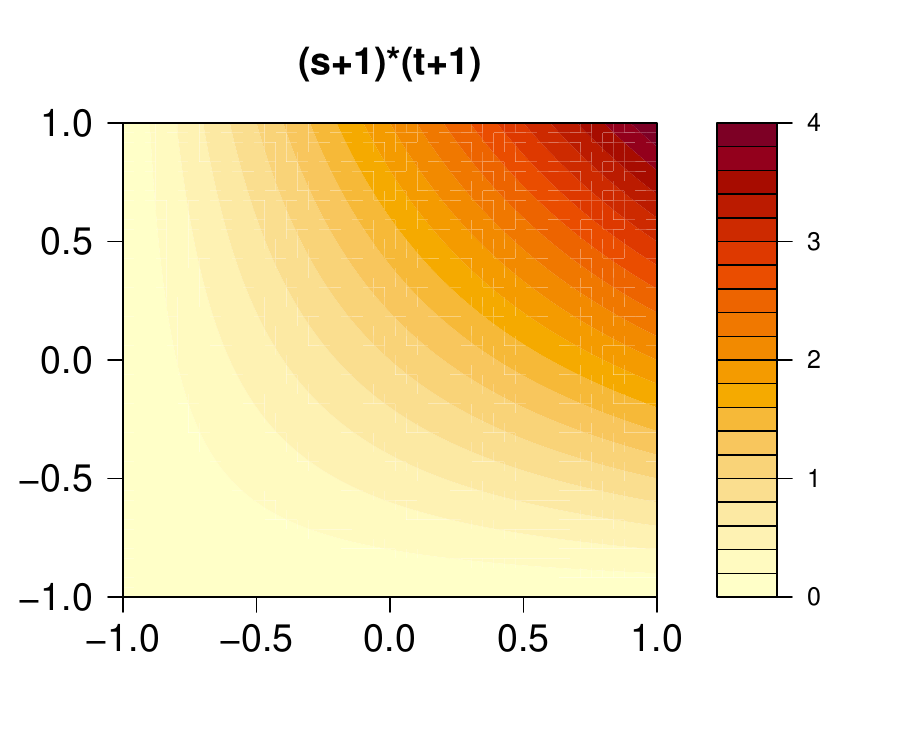}}
    \caption{Simulation V: Contour plots of prior covariance functions. \label{fig:sim5_contour}}
\end{figure}

We tested the performance of our proposed RB-FPCA method with different prior covariance functions. When the true covariance function is $\text{Cov}_{truth}(s,t) = \exp\{-3(t-s)^2\}$, we set prior covariance functions with the following forms:
\begin{itemize}
    \item $\text{Cov}_{prior1}(s,t) = \exp\{-3(t-s)^2\}$ (same as truth)
    \item $\text{Cov}_{prior2}(s,t) = \min\left(s+1, t+1\right)$
    \item $\text{Cov}_{prior3}(s,t) = \exp\{-(t-s)^2\}$
    \item $\text{Cov}_{prior4}(s,t) = \text{covariance estimation from PACE method}$
\end{itemize}
When the true covariance function is $\text{Cov}_{truth}(s,t) = \min\left(s+1, t+1\right)$, we set prior covariance functions with the follwing forms:
\begin{itemize}
    \item $\text{Cov}_{prior1}(s,t) = \exp\{-3(t-s)^2\}$
    \item $\text{Cov}_{prior2}(s,t) = \min\left(s+1, t+1\right)$ (same as truth)
    \item $\text{Cov}_{prior3}(s,t) = \left(s+1\right)*\left(t+1\right)$
    \item $\text{Cov}_{prior4}(s,t) = \text{covariance estimation from PACE method}$
\end{itemize}
Figure \ref{fig:sim5_contour} shows the contour plots of covariance functions. For each setting, we selected four prior covariance functions with different characteristics that may have an effect on the performance of the RB-FPCA method. Specifically, except for using the true covariance function as the prior covariance function, we also included one prior covariance function, which has a similar surface structure as the true covariance function and one prior covariance function with a different shape. In addition, we also involved the smoothed covariance surface estimated from the PACE method.

\begin{figure}[!htb]
    \centering
    \subfloat[]{\includegraphics[width=0.44\textwidth]{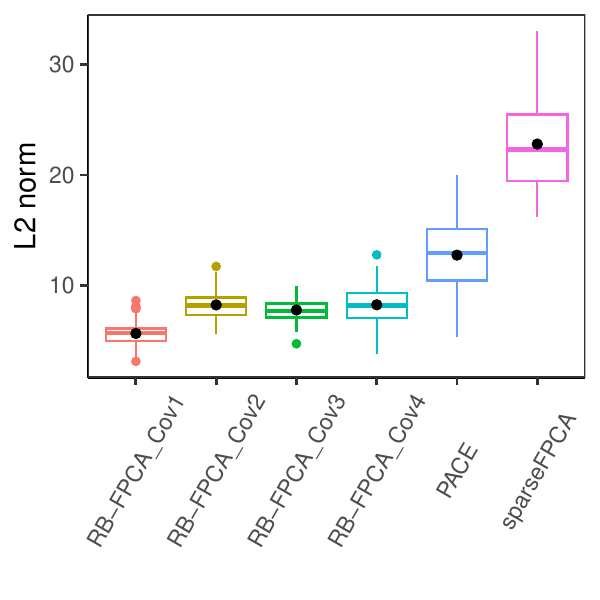}\label{fig:sim5_boxplot_n_a}}
    \subfloat[]{\includegraphics[width=0.44\textwidth]{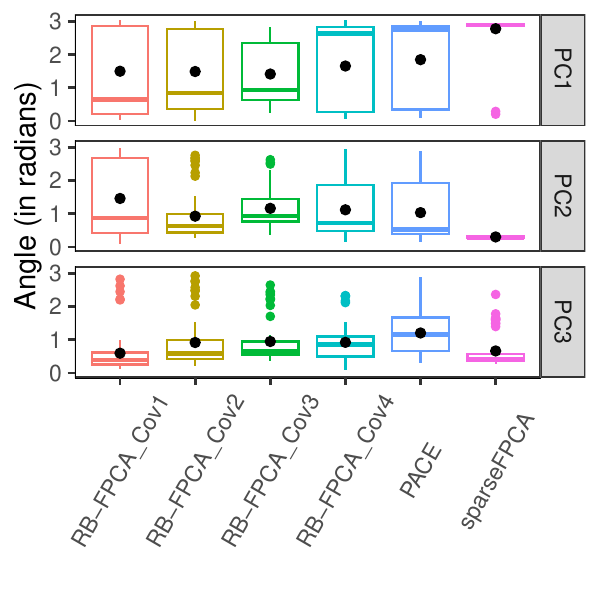}\label{fig:sim5_boxplot_n_b}} \\
    \subfloat[]{\includegraphics[width=0.44\textwidth]{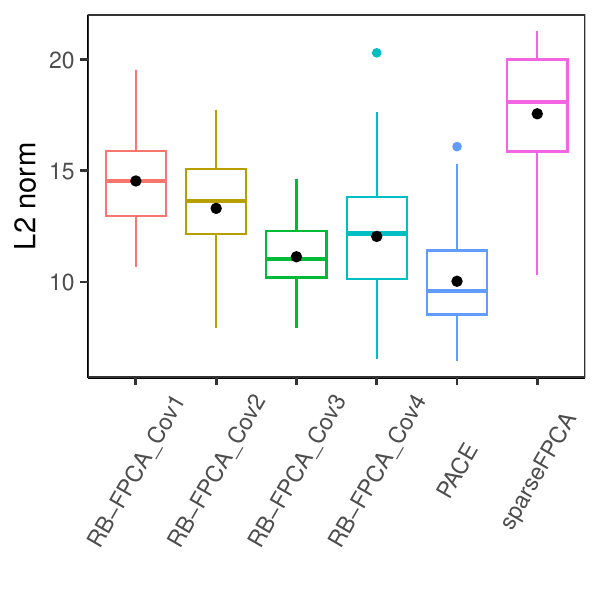}\label{fig:sim5_boxplot_n_c}} 
    \subfloat[]{\includegraphics[width=0.44\textwidth]{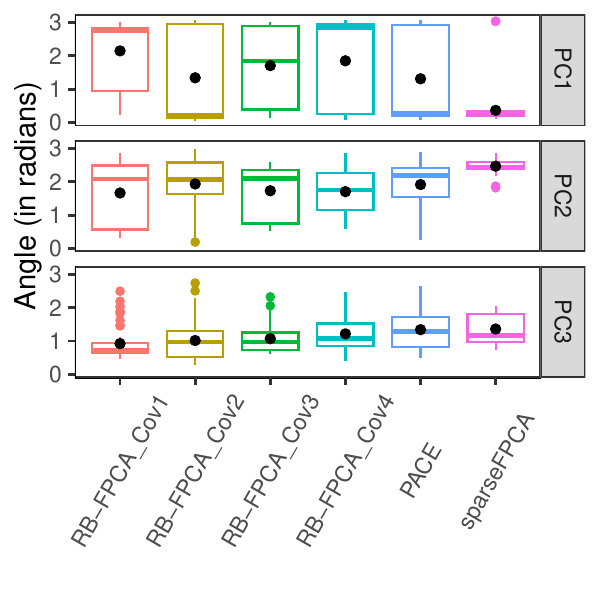}\label{fig:sim5_boxplot_n_d}}
    \caption{Simulation V: \textcolor{black}{Comparison of RB-FPCA method with PACE and sparseFPCA methods with $N(0,3)$ noises.} (a) and (c) show boxplots of distances (in $L_2$ norms) between the estimates and the true correlation function. (b) and (d) show boxplots of the angle (in radians) of the estimations for the first 3 principal components. The top row shows the results given the true covariance function is $\text{Cov}_{truth}(s,t) = \exp\{-3(t-s)^2\}$. The bottom row shows the results when the true covariance function is $\text{Cov}_{truth}(s,t) = \min\left(s+1, t+1\right)$. The black dot in each box corresponds to the value of the mean. \label{fig:sim5_boxplot_n}}
\end{figure}

\begin{figure}[!htb]
    \centering
    \subfloat[]{\includegraphics[width=0.44\textwidth]{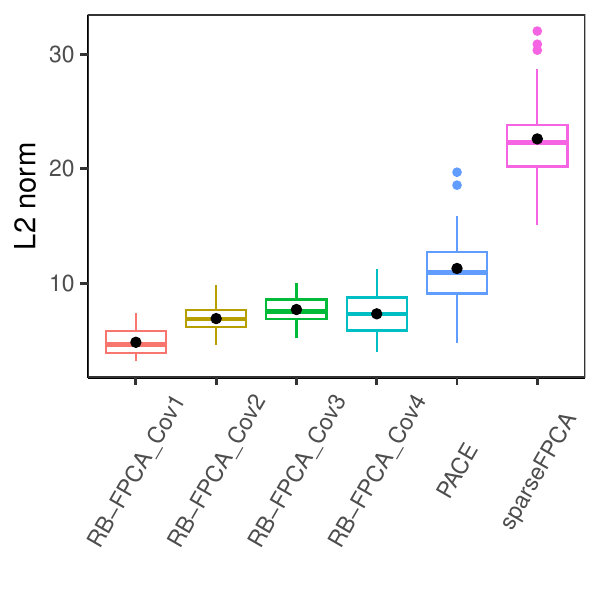}\label{fig:sim5_boxplot_sn_a}}
    \subfloat[]{\includegraphics[width=0.44\textwidth]{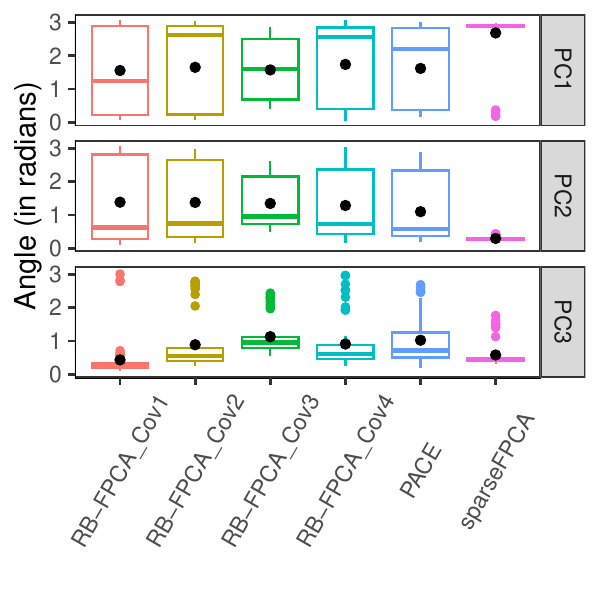}\label{fig:sim5_boxplot_sn_b}} \\
    \subfloat[]{\includegraphics[width=0.44\textwidth]{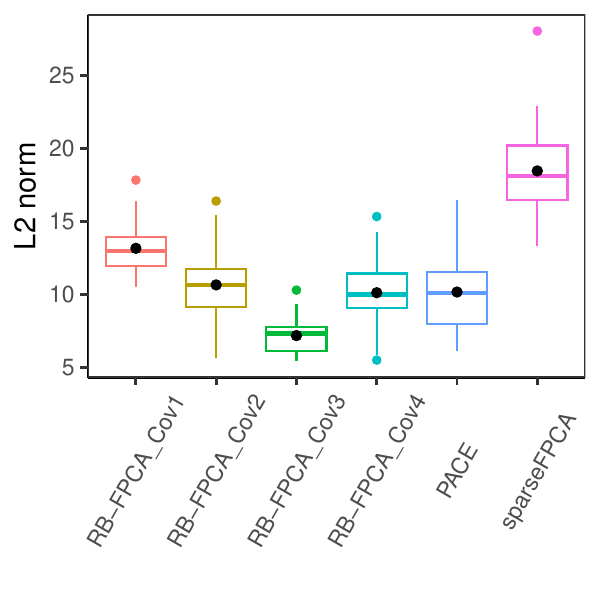}\label{fig:sim5_boxplot_sn_c}} 
    \subfloat[]{\includegraphics[width=0.44\textwidth]{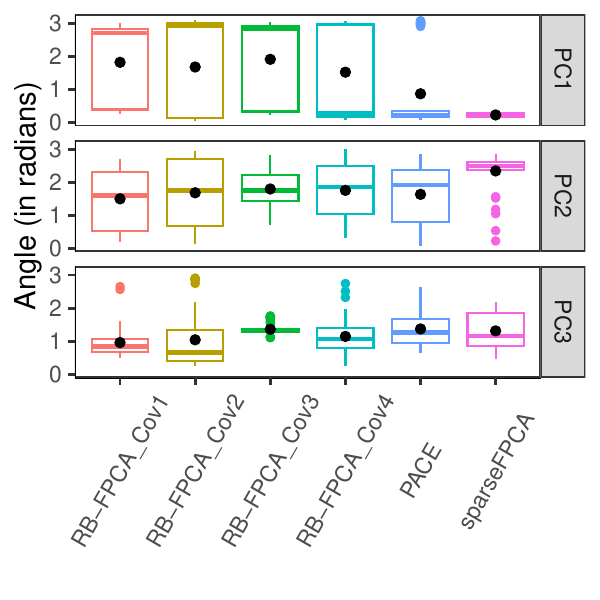}\label{fig:sim5_boxplot_sn_d}}
    \caption{Simulation V: \textcolor{black}{Comparison of RB-FPCA method with PACE and sparseFPCA methods with $SN(0,1,5)$ noises. (a) and (c) show boxplots of distances (in $L_2$ norms) between the estimates and the true correlation function. (b) and (d) show boxplots of the angle (in radians) of the estimations for the first 3 principal components. The top row shows the results given the true covariance function is $\text{Cov}_{truth}(s,t) = \exp\{-3(t-s)^2\}$. The bottom row shows the results when the true covariance function is $\text{Cov}_{truth}(s,t) = \min\left(s+1, t+1\right)$. The black dot in each box corresponds to the value of the mean.}\label{fig:sim5_boxplot_sn}}
\end{figure}

The numbers of basis functions and eigenfunctions are preset as $P = 5$ and $K = 5$. Each setting was repeated 50 times with different seeds and noise levels, and the results are the average of the estimated means from 50 runs. We compared the performance of the proposed RB-FPCA method with two frequentist methods, which are the PACE method of \cite{wang2016functional}, implemented in \texttt{fdapace} package in \texttt{R}, and a robust FPCA (sparseFPCA) method of \cite{boente2021robust}, implemented in \texttt{sparseFPCA} package in \texttt{R}. The performance of each model is compared in terms of estimations of the correlation surface evaluated by $L_2$ norms and estimations of the first three principal components assessed by the angle (in radians) between the truth and estimates. Results are presented in Figures \ref{fig:sim5_boxplot_n} and \ref{fig:sim5_boxplot_sn}.

Figures \ref{fig:sim5_boxplot_n_a} and \ref{fig:sim5_boxplot_n_b} demonstrate the advantage of using the RB-FPCA method over the other two frequentist methods by comparing the estimations of the correlation functions and the PCs. The proposed RB-FPCA method produces better measures of the correlation functions with smaller $L_2$ norms and admits small variations. Figures \ref{fig:sim5_boxplot_n_c} and \ref{fig:sim5_boxplot_n_d} show different methods' performances under the other true covariance function. The proposed RB-FPCA method outperforms one of the frequentist methods, namely sparseFPCA, in terms of the estimations of the correlation functions. PACE performs well for estimating the correlation function, whereas the RB-FPCA method still has a potential advantage if considering the variations of the estimations. Note that even though the sparseFPCA method produces estimations for PCs with the smallest variations, it does not outperform PACE and RB-FPCA methods in estimating the correlation functions and PCs. The robustness of the sparseFPCA method may explain the small variations of the estimations of PCs. 
\textcolor{black}{Figure \ref{fig:sim5_boxplot_sn} illustrates patterns analogous to those shown in Figure \ref{fig:sim5_boxplot_n}, further demonstrating the robustness of the RB-FPCA method across various noise distributions. Additional results involving the $t(5)$ and $ST(0,1,5,5)$ noises are available in the supplementary material.}
The overall picture is that the proposed RB-FPCA method has the ability to deal with sparse longitudinal data, perform competitively with some well-known frequentist methods, and provide a Bayesian framework to enable incorporating the domain knowledge into the analysis via flexible prior settings.

\section{Data Applications}
\label{sec:data_application}

\subsection{Hawaii Ocean Oxygen Data} \label{sec:app1}

We first illustrate our method on the Hawaii ocean oxygen data collected from the Hawaii Ocean Time-series Data Organization \& Graphical System \citep{HOT-DOGS}. Scientists participating in the Hawaii Ocean Time-series (HOT) program have been monitoring and making continuous measurements of the water column's hydrography, chemistry and biology at a station near Oahu, Hawaii, since October 1988. The primary objective of this research is to monitor and interpret the variability of physical and biogeochemical processes at deep-water hydrostations and deliver a comprehensive overview of the ocean at a representative site in the North Pacific subtropical region. The deep-water station is visited approximately once a month by cruises to obtain water samples from desired ocean depths. Observations of ocean data over long periods are extremely valuable for climate studies. Researchers require repeated measurements of oceanographic data to investigate slow or irregular changes in natural processes or phenomena and some rapid event-driven variations. Oceanic data are obtained via the Hawaii Ocean Time-series HOT-DOGS application, University of Hawai'i at Mānoa, National Science Foundation Award \# 1756517, and can be downloaded from the Hawaii Ocean Time-series Data Organization \& Graphical System website (\url{https://hahana.soest.hawaii.edu/hot/hot-dogs/cextraction.html}). 

In this study, we analyzed the oxygen concentrations in units of $\mu$mol/kg measuring at different depths below the sea surface from January 1, 2008, to December 31, 2021. The oxygen concentration is measured every 2 meters at a depth of 0 to 200 meters under the ocean's surface. Such data with a shorter time frame have been analyzed by \cite{shi2022robust}. We obtained 133 trajectories to study the functional relationship between the oxygen concentrations and depth below the sea surface. Each trajectory has 100 data points with no missing values. All trajectories have measurements at the same depths. We applied the RB-FPCA method with three candidate models to the Hawaii ocean oxygen data with a predetermined set of numbers of basis functions and eigenfunctions, i.e., $P = 15$ and $K = 5$. The data were first detrended, and the prior covariance function was chosen as $\exp\{-3(t-s)^2\}$. \textcolor{black}{For the tuning parameters in the ASMC, we employed 200 particles with a resampling threshold of 0.5 and an annealing parameter sequence threshold of 0.9.} This study aims to find patterns in data of high dimensions, capture the primary mode of variation, and flag potential outliers. 

\begin{figure}[htbp]
  \centering
  \includegraphics[width=0.85\textwidth]{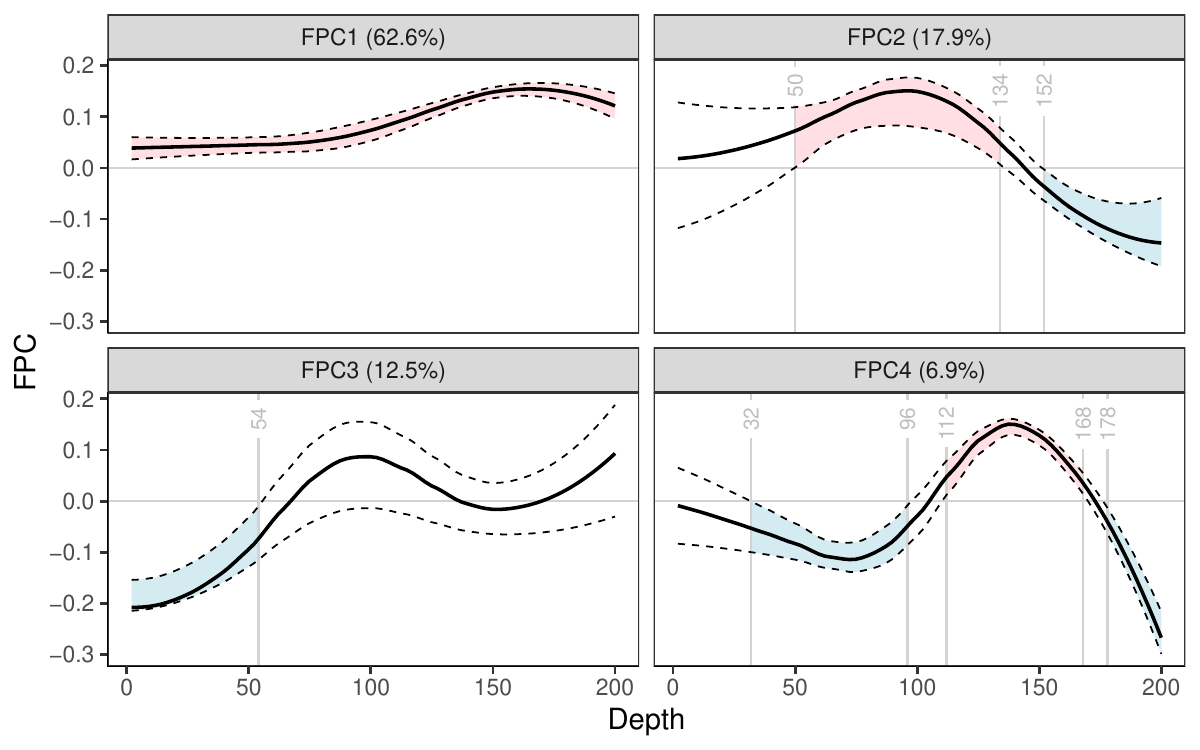}
  \caption{Hawaii Ocean Oxygen Data: The top four estimated FPCs from the full posterior mean of the covariance function with the proportion of variance explained by each FPC. The 95\% credible intervals are represented by the dashed lines. The shaded areas denote credible intervals that do not contain zeros. The numeric labels along the vertical lines correspond to the depths associated with the boundaries of each respective shaded area.} \label{fig:ex1_fpc}
\end{figure}

\textcolor{black}{All three models were evaluated, and the estimated log marginal likelihoods were used to select the optimal model. In this study, RB-FPCA-SN was identified as optimal, with a log marginal likelihood estimate of -31810.56. The log marginal likelihood estimates for the RB-FPCA-ST and RB-FPCA-MM models were -32864.49 and -34131.17, respectively. All subsequent analyses are based on the optimal RB-FPCA-SN model.} Figure \ref{fig:ex1_fpc} shows the first four FPCs from the RB-FPCA-SN model with variance explained by each FPC. The 95\% credible intervals are included with the dashed lines. The credible intervals generated within the Bayesian approach offer a direct interpretation, reflecting the probability of the true parameter being contained within a specific range, for instance, the 95\% credible interval indicating a 95\% probability of encompassing the true parameter. The credible intervals provide valuable information for interpreting the significance of each FPCs in the analysis. In general, the credible interval for the first FPC is narrow and relatively narrow for the remaining FPCs except at the boundaries. The first FPC is positive and significant over the whole range at a level of 0.05. It represents the weighted average of the oxygen level of each visit to the deep-water station over a depth from 0 to 200 meters. Most of the variation is due to the amount of horizontal shift from the mean function. The second FPC explains about 18\% of the total variability. The second FPC can be interpreted as a change in the oxygen level over two depth intervals. The second FPC is positive when the depth is between 0 to 150 meters and negative when the depth is above 150 meters, indicating the relative difference in oxygen levels between 0 to 150 meters and 150 to 200 meters. The credible intervals indicate significant variations over the depth intervals [50, 134] and [152, 200]. FPCs at a higher level represent more complex phenomena. The top 3 FPCs capture over 90 \% of the total variation in the data. 

\begin{figure}[htbp]
  \centering
  \includegraphics[width=0.8\textwidth]{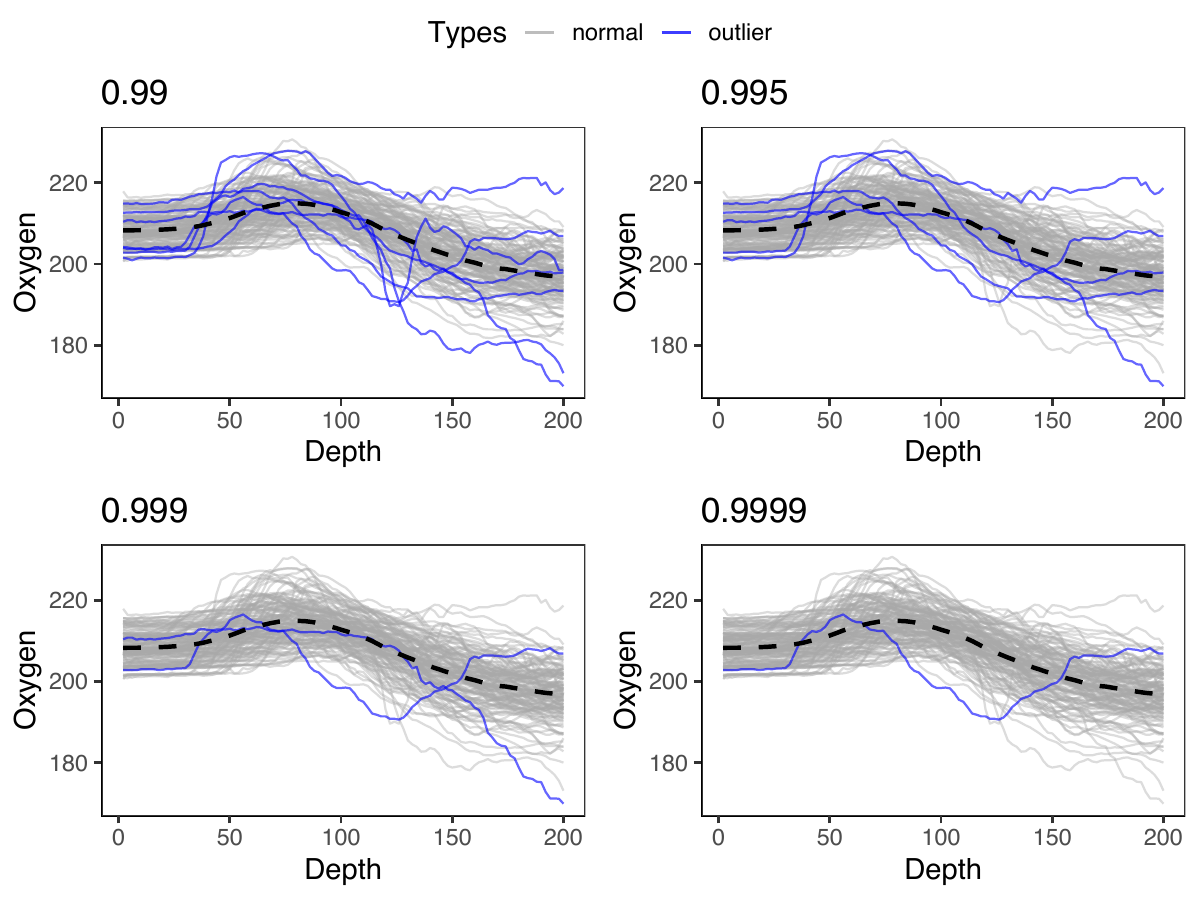}
  \caption{Hawaii Ocean Oxygen Data: Detected outliers and the oxygen trajectories. The dashed curve is the mean function calculated with all trajectories. The values on the top left corner are the threshold values used in the quantile of the $\chi^2_2$ distribution to identify possible outliers. For example, the top left plot shows outliers with FPC scores have distances larger than the 99\% quantile of a $\chi^2_2$ distribution.} \label{fig:ex1_outlier}
\end{figure}

We implemented the proposed approach of identifying the outliers using the FPC scores. Different threshold values result in different trajectories being flagged as abnormal. We investigated four threshold values, and the results are presented in Figure \ref{fig:ex1_outlier}. The result follows the expectation that fewer outliers are detected as the threshold value increases. The detected atypical trajectories lie near the top or bottom. Compared with other curves, abnormal trajectories appear to have different curvature, or maintain high values over depth levels, or decline rapidly.

\subsection{Annual Sea Surface Temperature Data}

The second example considers the annual sea surface temperature data. The sea surface temperature is the temperature of the ocean's surface at its top millimeter. As a fundamental measure of global climate change, sea surface temperature provides a glimpse into the overall trend in the climate system. Understanding the behaviour of the sea surface temperature changes and discovering potential anomalies are essential in studying global climate change. Sea surface temperature anomalies are characteristic of El Niño and La Niña climate cycles, which can affect weather patterns worldwide. Strong and localized anomalies may identify ocean currents in sea surface temperature. Anomalies in sea surface temperature over many years can be signs of regional or global climate change, for example, global warming. In addition to their scientific value, sea surface temperature anomalies are also useful for practical purposes. For instance, an anomalous temperature (warm or cool) in coastal areas can favor certain organisms in an ecosystem over others, resulting in a thriving or declining population of bacteria, algae, or fish. Such data have been analyzed by \cite{hyndman2010rainbow}, \cite{sun2011functional}, \cite{xie2017geometric}, and \cite{dai2020functional}. Annual sea surface temperature data can be downloaded from the Climate Prediction Center website (\url{https://www.cpc.ncep.noaa.gov/data/indices/}).

The dataset consists of monthly observations from Niño 1+2 region from January 1950 to December 2021. We obtained 72 functional observations observed on a common grid of 12 time points. They correspond to the monthly sea surface temperature over 72 years. The RB-FPCA method with three candidate models for dense data was applied to find projections of maximum variance. The numbers of basis functions and eigenfunctions are fixed with $P = 10$ and $K = 5$. The prior covariance function is chosen as $\exp\{-3(t-s)^2\}$. \textcolor{black}{For the ASMC parameters, we used 200 particles, with a resampling threshold set at 0.5 and an annealing parameter sequence threshold set at 0.9.}  The estimations are only calculated for the time points that have been sampled.

\begin{figure}[htbp]
    \centering
    \subfloat[]{\includegraphics[scale = 0.36]{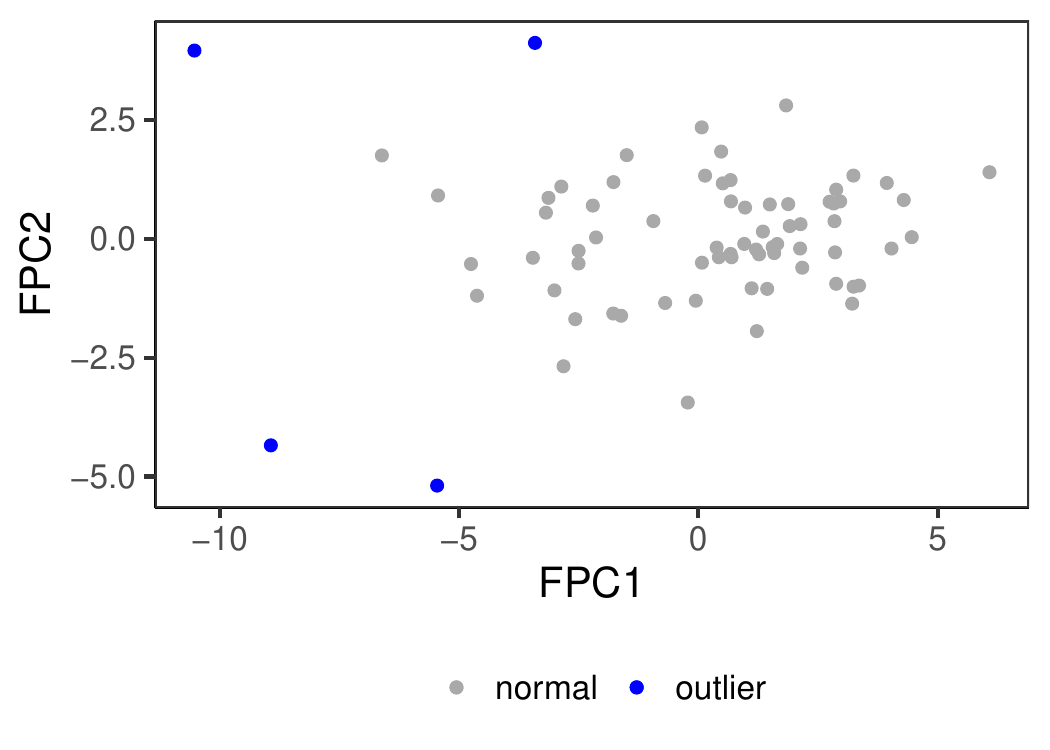}\label{fig:ex2_FPC1}}
    \subfloat[]{\includegraphics[scale = 0.36]{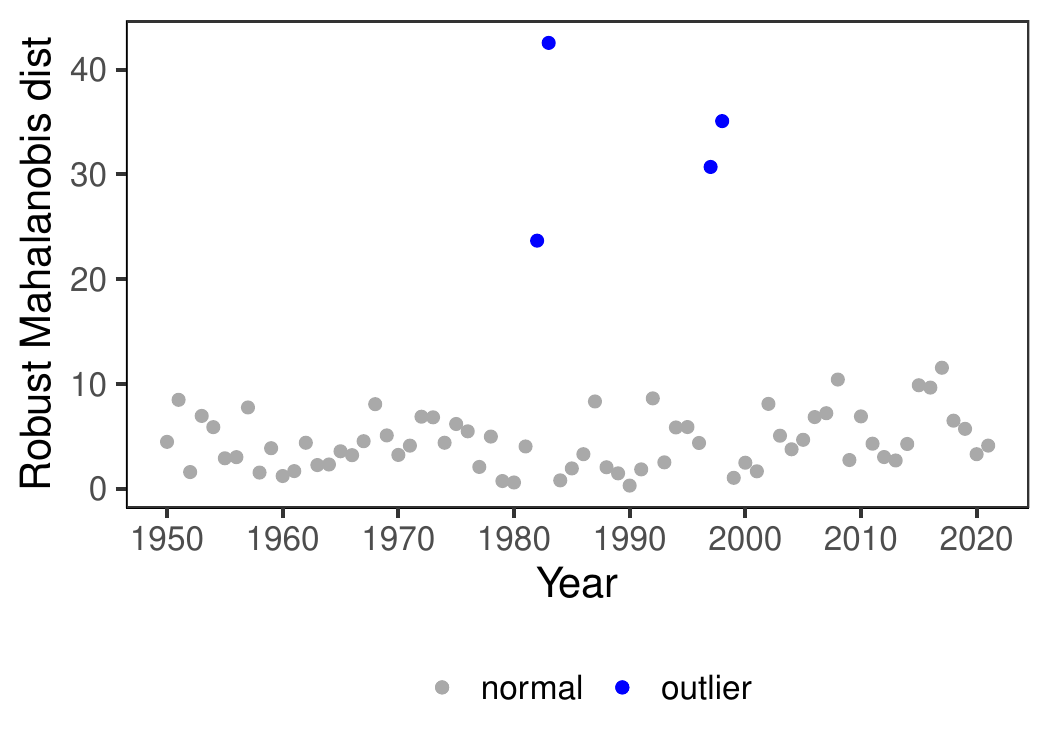}\label{fig:ex2_FPC2}}
    \caption{Annual sea surface temperature data: (a) FPC scores for the first two FPCs. Possible outliers and normal observations are differentiated by colors. (b) Plot of the robust Mahalanobis distances for observations. Potential outliers are highlighted. These four suspected outliers correspond to the same observations identified in both (a) and (b). \label{fig:ex2_fpc_dist}}
\end{figure}

\textcolor{black}{We evaluated all three models and selected the optimal one based on their estimated log marginal likelihoods. The log marginal likelihood estimates are -610.96 for the RB-FPCA-SN model, -742.89 for the RB-FPCA-ST model, and -651.18 for the RB-FPCA-MM model. Therefore, the RB-FPCA-SN model, with the highest log marginal likelihood estimate, is identified as the optimal choice, and all subsequent analyses are performed using this model.} The first two FPCs explain 94\% of the variance. The first FPC is almost constant below the zero axis, representing that the major variation is the degree of horizontal shift below the mean function. The second FPC crosses the zero axis once near July, which can be interpreted as the relative change of sea surface temperature between spring and fall months. To investigate the existence of potential outliers, the plot of the FPC scores for the first two FPCs is shown in Figure \ref{fig:ex2_FPC1}. We have observed that four observations deviate significantly from the majority of the observations and have been identified as outliers. The graphical representation of robust Mahalanobis distances in Figure \ref{fig:ex2_FPC2} provides further evidence of the outlying behaviors exhibited by these four observations.

\begin{figure}[htbp]
    \centering
    \subfloat[]{\includegraphics[scale = 0.36]{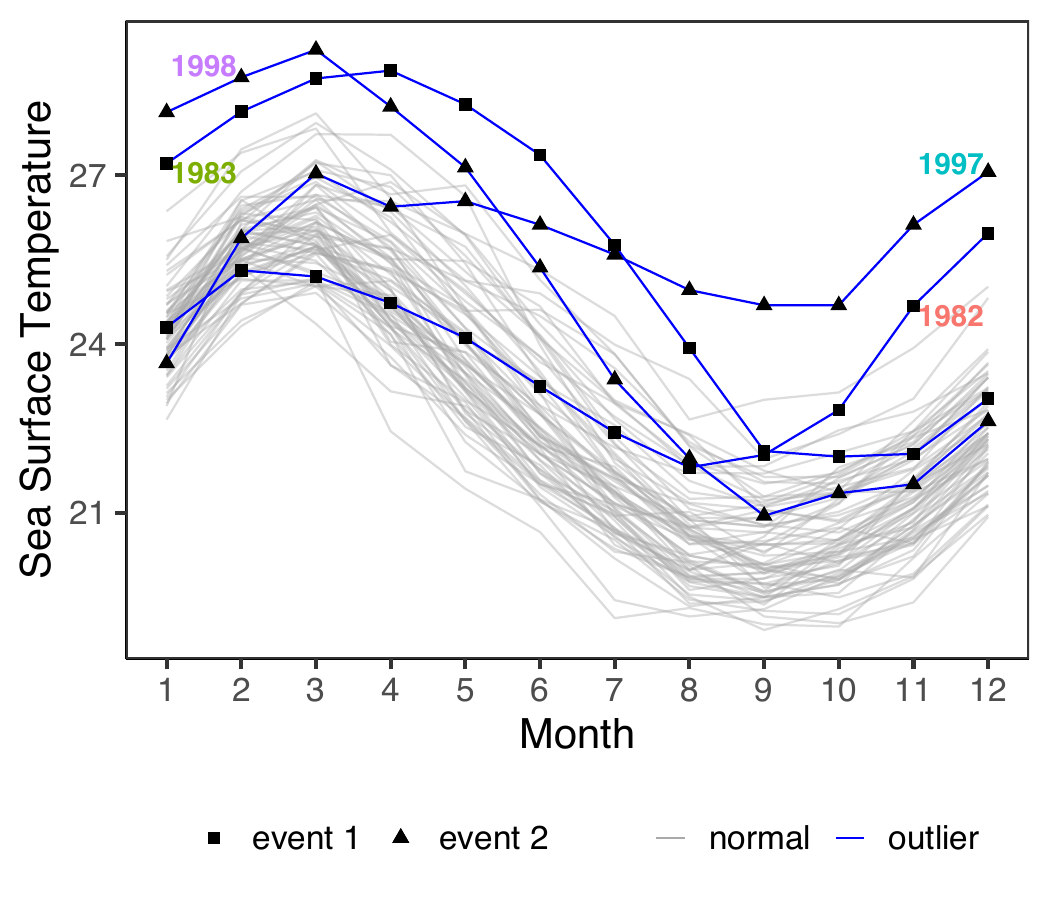}\label{fig:ex2_outlier1}}
    \subfloat[]{\includegraphics[scale = 0.36]{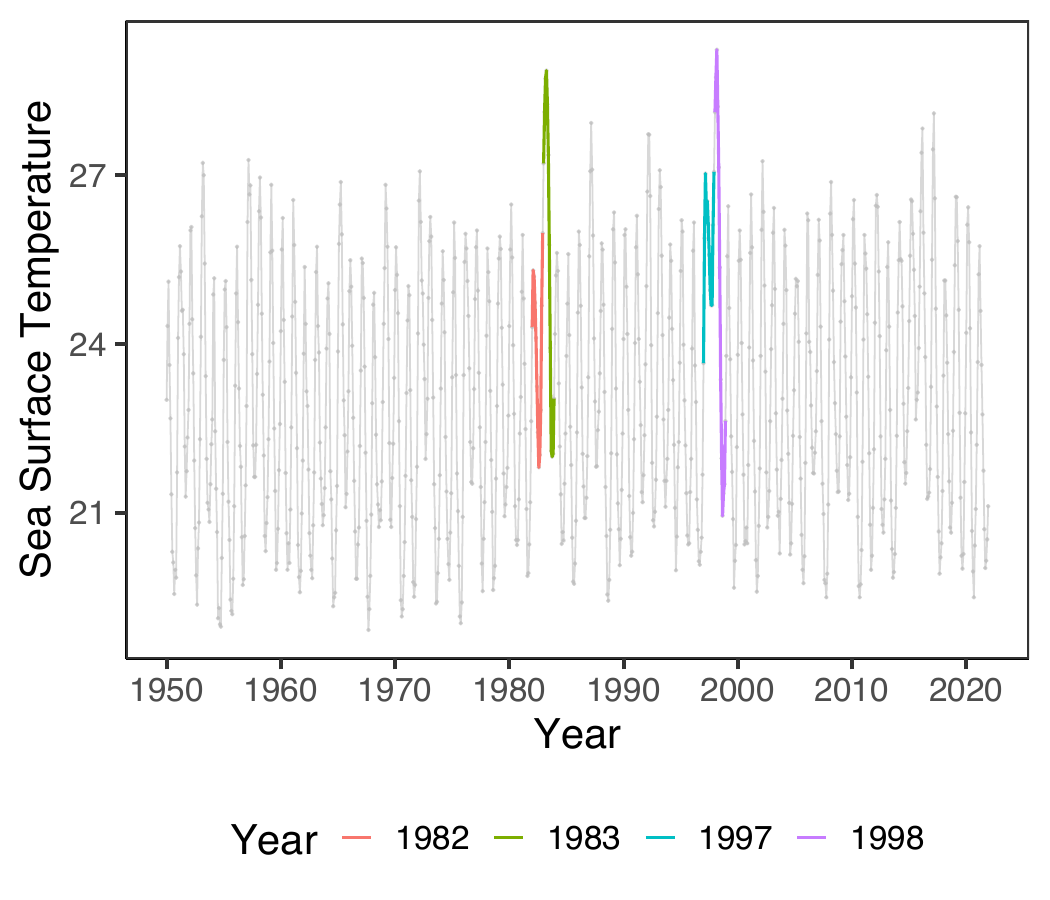}\label{fig:ex2_outlier2}}
    \caption{Annual sea surface temperature data: (a) Detected outliers and annual temperature trajectories. The four outliers (blue) are annotated with years, and the colors of the annotations are in line with the legends in panel (b). The detected El Niño events are differentiated by square and triangle. (b) Plot of the whole history of monthly sea surface temperature from January 1950 to December 2021. The locations of the four detected outliers are colored. \label{fig:ex2_outlier}}
\end{figure}

From a practical aspect, detecting sea surface temperature anomalies is essential to identify El Niño and La Niña events. El Niño events are widely used to describe the warming of sea surface temperature that occurs every few years. In contrast, La Niña events correspond to years with abnormally low sea surface temperatures. We can determine the abnormal annual sea surface temperature curves by FPC scores with robust Mahalanobis distances larger than 99.5\% quantile of a $\chi^2_2$ distribution. Four trajectories (1982, 1983, 1997, and 1998) are identified as anomalies and shown in Figure \ref{fig:ex2_outlier}. This finding confirms the potential outliers found from the preliminary investigation of the FPC scores, as depicted in Figure \ref{fig:ex2_fpc_dist}. The winters from 1982 to 1983 and 1997 to 1998 have the highest temperatures among all years, and the temperatures remain at high values over the summers of 1983 and 1998. According to National Climatic Data Center reports, these periods correspond to two of the strongest El Niño events in history and are illustrated by squares and triangles in \ref{fig:ex2_outlier1}. Although our method cannot flag all significant El Niño or La Niña years, we can detect some of the strongest activities. By decreasing the threshold value in our outlier detection procedure, more outlying curves are expected to appear.

\subsection{Sparse CD4 Data} \label{sec:app2}

\begin{figure}[htbp]
  \centering
  \subfloat[]{\includegraphics[scale = 0.36]{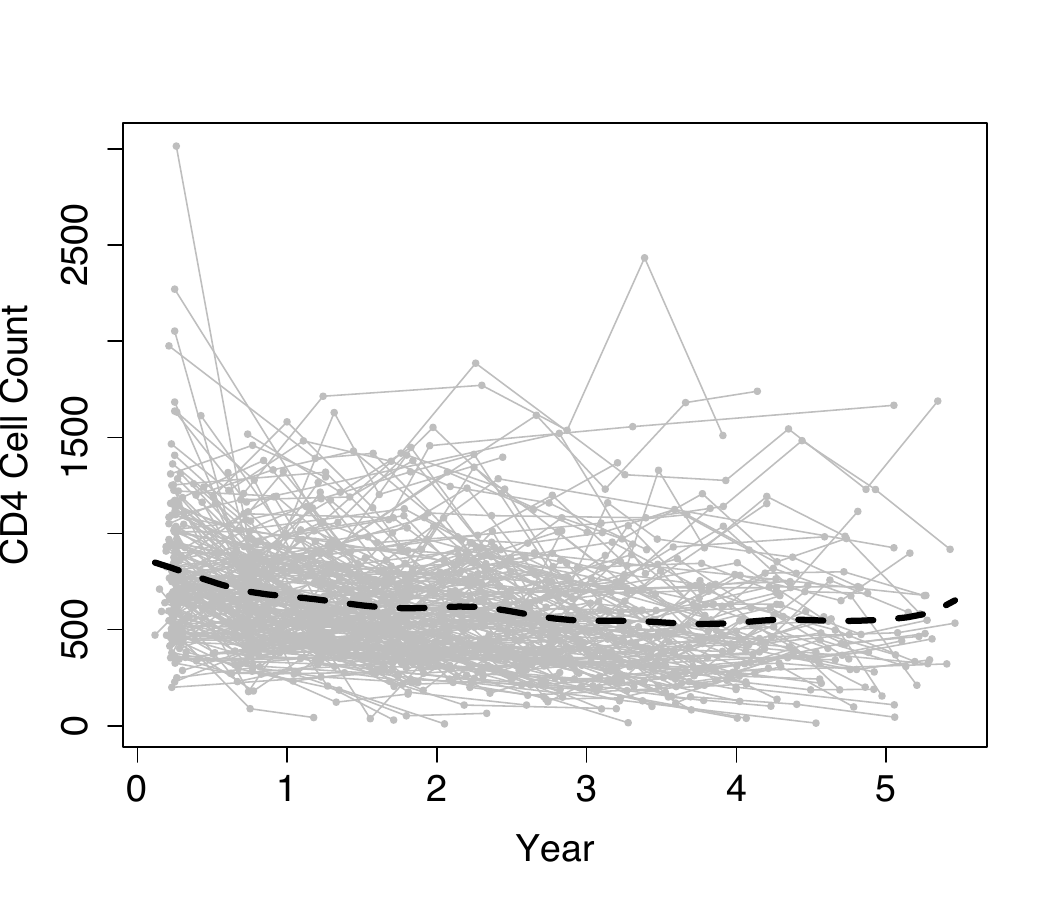}\label{fig:ex3_data}}
  \subfloat[]{\includegraphics[scale = 0.36]{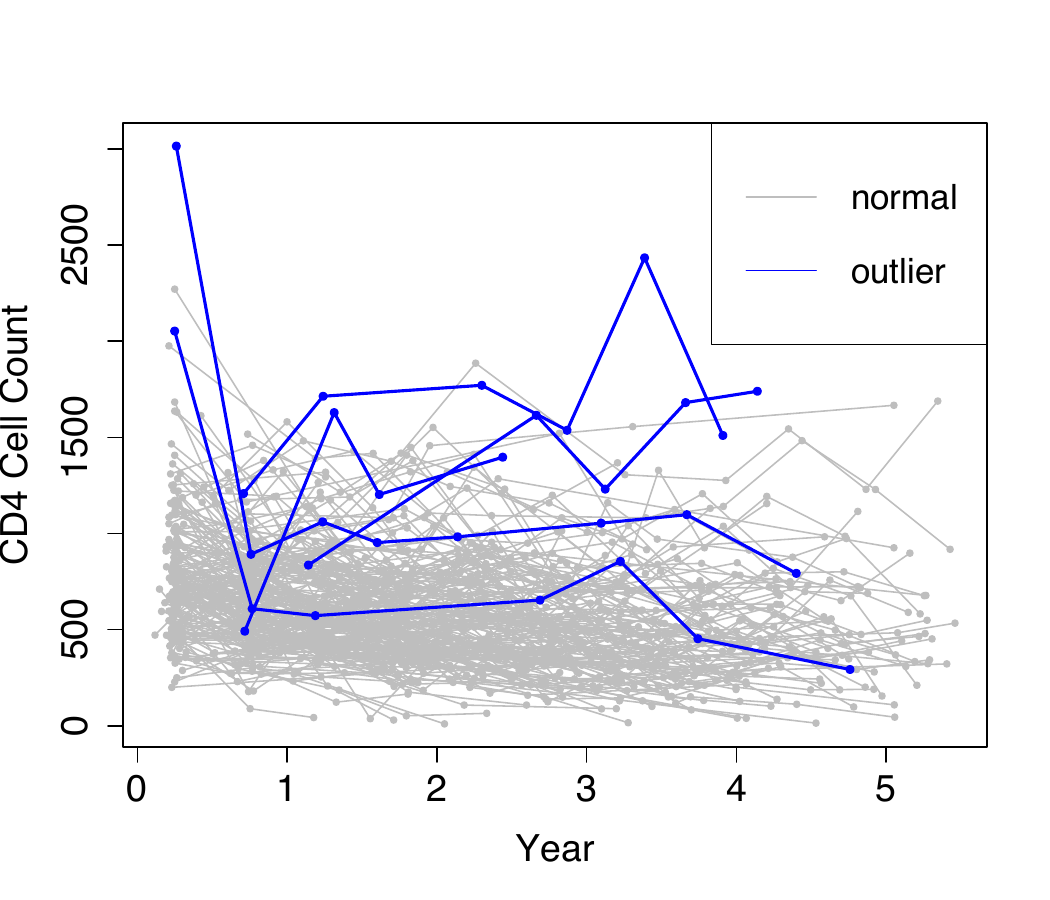}\label{fig:ex3_outlier}}
  \caption{Sparse CD4 Data: (a) Data contain 241 observed trajectories for CD4 cell counts. The dashed line is the smooth estimate of the mean function. (b) Detected five most outlying trajectories in CD4 data.} \label{fig:ex3_data_outlier}
\end{figure}

The last example considers the CD4 data, a sparse longitudinal dataset collected within the Multicenter AIDS Cohort Study (MACS). Study participants include 1809 HIV-infected men at the start of the study and 371 men who were seronegative at entry and seroconverted during the study period (see \cite{zeger1994semiparametric}). In our study, we used the data from the \texttt{catdata} package in \texttt{R}, which includes 2376 measurements of the number of CD4 cells taken over time on 369 seroconverters. The variable of interest is the CD4 cell counts measured in years since seroconversion, which can be used to assess disease progression. Specifically, we are interested in the typical decay of CD4 cell counts over time and the variability across subjects. The CD4 data are sparse, owing to unequal numbers of repeated measurements for each subject and the different timing of measurements for each subject. We obtained 241 trajectories after removing observations with less than three measurements, and the number of measurements per subject ranges from 3 to 11, with a median of 5. Figure \ref{fig:ex3_data} shows the data together with a smooth estimate of the mean function from the PACE method. The overall trend in the CD4 cell counts is decreasing.

We applied the model described in Section \ref{sec:model_sparse} to fit the CD4 data and compare the estimates of the covariance function with the PACE method. We used a fixed number of basis functions and eigenfunctions, i.e., $P = 3$ and $K = 3$. We utilized the PACE method's estimates of the covariance function as the prior covariance function because such an informative prior expresses one's beliefs about this quantity. The tuning parameters for the PACE method were set to default values as described in \cite{wang2016functional}. \textcolor{black}{For the tuning parameters in the ASMC, we employed 100 particles with a resampling threshold of 0.5 and an annealing parameter sequence threshold of 0.9.}

\begin{figure}[htbp]
    \centering
    \subfloat[RB-FPCA-MM (complete data)]{\includegraphics[scale = 0.09]{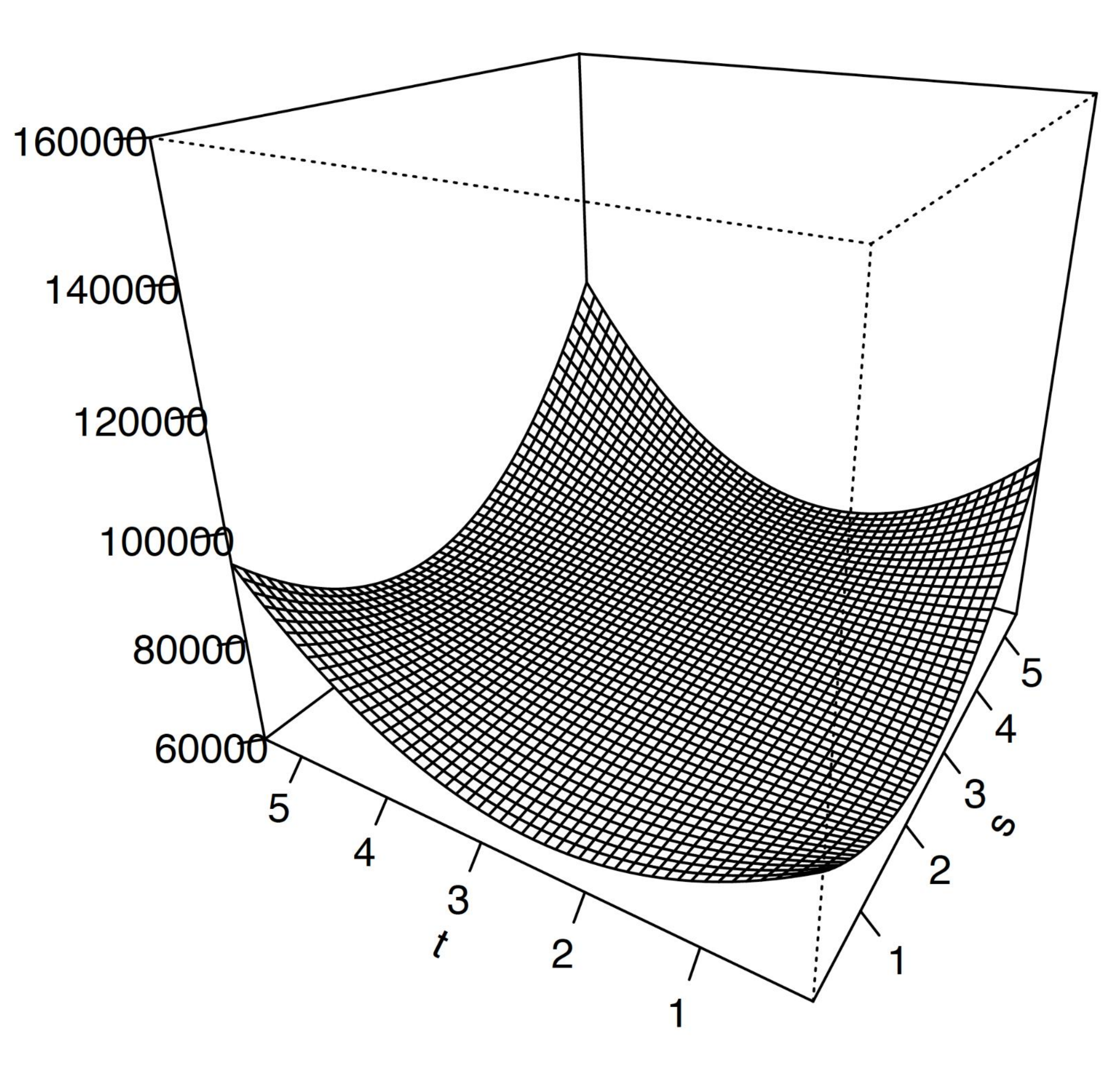}\label{fig:ex3_cov_RBFPCA}} \hspace{0.15\textwidth}
    \subfloat[PACE (complete data)]{\includegraphics[scale = 0.09]{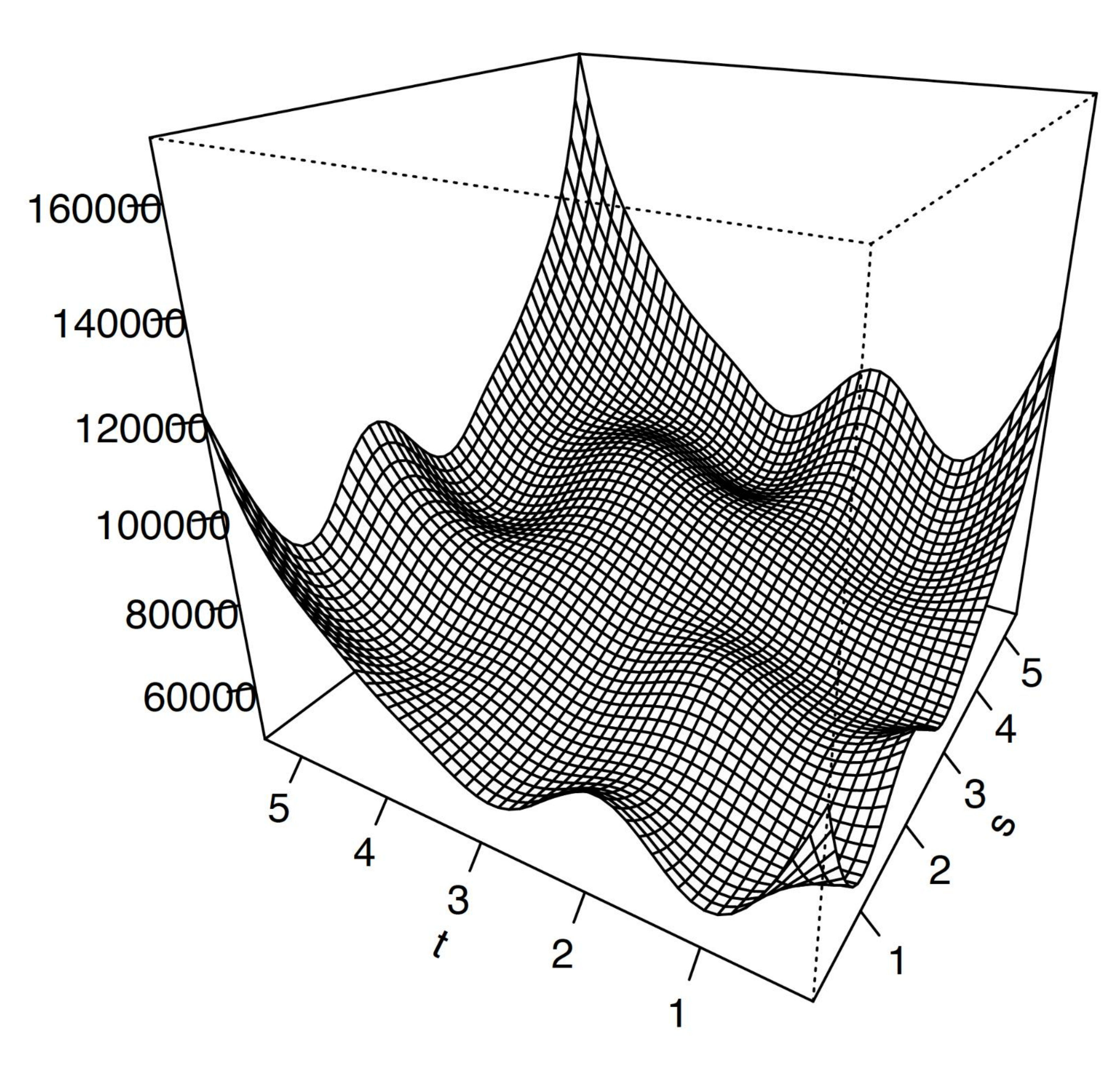}\label{ex3_cov_PACE}} \\
    \subfloat[RB-PFCA-MM (outliers removed)]{\includegraphics[scale = 0.09]{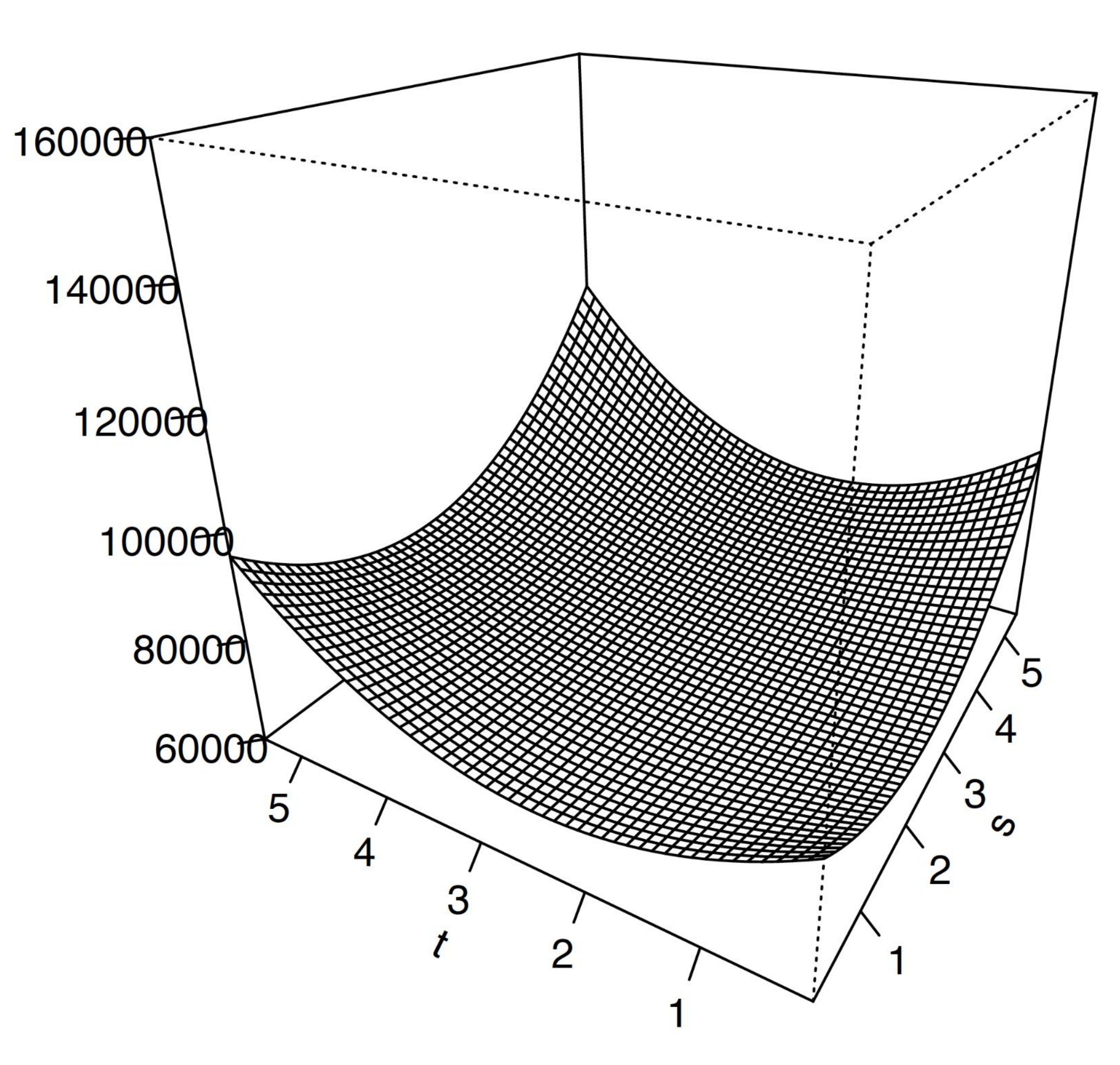}\label{fig:ex3_cov_RBFPCA_outlier_removed}} \hspace{0.15\textwidth}
    \subfloat[PACE (outliers removed)]{\includegraphics[scale = 0.09]{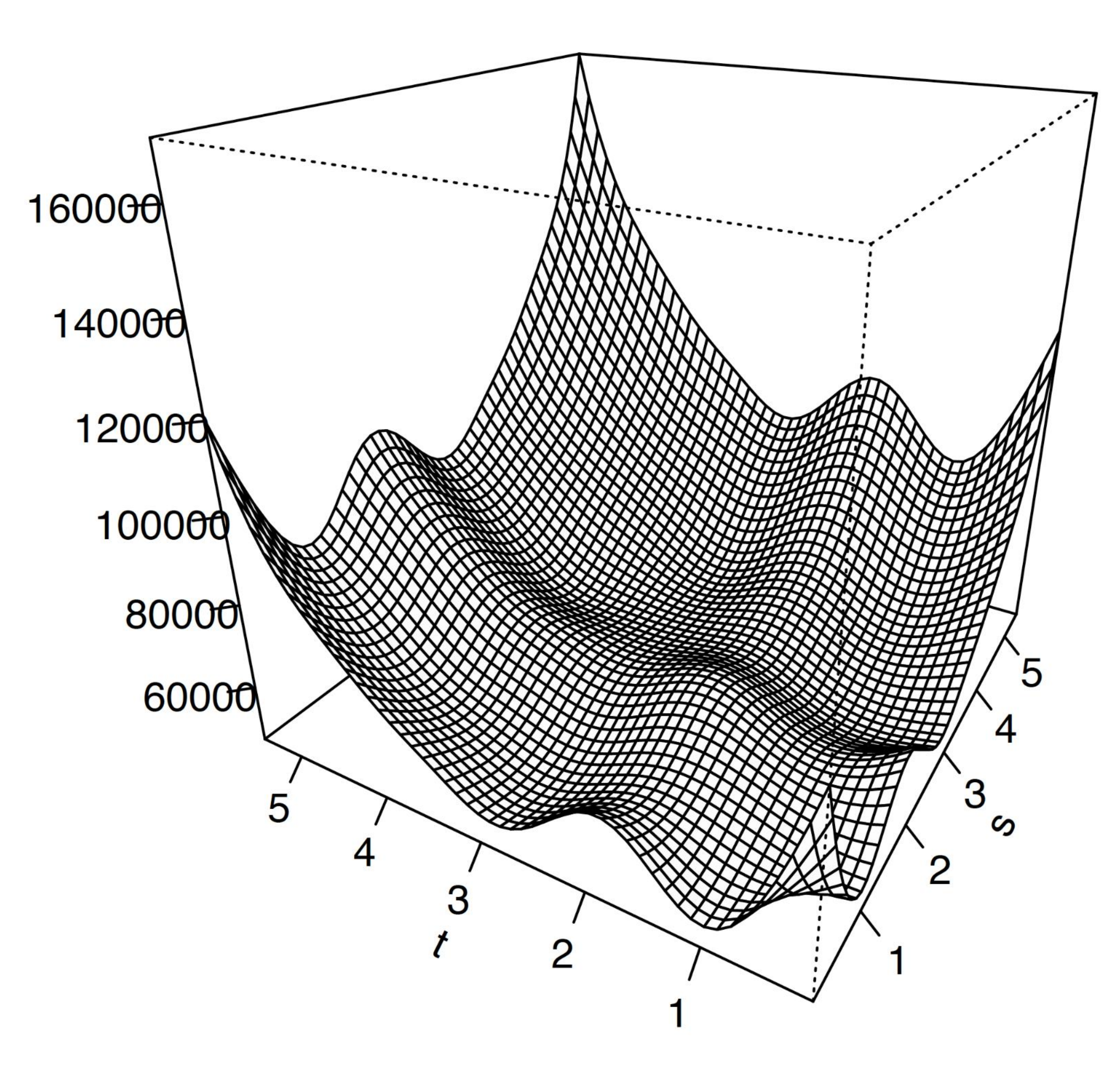}\label{fig:ex3_cov_PACE_outlier_removed}} \vspace{5mm}
    \caption{Sparse CD4 Data: Comparison of smooth estimates of the covariance functions under RB-FPCA-MM and PACE. The covariance surfaces are estimated with all trajectories in \textcolor{black}{(a)} and (b) and with the five most outlying trajectories removed in \textcolor{black}{(c)} and (d). \label{fig:ex3_cov}}
\end{figure}

\textcolor{black}{All three models were evaluated, and the RB-FPCA-MM model was identified as the optimal model, with a log marginal likelihood estimate of -16607.56, compared to -23390.66 for the RB-FPCA-SN model and -23277.29 for the RB-FPCA-ST model. With the optimal model,} we identified some potential outliers using the estimated FPC scores from conditional expectations. The top five most outlying trajectories are identified and highlighted in Figure \ref{fig:ex3_outlier}. These outlying trajectories appear to have either some rapid decreases or increases compared to the rest. They also tend to have overall CD4 cell counts very high compared to the others during the whole period. We estimated the covariance functions using all the trajectories and only normal trajectories. The comparisons of the covariance function estimates under RB-FPCA-MM and PACE are displayed in Figure \ref{fig:ex3_cov}. The RB-FPCA-MM and PACE estimators have similar overall shapes, whereas RB-FPCA-MM produces a more smooth surface. When outliers are removed, RB-FPCA-MM induces a similar covariance surface compared with the estimation from complete data except near the boundary, demonstrating the robustness of the RB-FPCA-MM model.

\section{Discussion}
\label{sec:discussion}

In this work, we developed a robust Bayesian functional principal components analysis method that utilizes the class of skew elliptical distributions in the modelling. The general and flexible class of skew elliptical distributions can serve as an alternative to the symmetric distribution commonly assumed in the previous works of Bayesian FPCA. The proposed method is able to handle the sparse longitudinal data in which only a few observations per trajectory (possibly sampled at irregular intervals) are available. Such data are commonly seen and attracted interest in the area of functional data analysis. We have shown in simulation studies and data applications that the proposed method can effectively capture the major variation and provide useful information in the presence of noises and outliers in the data.

\textcolor{black}{The number of eigenfunctions $K$ is selected to adequately approximate the infinite-dimensional functional data. One straightforward approach is to choose the number of eigenfunctions $K$ such that it explains a sufficiently large portion of the total variation. In this work, we predetermined the value of $K$ in all simulation studies and data analysis to improve computational efficiency. Our results demonstrated that the first few eigenfunctions consistently explained at least 95\% of the total variability in both the simulations and real data, which aligns with the findings in \cite{Sang2017}.
 An alternative way for choosing $K$ in the Bayesian framework is through model comparison using Bayes factors \citep{jeffreys1935some, han2001markov}. The calculation of the Bayes factor requires estimating the marginal likelihood, the normalization factor of the posterior density, which can be obtained without additional computational effort using the ASMC method. Various values of $K$ can be processed in parallel to mitigate the computational cost of evaluating multiple models.}

\textcolor{black}{Several approaches can be employed to select the prior covariance functions $\boldsymbol{\Omega}^{*}$ in the specification of $\boldsymbol{\Psi} = (\boldsymbol{H}_P'\boldsymbol{H}_P)^{-1} \boldsymbol{H}_P' \boldsymbol{\Omega}^{*} \boldsymbol{H}_P (\boldsymbol{H}_P'\boldsymbol{H}_P)^{-1}$, which can be interpreted as a least-squares projection. Various forms of covariance functions---such as exponentiated quadratic, Brownian, and product covariance---can be implemented and compared using the Bayes factor within a Bayesian framework. Alternatively, empirical Bayes methods that estimate the prior covariance from the data may also be incorporated into the analysis to improve estimations. Overall, the significance of prior information is evident in improving the reconstruction accuracy, particularly  when the true covariance or a similar covariance structure is used to construct the prior. As is common practice in the literature, other hyperparameters are set to  noninformative prior distributions.}

\textcolor{black}{The ASMC framework proposed in this work provides several advantages over traditional MCMC algorithms. First, ASMC offers more efficient
exploration of  multimodal posterior surfaces. Second, it produces unbiased estimators of marginal likelihoods as a byproduct of the sampling process, facilitating straightforward model comparison. Third, ASMC is an embarrassingly parallel method, enabling significant computational efficiency by distributing a large number of particles across multiple CPUs or GPUs. This framework has been demonstrated to be efficient in previous studies, including
applications in phylogenetics and nonlinear differential equations 
\citep{wang2020annealed, wang2022adaptive}, as well as in the review by \cite{Dai2022}. These results collectively support the applicability and efficiency of the ASMC framework in the present study.
}

Multiple future directions of research are presented as follows. \textcolor{black}{The finite mixture model can be extended to the infinite mixture model, where the number of mixture components is not predetermined but instead inferred from the data. This is useful when the number of underlying groups is unknown.} Furthermore, the standard FPCA operates in a static way, therefore, does not provide an adequate dimension reduction when considering the functional time series data. \cite{hormann2015dynamic} proposed a dynamic FPCA which takes into account the serial dependence in the functional time series. A Bayesian version of the robust FPCA for functional time series is worth investigating. Regarding posterior inference, there exist popular alternative algorithms, such as reversible jump Markov Chain Monte Carlo, which could provide some advantages in computation. Further research into the computational aspects of the robust Bayesian FPCA methods is necessary.

\begin{appendices}






\end{appendices}



\bibliography{bibliography}

\end{document}


\maketitle

\section{Conditional posterior distributions for finite mixture model}

In this section, we present the conditional posterior distributions for the parameters of the two components finite mixture model in Section 2.4. We construct the mixtures so that $\boldsymbol{\theta}_{1}  = \{ \boldsymbol{\beta}_{i,1}, \boldsymbol{D}_1, \boldsymbol{z}_{i,1}, \boldsymbol{\Sigma}_1 \}$ and $\boldsymbol{\theta}_{2}  = \{ \boldsymbol{\beta}_{i,2}, \boldsymbol{D}_2, \boldsymbol{z}_{i,2}, \boldsymbol{\Sigma}_2, w_{i,2} \}$ are parameters in the multivariate skew normal and multivariate skew t distributions, respectively. Subscripts 1 and 2 are used in the model parameters to denote their corresponding underlying components. We use $| \cdots$ in each conditional distribution to denote conditioning on the data and all other parameters within the specific component.

\begin{itemize}
   \item For $\pi_{1}$, the full conditional distribution is
   \begin{equation}\label{eq:cond_dist_pi1}
       \pi_{1} | \cdots \sim \text{Beta}(1+n_1, 1+n_2),
   \end{equation}
   where $n_1$ and $n_2$ denote the number of observations allocated to components 1 and 2, respectively.

    \item For $i = 1,\ldots,n_1$, the full conditional distribution for $\boldsymbol{\beta}_{i,1}$ is
    \begin{equation}\label{eq:cond_dist_beta1}
        \begin{split}
            \boldsymbol{\beta}_{i,1} | \cdots \sim \text{MVN}_{K}((\boldsymbol{U}_K' \boldsymbol{H}_P' & \boldsymbol{\Sigma}_1^{-1} \boldsymbol{H}_P \boldsymbol{U}_K + \boldsymbol{\Omega}_1^{-1})^{-1}(\boldsymbol{U}_K' \boldsymbol{H}_P' \boldsymbol{\Sigma}_1^{-1} \boldsymbol{Y}_i), \\
            & (\boldsymbol{U}_K' \boldsymbol{H}_P' \boldsymbol{\Sigma}_1^{-1} \boldsymbol{H}_P \boldsymbol{U}_K + \boldsymbol{\Omega}_1^{-1})^{-1}).
        \end{split}
    \end{equation}

    For $i = 1,\ldots,n_2$, the full conditional distribution for $\boldsymbol{\beta}_{i,2}$ is
    \begin{equation}\label{eq:cond_dist_beta2}
        \begin{split}
            \boldsymbol{\beta}_{i,2} | \cdots \sim \text{MVN}_{K}((\boldsymbol{U}_K' \boldsymbol{H}_P' & \boldsymbol{\Sigma}_2^{-1} \boldsymbol{H}_P \boldsymbol{U}_K + \boldsymbol{\Omega}_2^{-1})^{-1}(\boldsymbol{U}_K' \boldsymbol{H}_P' \boldsymbol{\Sigma}_2^{-1} \boldsymbol{Y}_i), \\
            & (\boldsymbol{U}_K' \boldsymbol{H}_P' \boldsymbol{\Sigma}_2^{-1} \boldsymbol{H}_P \boldsymbol{U}_K + \boldsymbol{\Omega}_2^{-1})^{-1}),
        \end{split}
    \end{equation}
    where $\boldsymbol{H}_P$ is picked to be the Legendre polynomial basis functions, and the choice of $\boldsymbol{U}_K$ is discussed in the conditional distribution for $\boldsymbol{\Omega}_1^{-1}$.

    \item For $\boldsymbol{\Omega}_1^{-1}$, the full conditional distribution is
    \begin{equation} \label{eq:cond_dist_Omega1}
         \boldsymbol{\Omega}_1^{-1} | \cdots \sim \text{Wishart}_K(\nu + n_1 + 1, (\boldsymbol{L}_k + \sum^{n_1}_{i = 1} \boldsymbol{\beta}_{i,1} \boldsymbol{\beta}_{i,1}')^{-1}).
    \end{equation}
    
    For $\boldsymbol{\Omega}_2^{-1}$, the full conditional distribution is
    \begin{equation} \label{eq:cond_dist_Omega2}
         \boldsymbol{\Omega}_2^{-1} | \cdots \sim \text{Wishart}_K(\nu + n_2 + 1, (\boldsymbol{L}_k + \sum^{n_2}_{i = 1} \boldsymbol{\beta}_{i,2} \boldsymbol{\beta}_{i,2}')^{-1}),
    \end{equation}
    
    where $\nu$ is the number of degrees of freedom. $\boldsymbol{L}_K$ and $\boldsymbol{U}_K$ are derived by first decomposing $\boldsymbol{\Psi}= \boldsymbol{U} \boldsymbol{L} \boldsymbol{U}'$, where $\boldsymbol{\Psi}$ is the inverse of the scale matrix in Wishart prior for $\boldsymbol{\Omega}^{-1}$ when $K=P$. The matrices $\boldsymbol{L}$ and $\boldsymbol{U}$ are then subsetted to obtain the desired $\boldsymbol{L}_K$ and $\boldsymbol{U}_K$ for $K \leq P$. A sensible choice of $\boldsymbol{\Psi}$ is given as $\boldsymbol{\Psi} = (\boldsymbol{H}_P'\boldsymbol{H}_P)^{-1} \boldsymbol{H}_P' \boldsymbol{\Omega}^{*} \boldsymbol{H}_P (\boldsymbol{H}_P'\boldsymbol{H}_P)^{-1}$, where $\boldsymbol{\Omega}^{*}$ is the prior covariance function corresponding to the time grid being used. 
    
    \item For $i = 1,\ldots,n_1$, the full conditional distribution for $\boldsymbol{z}_{i,1}$ is
    \begin{equation} \label{eq:cond_dist_z1}
        \boldsymbol{z}_{i,1} | \cdots \sim \text{MVN}_{m}(\boldsymbol{A}_{i,1}^{-1} \boldsymbol{a}_{i,1}, \boldsymbol{A}_{i,1}^{-1})I(\boldsymbol{z}_{i,1} > 0),
    \end{equation}
    where $\boldsymbol{A}_{i,1} = \boldsymbol{I} + \boldsymbol{D}_1 \boldsymbol{\Sigma}_1^{-1} \boldsymbol{D}_1$ and $\boldsymbol{a}_{i,1} = \boldsymbol{D}_1 \boldsymbol{\Sigma}_1^{-1}(\boldsymbol{Y}_i - \boldsymbol{H}_P \boldsymbol{U}_K \boldsymbol{\beta}_{i,1})$. 

    For $i = 1,\ldots,n_2$, the full conditional distribution for $\boldsymbol{z}_{i,2}$ is
    \begin{equation} \label{eq:cond_dist_z2}
        \boldsymbol{z}_{i,2} | \cdots \sim \text{MVN}_{m}(\boldsymbol{A}_{i,2}^{-1} \boldsymbol{a}_{i,2}, \boldsymbol{A}_{i,2}^{-1})I(\boldsymbol{z}_{i,2} > 0),
    \end{equation}
    where $\boldsymbol{A}_{i,2} = \boldsymbol{I} + w_{i,2} \boldsymbol{D}_2 \boldsymbol{\Sigma}_2^{-1} \boldsymbol{D}_2$ and $\boldsymbol{a}_{i,2} = w_{i,2} \boldsymbol{D}_2 \boldsymbol{\Sigma}_2^{-1}(\boldsymbol{Y}_i - \boldsymbol{H}_P \boldsymbol{U}_K \boldsymbol{\beta}_{i,2})$. 
    
    \item For $\text{vec}(\boldsymbol{D}_1)$, the full conditional distribution is
    \begin{equation} \label{eq:cond_dist_vecD1}
        \text{vec}(\boldsymbol{D}_1) | \cdots \sim \text{MVN}_{m}(\boldsymbol{B}_1^{-1}\boldsymbol{b}_1, \boldsymbol{B}_1^{-1}),
    \end{equation}
    with $\boldsymbol{B}_1 = \boldsymbol{\Gamma}^{-1} + \sum^{n_1}_{i = 1} \text{diag}(\boldsymbol{z}_{i,1}) \boldsymbol{\Sigma}_1^{-1} \text{diag}(\boldsymbol{z}_{i,1})$ and \\ $\boldsymbol{b}_1 = \sum^{n_1}_{i = 1} \text{diag}(\boldsymbol{z}_{i,1}) \boldsymbol{\Sigma}_1^{-1} (\boldsymbol{Y}_i - \boldsymbol{H}_P \boldsymbol{U}_K \boldsymbol{\beta}_{i,1})$.

    For $\text{vec}(\boldsymbol{D}_2)$, the full conditional distribution is
    \begin{equation} \label{eq:cond_dist_vecD2}
        \text{vec}(\boldsymbol{D}_2) | \cdots \sim \text{MVN}_{m}(\boldsymbol{B}_2^{-1}\boldsymbol{b}_2, \boldsymbol{B}_2^{-1}),
    \end{equation}
    with $\boldsymbol{B}_2 = \boldsymbol{\Gamma}^{-1} + \sum^{n_2}_{i = 1} \text{diag}(\boldsymbol{z}_{i,2}) \boldsymbol{\Sigma}_2^{-1} \text{diag}(\boldsymbol{z}_{i,2})$ and \\ $\boldsymbol{b}_2 = \sum^{n_2}_{i = 1} \text{diag}(\boldsymbol{z}_{i,2}) \boldsymbol{\Sigma}_2^{-1} (\boldsymbol{Y}_i - \boldsymbol{H}_P \boldsymbol{U}_K \boldsymbol{\beta}_{i,2})$.
    
    \item For $\boldsymbol{\Sigma}_1^{-1}$, the full conditional distribution is
    \begin{equation} \label{eq:cond_dist_Sigma1}
    \begin{split}
        \boldsymbol{\Sigma}_1^{-1} | & \cdots \sim \text{Wishart}_{m}(2r + n_1, ((2\boldsymbol{\kappa})^{-1} + \\
        & \sum^{n_1}_{i = 1} (\boldsymbol{Y}_i - \boldsymbol{H}_P \boldsymbol{U}_K \boldsymbol{\beta}_{i,1} - \boldsymbol{D}_1\boldsymbol{z}_{i,1}) (\boldsymbol{Y}_i - \boldsymbol{H}_P \boldsymbol{U}_K \boldsymbol{\beta}_{i,1} - \boldsymbol{D}_1\boldsymbol{z}_{i,1})')^{-1}.
    \end{split}
    \end{equation}

    For $\boldsymbol{\Sigma}_2^{-1}$, the full conditional distribution is
    \begin{equation} \label{eq:cond_dist_Sigma2}
    \begin{split}
        \boldsymbol{\Sigma}_2^{-1} | & \cdots \sim \text{Wishart}_{m}(2r + n_2, ((2\boldsymbol{\kappa})^{-1} + \\
        & \sum^{n_2}_{i = 1} (\boldsymbol{Y}_i - \boldsymbol{H}_P \boldsymbol{U}_K \boldsymbol{\beta}_{i,2} - \boldsymbol{D}_2\boldsymbol{z}_{i,2}) (\boldsymbol{Y}_i - \boldsymbol{H}_P \boldsymbol{U}_K \boldsymbol{\beta}_{i,2} - \boldsymbol{D}_2\boldsymbol{z}_{i,2})')^{-1}.
    \end{split}
    \end{equation}

    \item For $i = 1,\ldots,n_2$, the full conditional distribution for $w_{i,2}$ is
    \begin{equation} \label{eq:cond_dist_wi}
        w_{i,2} | \cdots \sim \text{Gamma}(\nu_{w_{i,2}}/2 + m/2, \nu_{w_{i,2}}/2 + m \times s_i / 2),
    \end{equation}
    where $s_i$ is the sample variance of the measurements in the $i$th curve in component 2.

    \item For $i = 1,\ldots,n$, the full conditional distribution for $\boldsymbol{\tau}_i$ is 
    \begin{equation} \label{eq:cond_dist_tau}
        \boldsymbol{\tau}_i | \cdots \sim \text{Bernoulli}\left(\frac{A}{A+B}\right),
    \end{equation}
    where \\ $A = \pi_1 (2\pi)^{-m/2}\text{det}(\boldsymbol{\Sigma}_1)^{-1/2} \exp\{-(\boldsymbol{Y}_i - \boldsymbol{H}_P \boldsymbol{U}_K \boldsymbol{\beta}_{i,1} - \boldsymbol{D}_1\boldsymbol{z}_{i,1})' \boldsymbol{\Sigma}_1^{-1} (\boldsymbol{Y}_i - \boldsymbol{H}_P \boldsymbol{U}_K \boldsymbol{\beta}_{i,1} - \boldsymbol{D}_1\boldsymbol{z}_{i,1})/2 \} $ and \\ $B = (1-\pi_1) (2\pi)^{-m/2}\text{det}(\boldsymbol{\Sigma}_2 / w_{i,2})^{-1/2} \exp\{-(\boldsymbol{Y}_i - \boldsymbol{H}_P \boldsymbol{U}_K \boldsymbol{\beta}_{i,2} - \boldsymbol{D}_2\boldsymbol{z}_{i,2})' \\(\boldsymbol{\Sigma}_2 / w_{i,2})^{-1} (\boldsymbol{Y}_i - \boldsymbol{H}_P \boldsymbol{U}_K \boldsymbol{\beta}_{i,2} - \boldsymbol{D}_2\boldsymbol{z}_{i,2})/2 \} $.

\end{itemize}

\section{Annealed sequential Monte Carlo}
Algorithm \ref{algo:adapt} describes the annealed sequential Monte Carlo for Bayesian inference of the target distribution $\pi(\boldsymbol{\theta})\propto p(\ybold|\boldsymbol{\theta})\pi_0(\boldsymbol{\theta})$, where  $\boldsymbol{\theta}$ denotes all the unknown parameters, $\ybold$ represents all the data,   $p(\ybold|\boldsymbol{\theta})$ is the likelihood function, and $\pi_0(\boldsymbol{\theta})$ is the prior distribution for $\boldsymbol{\theta}$. We consider the following sequence of intermediate distributions 
$$\pi_r(\boldsymbol{\theta})\propto [p(\ybold|\boldsymbol{\theta})]^{\alpha_r}\pi_0(\boldsymbol{\theta}),$$
where $\alpha_r$ is a sequence of annealing parameters ranging from 0 to 1. Let $\pi_R$ denote the last intermediate distribution, which is the posterior distribution.  

\begin{algorithm}
   \caption{\bf{ASMC algorithm for Bayesian inference}}
  \label{algo:adapt}
{\fontsize{10pt}{10pt}\selectfont
\begin{algorithmic}[1]
  \State {\bfseries Inputs:} (a) The prior distribution $\pi_{0}(\cdot)$ over model parameters $\boldsymbol{\theta}$;  (b) relative CESS (rCESS) {\color{black} threshold} $\iota$; (c) resampling threshold $\varsigma$; (d) the number of particles $\mathcal{K}$. 

   \State {\bfseries Outputs:}   (a) Posterior approximation, $\hat{\pi}(\boldsymbol{\theta}) = \sum_{k = 1}^{\mathcal{K}}W_{R}^{(k)}\cdot \delta_{\boldsymbol{\theta}_{R}^{(k)}}(\boldsymbol{\theta})$; (b) Approximation $\hat{p}(\ybold)$ of the marginal likelihood, $p(\ybold)=\int p(\ybold|\boldsymbol{\theta})\pi_0(\boldsymbol{\theta}) d\boldsymbol{\theta}$.

	\State  Initialize the SMC iteration index and annealing parameter: $r \leftarrow 0$, $\alpha_0 \leftarrow 0$.  
   \For{$k \in \{1, 2, \dots, \mathcal{K}\}$}
    \State  Initialize particles, $\boldsymbol{\theta}_{0}^{(k)}$,  with independent samples from the prior distribution.
	\State  Initialize unnormalized weights $\omega_{0}^{(k)} \leftarrow 1$, normalized weights $W_{0}^{(k)} \leftarrow 1/\mathcal{K}$, and $\hat{p}(\ybold) \leftarrow 1$. 
\EndFor    
 \For{$r\in \{1, 2, \dots\}$}
    \State Compute annealing parameter $\alpha_{r}$ using {\color{black} a} bisection method with $$f(\alpha_{r}) = \text{rCESS}\left(W_{r-1}^{(\cdot)}, \left( p(\ybold|\boldsymbol{\theta}_{r-1}^{(\cdot)})\right)^{\alpha_{r} - \alpha_{r-1}} \right) =\iota.$$ 
    \For{$k \in \{1, \dots, \mathcal{K}\}$}
      \State \label{step:adapt-weigh} Compute unnormalized weights for $\boldsymbol{\theta}_{r}^{(k)}$: $$\omega_{r}^{(k)} =  \omega_{r-1}^{(k)}\cdot  \left( p(\ybold|\boldsymbol{\theta}_{r-1}^{(k)})\right)^{\alpha_{r} - \alpha_{r-1}}.$$
      \State Normalize weights: $W_{r}^{(k)}=\omega_{r}^{(k)} /(\sum_{k=1}^{\mathcal{K}}\omega_{r}^{(k)})$.
        \State Sample particles $\boldsymbol{\theta}_{r}^{(k)}$ with one step of the Gibbs sampler followed by one Metropolis-Hastings (MH) step admitting $\pi_r$ as the invariant distribution, using particles ${\boldsymbol{\theta}}_{r-1}^{(k)}$. 
    \EndFor    
    \State  $\hat{p}(\ybold) \leftarrow \hat{p}(\ybold) \sum_{k=1}^\mathcal{K} W_{r-1}^{k} \left( p(\ybold|\boldsymbol{\theta}_{r-1}^{(k)})\right)^{\alpha_{r} - \alpha_{r-1}}$. 
     \If{$\alpha_r = 1$} 
   \State \label{step:adapt-return} return the current particle population $\{(\boldsymbol{\theta}_{r}^{(k)}, W_{r}^{(k)})\}_{k = 1}^{\mathcal{K}}$ and $\hat{p}(\ybold)$.
   \Else
  \If{$\text{rESS} < \varsigma$}  
		\State Resample the particles. 
    \For{$k \in \{1, \dots, \mathcal{K}\}$}
    \State \label{step:adapt-reset-weights} Reset particle weights: $\omega_{r}^{(k)} = 1, W_{r}^{(k)} = 1/\mathcal{K}$.
    \EndFor
   \EndIf
      \EndIf
\EndFor 
 
\end{algorithmic}
}
\end{algorithm}

\clearpage
\section{Additional results for Simulation II}
\label{sec:supp_sim1}

In this simulation, we examine dense functional data with outliers from different noise distributions. We add independent noises to the sampling process in the form of different noise distributions, such as Student-$t$ distribution, skew Normal distribution and skew Student-$t$ distribution. Each setting is repeated 30 times.

\subsection{noise $\sim$ $t(5)$}

Student-$t$ distribution with df = 5

\begin{figure}[htbp]
  \centering
  \includegraphics[width=0.9\textwidth]{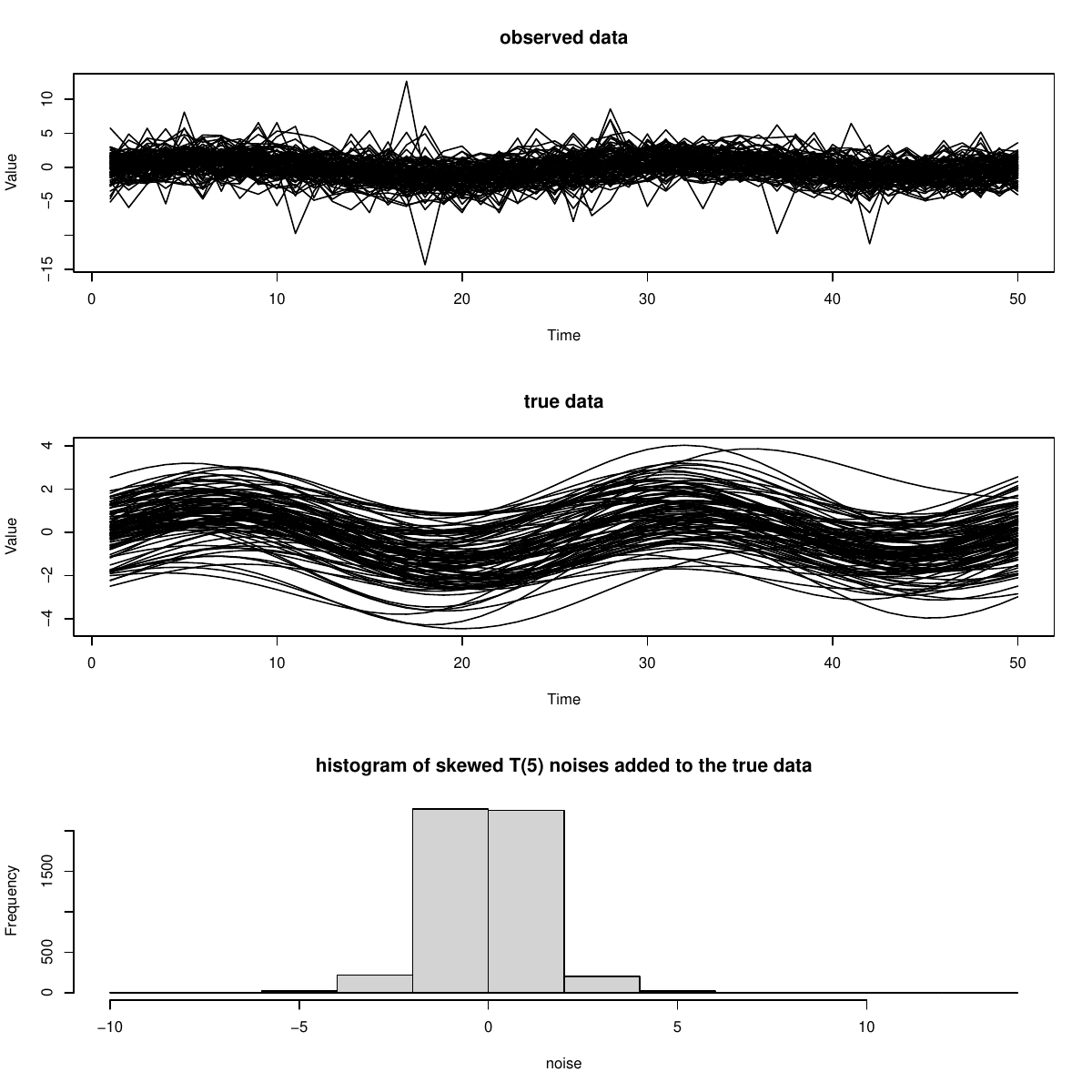}
  \caption{Simulation II with $t(5)$ noises: One visualization of data and noises when the true covariance is $\text{Cov}(s,t) = \exp\{-3(t-s)^2\}$.} \label{fig:sim2_data_t_1}
\end{figure}

\begin{figure}[htbp]
  \centering
  \includegraphics[width=0.9\textwidth]{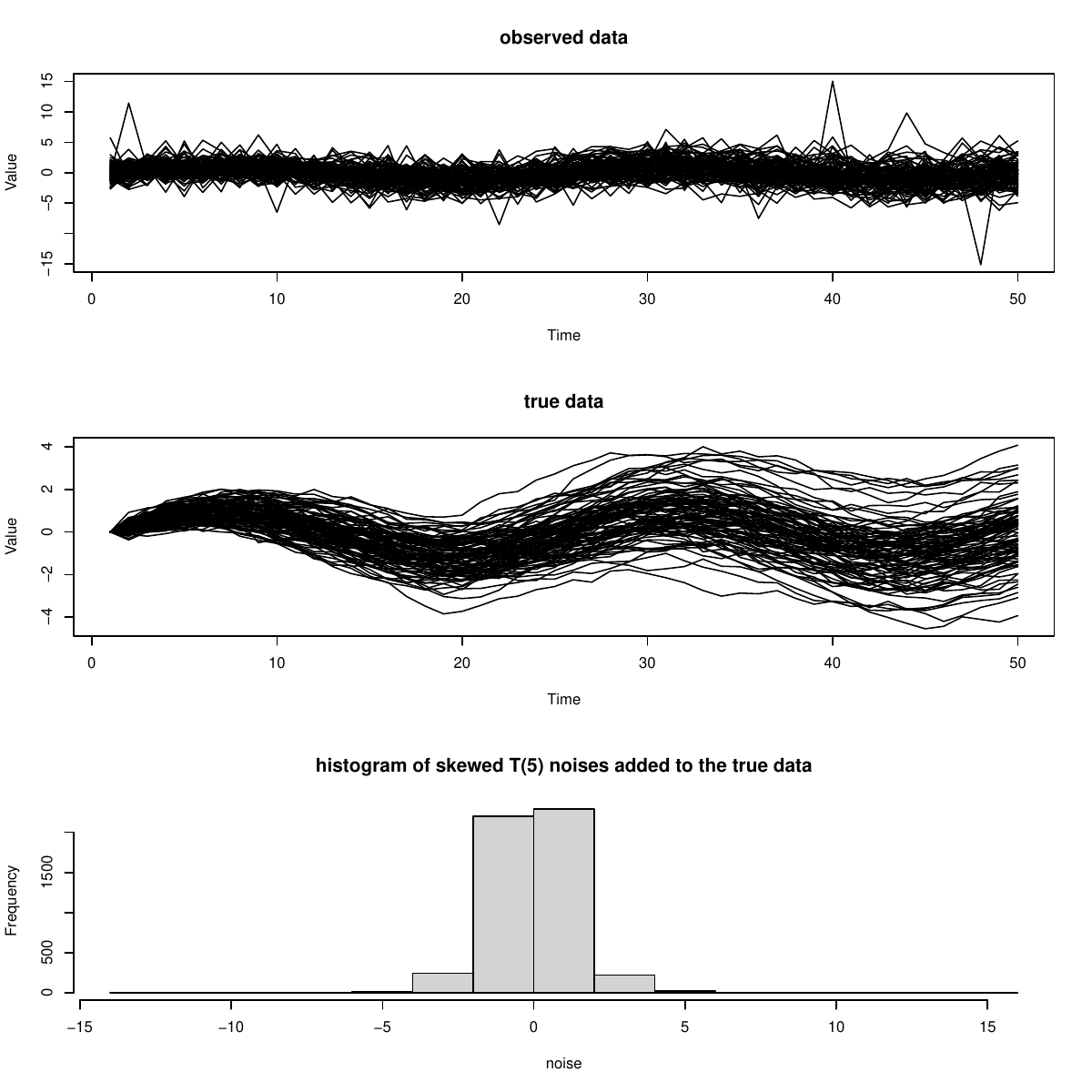}
  \caption{Simulation II with $t(5)$ noises: One visualization of data and noises when the true covariance is $\text{Cov}_2(s,t) = \text{min} \{s+1, t+1\}$.} \label{fig:sim2_data_t_2}
\end{figure}

\begin{table}[htbp]
\small
\begin{tabular}{|c|c|c|c|c|c|c|c|}
\hline
 & Truth & \begin{tabular}[c]{@{}c@{}}RB-FPCA\\ with \\ Prior \\ Cov1\end{tabular} & \begin{tabular}[c]{@{}c@{}}RB-FPCA \\ with \\ Prior \\ Cov2\end{tabular} & \begin{tabular}[c]{@{}c@{}}BFPCA \\ with \\ Prior \\Cov1\end{tabular} & \begin{tabular}[c]{@{}c@{}}BFPCA \\ with \\ Prior \\Cov2\end{tabular} & FACE & PACE \\ \hline
\multirow{2}{*}{Cov} & Cov1 & 5.291 & 7.562 & \textbf{4.726} & 7.760 & 5.371 & 8.964 \\ 
 & Cov2 & 8.991 & 8.007 & 8.429 & 8.224 & \textbf{7.085} & 11.601 \\ \hline
\end{tabular}
\caption{Simulation II with $t(5)$ noises: Estimations of the covariance function are evaluated by the distances between the estimates and the true covariance function using the $L_2$ norm.  \label{tab:sim1-cov_est_t}}
\end{table}

\begin{figure}[htbp]
  \centering
  \includegraphics[width=0.9\textwidth]{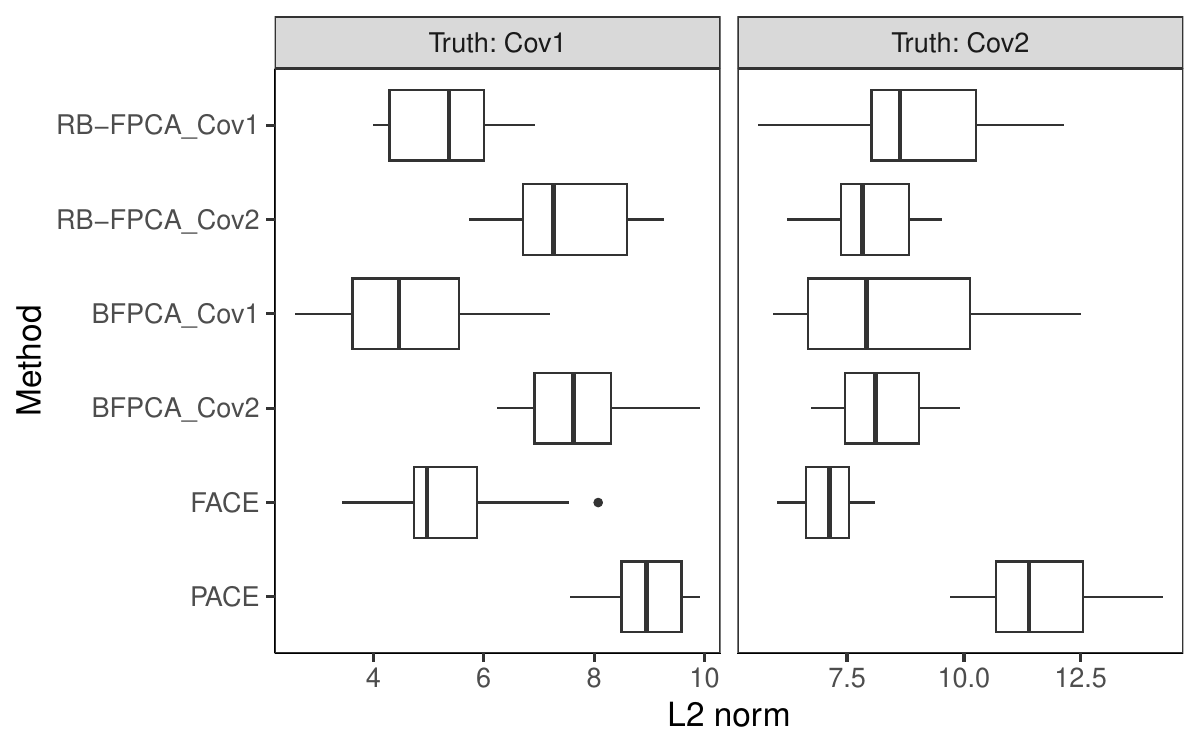}
  \caption{Simulation II with $t(5)$ noises: Boxplots of the distance in the $L_2$ norm between the estimates and the true covariance function.} \label{fig:sim1_cov_boxplot_t}
\end{figure}

\begin{table}[htbp]
\small
\begin{tabular}{|c|c|c|c|c|c|c|c|}
\hline
 & Truth & \begin{tabular}[c]{@{}c@{}}RB-FPCA\\ with \\ Prior \\Cov1\end{tabular} & \begin{tabular}[c]{@{}c@{}}RB-FPCA \\ with \\ Prior \\ Cov2\end{tabular} & \begin{tabular}[c]{@{}c@{}}BFPCA \\ with \\ Prior \\Cov1\end{tabular} & \begin{tabular}[c]{@{}c@{}}BFPCA \\ with \\ Prior \\Cov2\end{tabular} & FACE & PACE \\ \hline
\multirow{2}{*}{PC1} & Cov1 & 0.053 & 0.062 & 0.031 & 0.040 & \textbf{0.017} & 0.047 \\
 & Cov2 & 0.030 & \textbf{0.016} & 0.024 & 0.024 & 0.047 & 0.048 \\ \hline
\multirow{2}{*}{PC2} & Cov1 & 0.019 & 0.029 & 0.018 & 0.024 & \textbf{0.017} & 0.077 \\ 
 & Cov2 & 0.056 & 0.047 & 0.063 & 0.047 & 0.048 & \textbf{0.040} \\ \hline
\multirow{2}{*}{PC3} & Cov1 & 0.017 & 0.036 & \textbf{0.012} & 0.040 & 0.025 & 0.041 \\ 
 & Cov2 & 0.016 & \textbf{0.007} & 0.032 & 0.017 & 0.029 & 0.041 \\ \hline
\end{tabular}
\caption{Simulation II with $t(5)$ noises: Estimations of the first 3 principal components are compared with the Mean Squared Errors (MSEs) between the estimates and the true PCs. \label{tab:sim1-PCs_est_t}}
\end{table}

\clearpage

\subsection{noise $\sim$ $SN(0,1,5)$}

skewed normal with location = 0, scale = 1, shape = 5 (mean $\approx$ 0.78, variance $\approx$ 0.38)

\begin{figure}[htbp]
  \centering
  \includegraphics[width=0.9\textwidth]{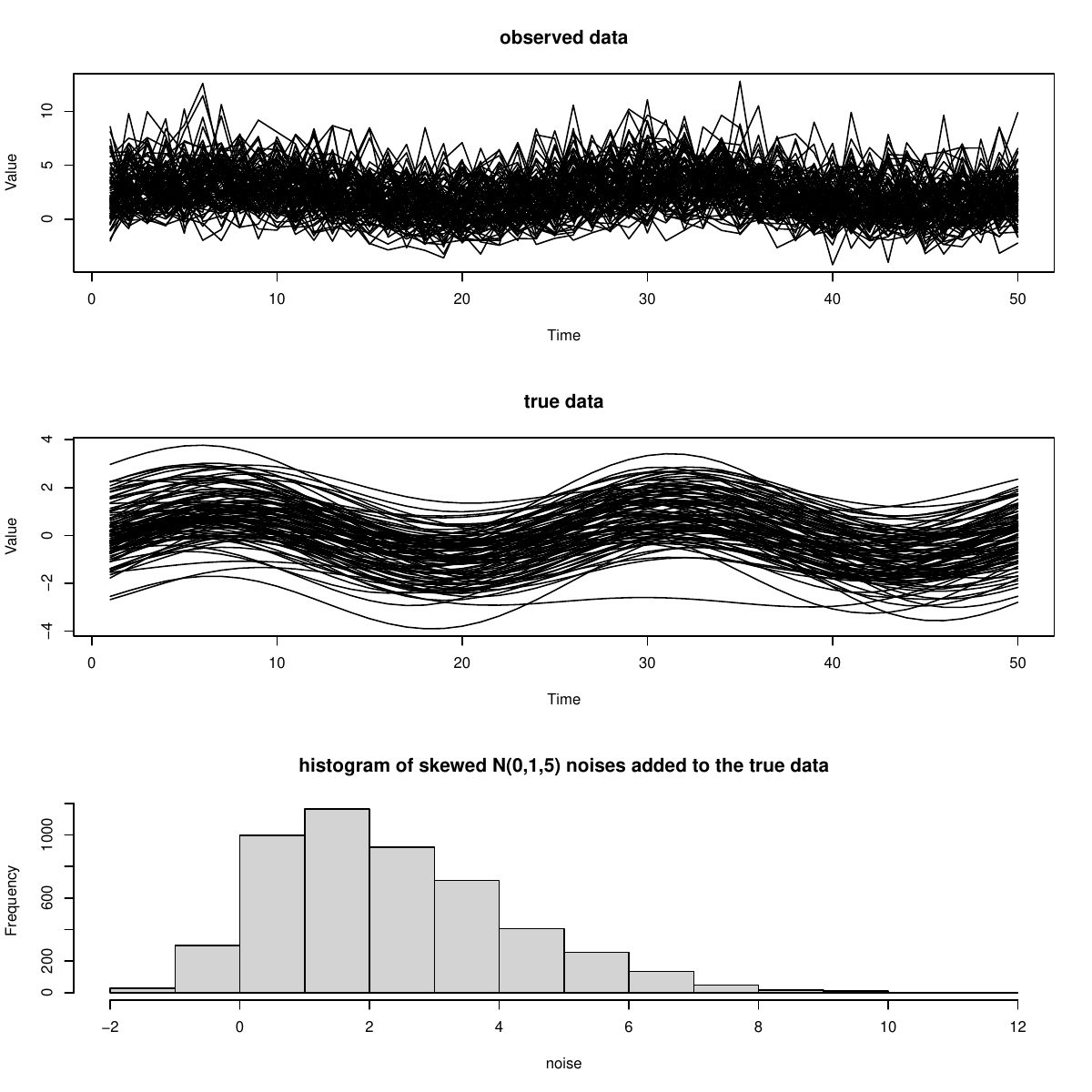}
  \caption{Simulation II with $SN(0,1,5)$ noises: One visualization of data and noises when the true covariance is $\text{Cov}(s,t) = \exp\{-3(t-s)^2\}$.} \label{fig:sim2_data_sn_1}
\end{figure}

\begin{figure}[htbp]
  \centering
  \includegraphics[width=0.9\textwidth]{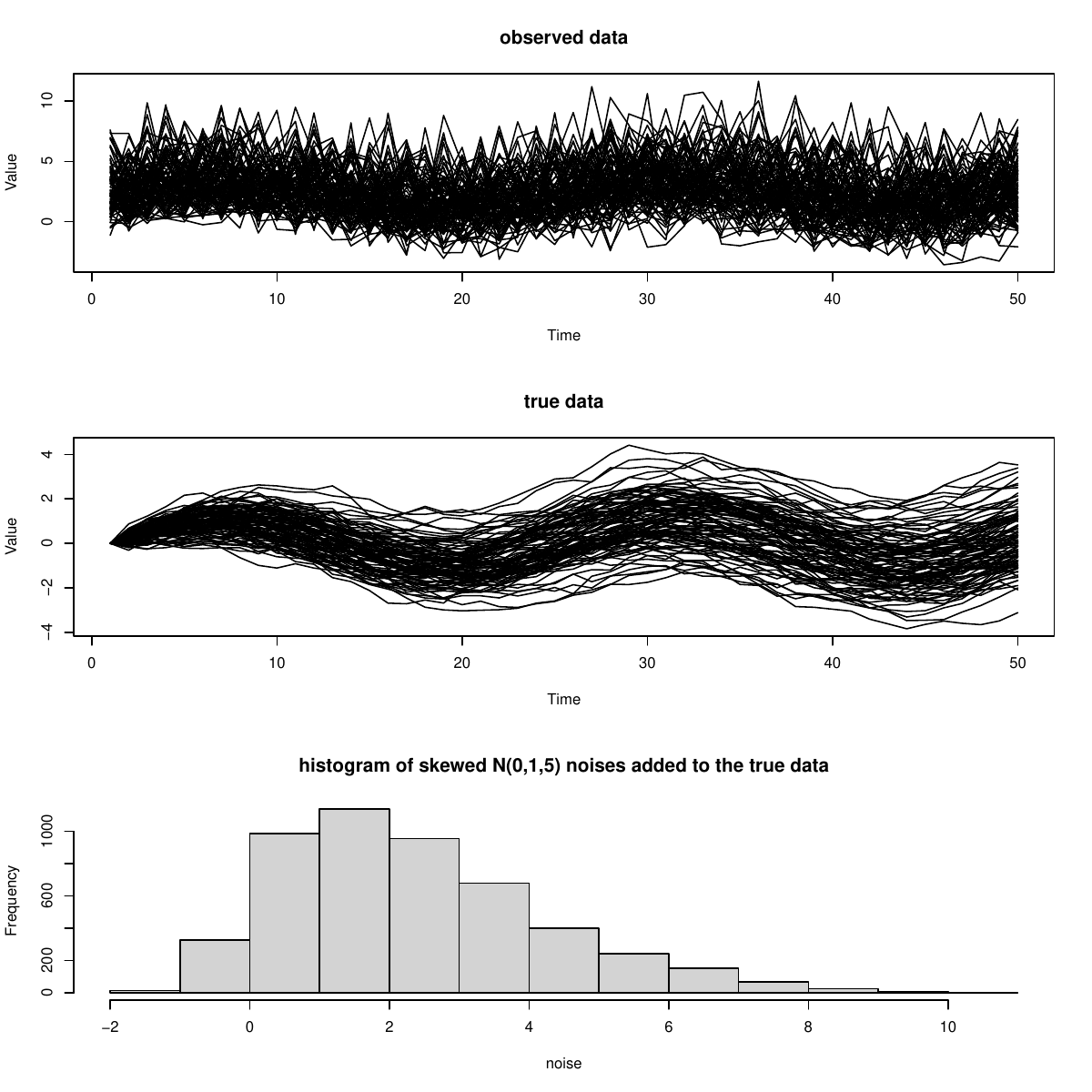}
  \caption{Simulation II with $SN(0,1,5)$ noises: One visualization of data and noises when the true covariance is $\text{Cov}_2(s,t) = \text{min} \{s+1, t+1\}$.} \label{fig:sim2_data_sn_2}
\end{figure}

\clearpage

\subsection{noise $\sim$ $ST(0,1,5,5)$}

skewed $t$ with location = 0, scale = 1, shape = 5, df = 5 (mean $\approx$ 0.93, variance $\approx$ 0.80)

\begin{figure}[htbp]
  \centering
  \includegraphics[width=0.9\textwidth]{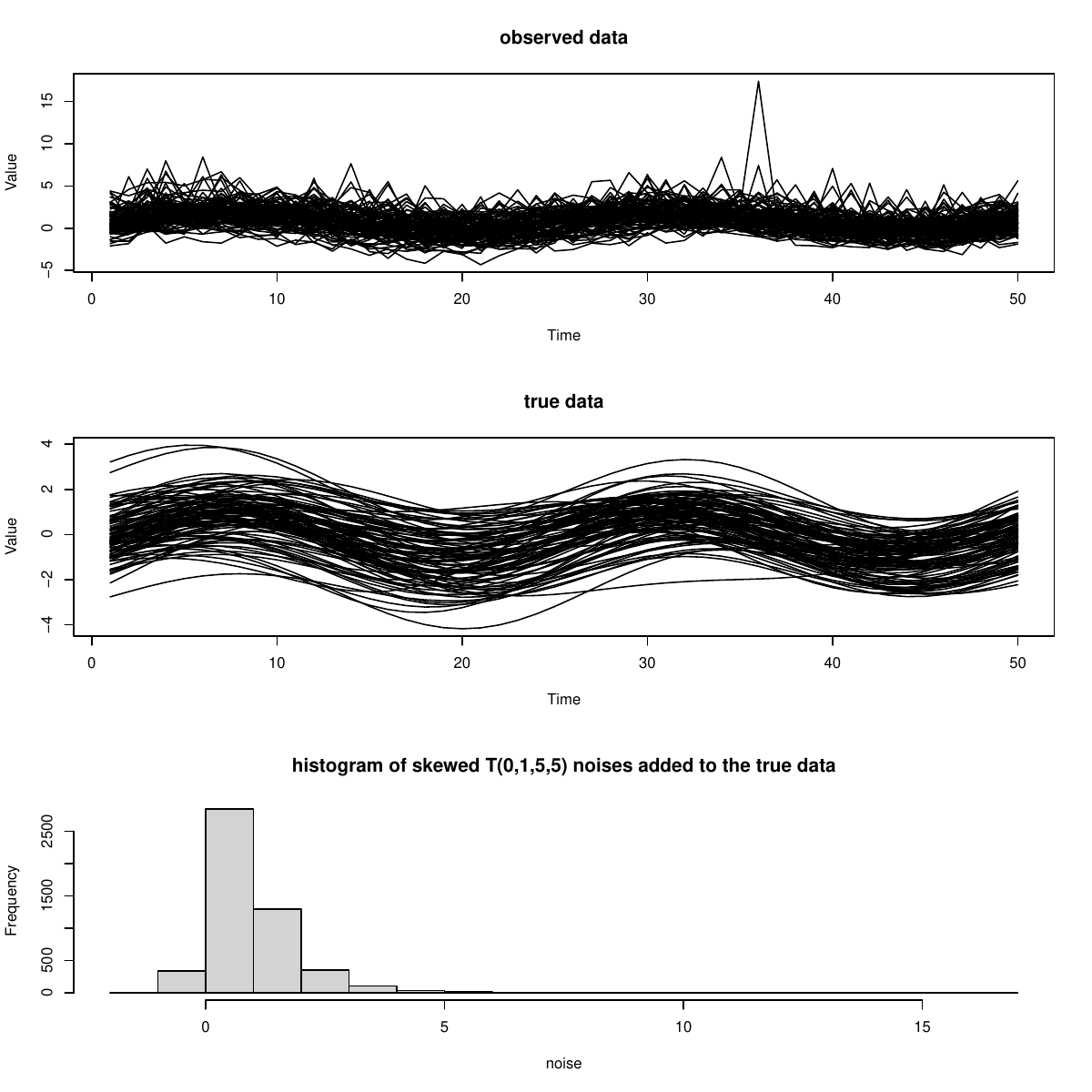}
  \caption{Simulation II with $ST(0,1,5,5)$ noises: One visualization of data and noises when the true covariance is $\text{Cov}(s,t) = \exp\{-3(t-s)^2\}$.} \label{fig:sim1_data_st_1}
\end{figure}

\begin{figure}[htbp]
  \centering
  \includegraphics[width=0.9\textwidth]{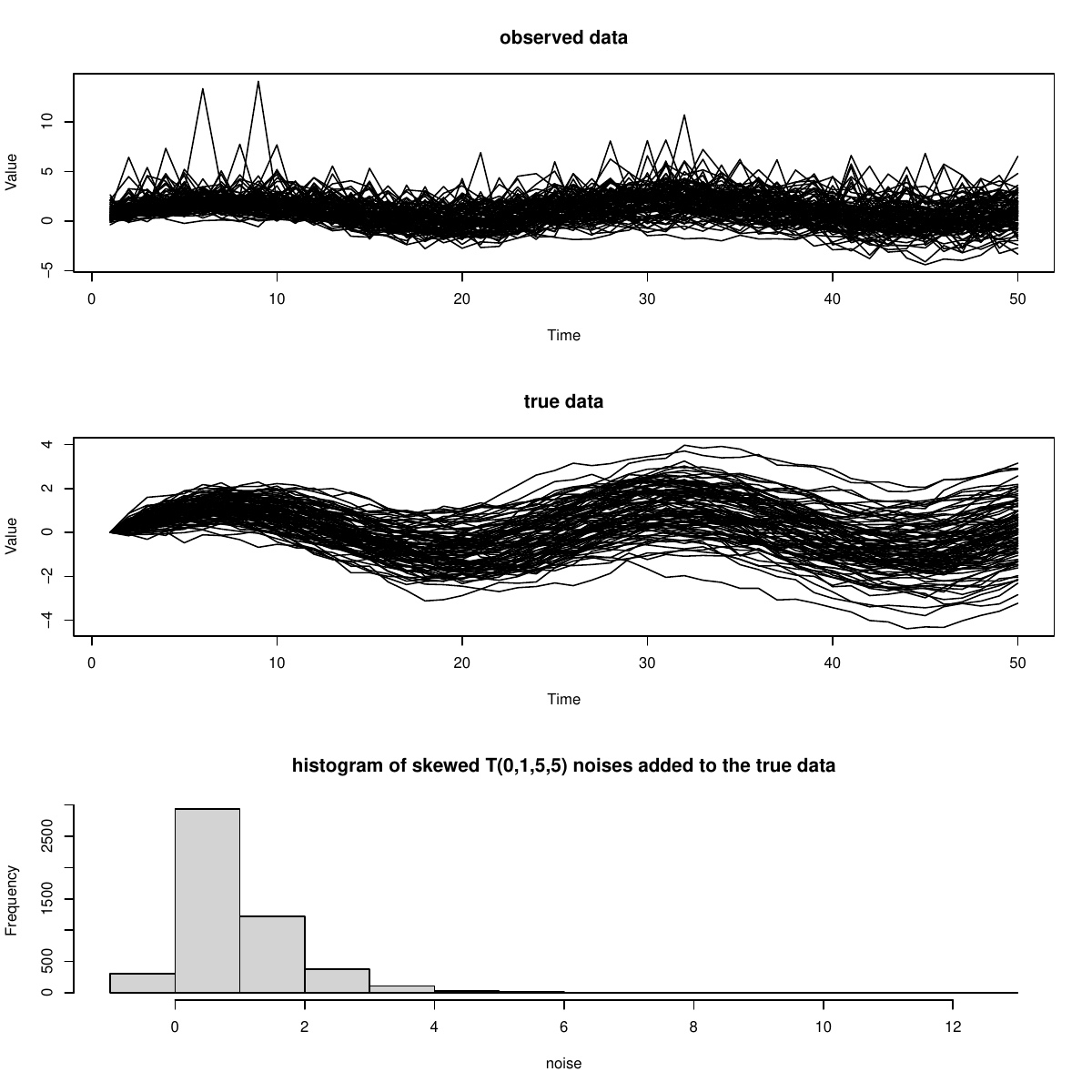}
  \caption{Simulation II with $ST(0,1,5,5)$ noises: One visualization of data and noises when the true covariance is $\text{Cov}_2(s,t) = \text{min} \{s+1, t+1\}$.} \label{fig:sim1_data_st_2}
\end{figure}

\begin{table}[htbp]
\small
\begin{tabular}{|c|c|c|c|c|c|c|c|}
\hline
 & Truth & \begin{tabular}[c]{@{}c@{}}RB-FPCA\\ with \\ Prior \\Cov1\end{tabular} & \begin{tabular}[c]{@{}c@{}}RB-FPCA \\ with \\ Prior \\Cov2\end{tabular} & \begin{tabular}[c]{@{}c@{}}BFPCA \\ with \\ Prior \\Cov1\end{tabular} & \begin{tabular}[c]{@{}c@{}}BFPCA \\ with \\ Prior \\Cov2\end{tabular} & FACE & PACE \\ \hline
\multirow{2}{*}{Cov} & Cov1 & 5.149 & 6.673 & 5.323 & 7.930 & \textbf{4.351} & 8.136 \\ 
 & Cov2 & 10.589 & \textbf{6.535} & 10.948 & 11.267 & 6.541 & 10.111 \\ \hline
\end{tabular}
\caption{Simulation II with $ST(0,1,5,5)$ noises: Estimations of the covariance function are evaluated by the distances between the estimates and the true covariance function using the $L_2$ norm.  \label{tab:sim1-cov_est_st}}
\end{table}

\begin{figure}[htbp]
  \centering
  \includegraphics[width=0.9\textwidth]{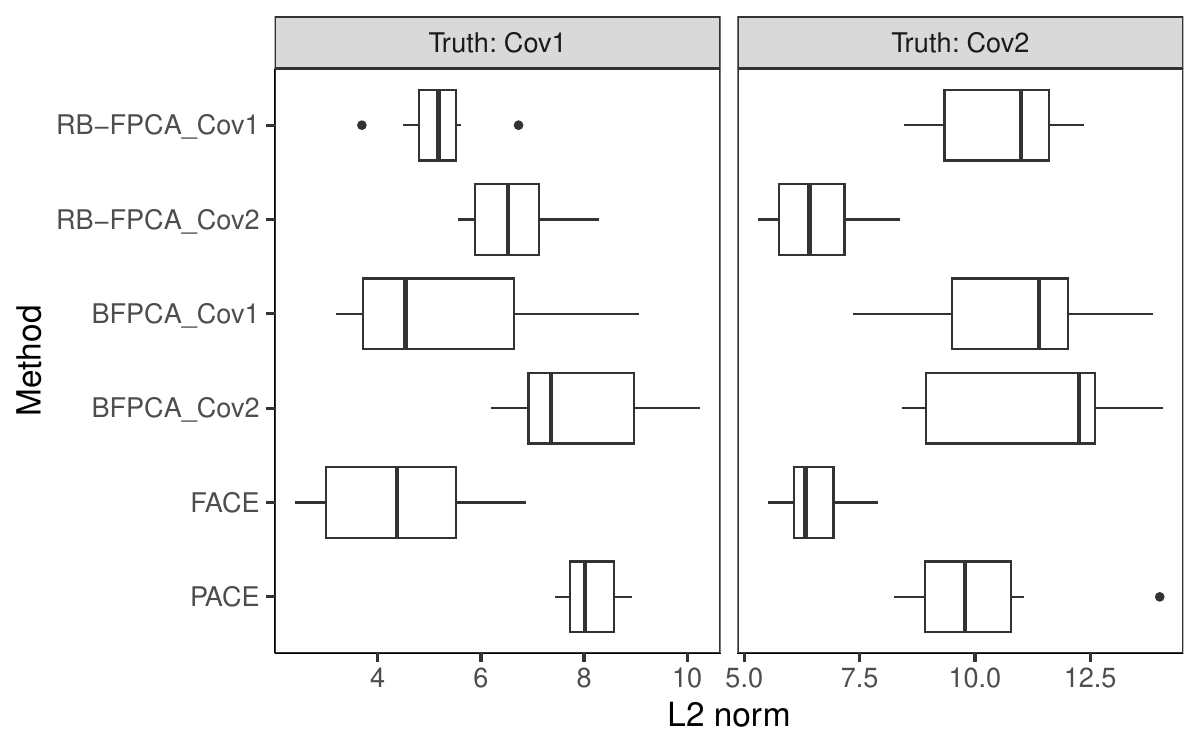}
  \caption{Simulation II with $ST(0,1,5,5)$ noises: Boxplots of the distance in the $L_2$ norm between the estimates and the true covariance function.} \label{fig:sim1_cov_boxplot_st}
\end{figure}

\begin{table}[htbp]
\small
\begin{tabular}{|c|c|c|c|c|c|c|c|}
\hline
 & Truth & \begin{tabular}[c]{@{}c@{}}RB-FPCA\\ with \\ Prior \\Cov1\end{tabular} & \begin{tabular}[c]{@{}c@{}}RB-FPCA \\ with \\ Prior \\Cov2\end{tabular} & \begin{tabular}[c]{@{}c@{}}BFPCA \\ with \\ Prior \\Cov1\end{tabular} & \begin{tabular}[c]{@{}c@{}}BFPCA \\ with \\ Prior \\Cov2\end{tabular} & FACE & PACE \\ \hline
\multirow{2}{*}{PC1} & Cov1 & \textbf{0.033} & 0.069 & 0.048 & 0.047 & 0.034 & 0.047 \\
 & Cov2 & 0.027 & 0.024 & 0.040 & \textbf{0.016} & 0.048 & 0.057 \\ \hline
\multirow{2}{*}{PC2} & Cov1 & 0.030 & 0.030 & \textbf{0.011} & 0.029 & 0.047 & 0.033 \\ 
 & Cov2 & 0.062 & 0.042 & 0.072 & 0.064 & \textbf{0.040} & 0.048 \\ \hline
\multirow{2}{*}{PC3} & Cov1 & \textbf{0.033} & 0.041 & \textbf{0.033} & 0.038 & 0.047 & 0.049 \\ 
 & Cov2 & 0.025 & \textbf{0.011} & 0.048 & 0.022 & 0.040 & 0.053 \\ \hline
\end{tabular}
\caption{Simulation II with $ST(0,1,5,5)$ noises: Estimations of the first 3 principal components are compared with the Mean Squared Errors (MSEs) between the estimates and the true PCs. \label{tab:sim1-PCs_est_st}}
\end{table}

\clearpage
\section{Additional results for Simulation IV}

In this simulation, we examine dense functional data with outliers from different noise distributions and principal components' weights. 

\subsection{noise $\sim$ $t(5)$}

Student-$t$ distribution with df = 5

\begin{figure}[htbp]
  \centering
  \includegraphics[width=0.7\textwidth]{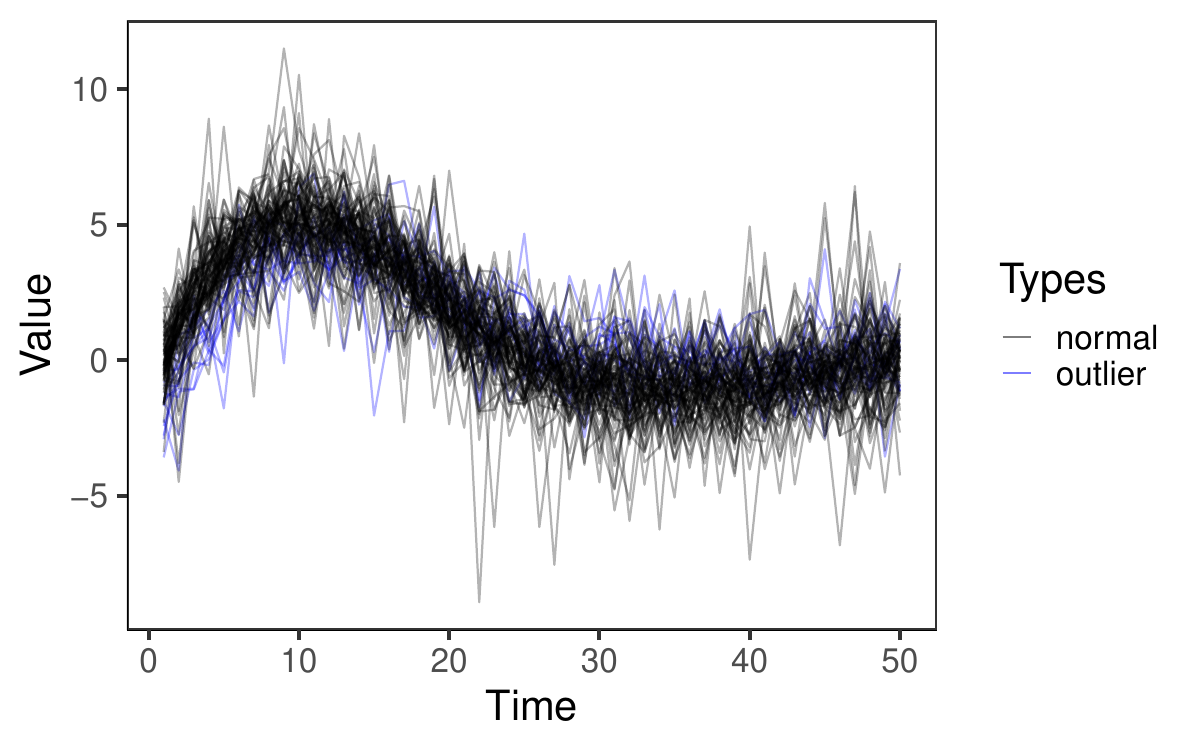}
  \caption{Simulation IV with $t(5)$ noises: Clean samples with contaminated samples from the data generation process. The outlier percentage is 10\%.} \label{fig:sim4-raw_data_t}
\end{figure}

\begin{table}[htbp]
\small
\begin{tabular}{|c|c|c|c|c|c|c|}
\hline
    & $p$           & RB-FPCA                         & BFPCA   & FACE    & PACE    & \begin{tabular}[c]{@{}c@{}}RB-FPCA \\         outperforms others\end{tabular} \\ \hline
    & 0.05        & \textbf{42.061} & 48.426 & 42.401 & 45.037 & 43$\%$ \\ 
    & 0.10        & \textbf{41.907} & 48.683 & 42.445 & 45.133 & 57$\%$ \\ 
    \multirow{-3}{*}{\begin{tabular}[c]{@{}c@{}}Cov\\ Function\end{tabular}} & 0.15 & {\textbf{41.951}} & 47.809 & 42.548 & 45.304 & 57$\%$ \\ \hline
\end{tabular}
\caption{Simulation IV with $t(5)$ noises: Estimations of the covariance function are evaluated by the distances between the estimates and the true covariance function with the $L_2$ norms.  \label{tab:sim4-cov_est_t}}
\end{table}

\begin{figure}[htbp]
    \centering
    \subfloat[]{\includegraphics[scale = 0.55]{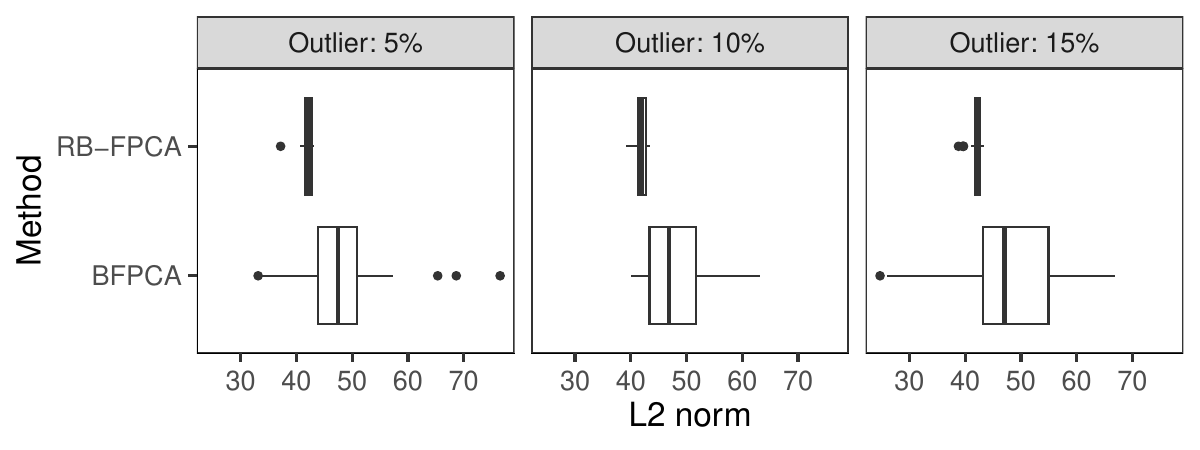}\label{fig:sim4-est_cov_1_t}} \\
    \subfloat[]{\includegraphics[scale = 0.55]{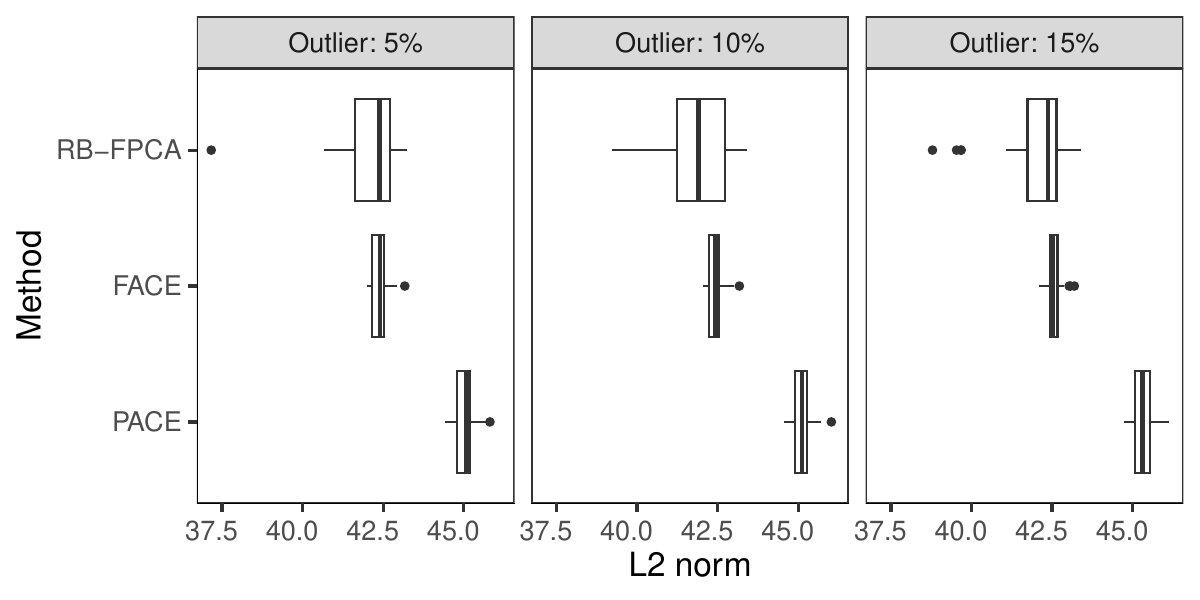}\label{fig:sim4-est_cov_2_t}}
    \caption{Simulation IV with $t(5)$ noises: (a) Boxplots of distances between the estimates and the true covariance function with the $L_2$ norms. Comparison is between the proposed RB-FPCA method and the other Bayesian methods. (b) Boxplots of distances between the estimates and the true covariance function with the $L_2$ norms. Comparison is between the proposed RB-FPCA method and the two frequentist methods. \label{fig:sim4-est_cov_t}}
\end{figure}

\begin{table}[htbp]
\small
\begin{tabular}{|c|c|c|c|c|c|c|}
\hline
    & $p$      & PCs & RB-FPCA     & BFPCA     & FACE     & PACE  \\ \hline
    &          & PC1 & {\textbf{1.354}} & 1.603     & 1.646     & 1.642   \\ 
    &          & PC2 & \textbf{1.322} & 1.462     & 1.538     & 1.443   \\ 
    & \multirow{-3}{*}{0.05} & PC3   & \textbf{1.326}     & 1.552     & 1.699    & 1.494 \\ 
    &                        & PC1   & 1.587     & 1.579     & 1.630    & \textbf{1.555} \\ 
    &                        & PC2   & \textbf{1.329} & 1.532     & 1.495     & 1.487 \\ 
    & \multirow{-3}{*}{0.10} & PC3   & \textbf{1.292}     & 1.423     & 1.895    & 1.528 \\ 
    &                        & PC1   & \textbf{1.330} & 1.580     & 1.546     & 1.573 \\ 
    &                        & PC2   & \textbf{1.467} & 1.542     & 1.556     & 1.516 \\ 
    \multirow{-9}{*}{PCs} & \multirow{-3}{*}{0.15} & PC3 & \textbf{1.279}    & 1.503     & 1.544    & 1.829 \\ \hline
\end{tabular}
\caption{Simulation IV with $t(5)$ noises: Estimations of the first 3 principal components are compared with angles (in radians) between the truth and estimates. \label{tab:sim4-PCs_est_t}}
\end{table}

\subsection{noise $\sim$ $SN(0,1,5)$}

skewed normal with location = 0, scale = 1, shape = 5 (mean $\approx$ 0.78, variance $\approx$ 0.38)

\begin{table}[htbp]
\small
\begin{tabular}{|c|c|c|c|c|c|c|}
\hline
    & $p$      & PCs & RB-FPCA     & BFPCA     & FACE     & PACE  \\ \hline
    &          & PC1 & \textbf{0.811} & 1.598     & 1.615     & 1.644   \\ 
    &          & PC2 & \textbf{1.121} & 1.599     & 1.620     & 1.559   \\ 
    & \multirow{-3}{*}{0.05} & PC3   & \textbf{1.099}     & 1.969     & 1.105    & 1.557 \\ 
    &                        & PC1   & \textbf{1.234}     & 1.642     & 1.569    & 1.587 \\ 
    &                        & PC2   & \textbf{1.122}     & 1.613     & 1.546     & 1.555 \\ 
    & \multirow{-3}{*}{0.10} & PC3   & 1.362     & 1.825     & \textbf{1.353}    & 1.543 \\ 
    &                        & PC1   & \textbf{1.193} & 1.416     & 1.565     & 1.568 \\ 
    &                        & PC2   & \textbf{1.096} & 1.479     & 1.563     & 1.575 \\ 
    \multirow{-9}{*}{PCs} & \multirow{-3}{*}{0.15} & PC3 & \textbf{1.402}    & 1.660     & 1.703    & 1.527 \\ \hline
\end{tabular}
\caption{Simulation IV with $SN(0,1,5)$ noises: Estimations of the first 3 principal components are compared with angles (in radians) between the truth and estimates. \label{tab:sim4-PCs_est_sn}}
\end{table}

\clearpage
\subsection{noise $\sim$ $ST(0,1,5,5)$}

skewed t with location = 0, scale = 1, shape = 5, df = 5 (mean $\approx$ 0.93, variance $\approx$ 0.80)

\begin{figure}[htbp]
  \centering
  \includegraphics[width=0.7\textwidth]{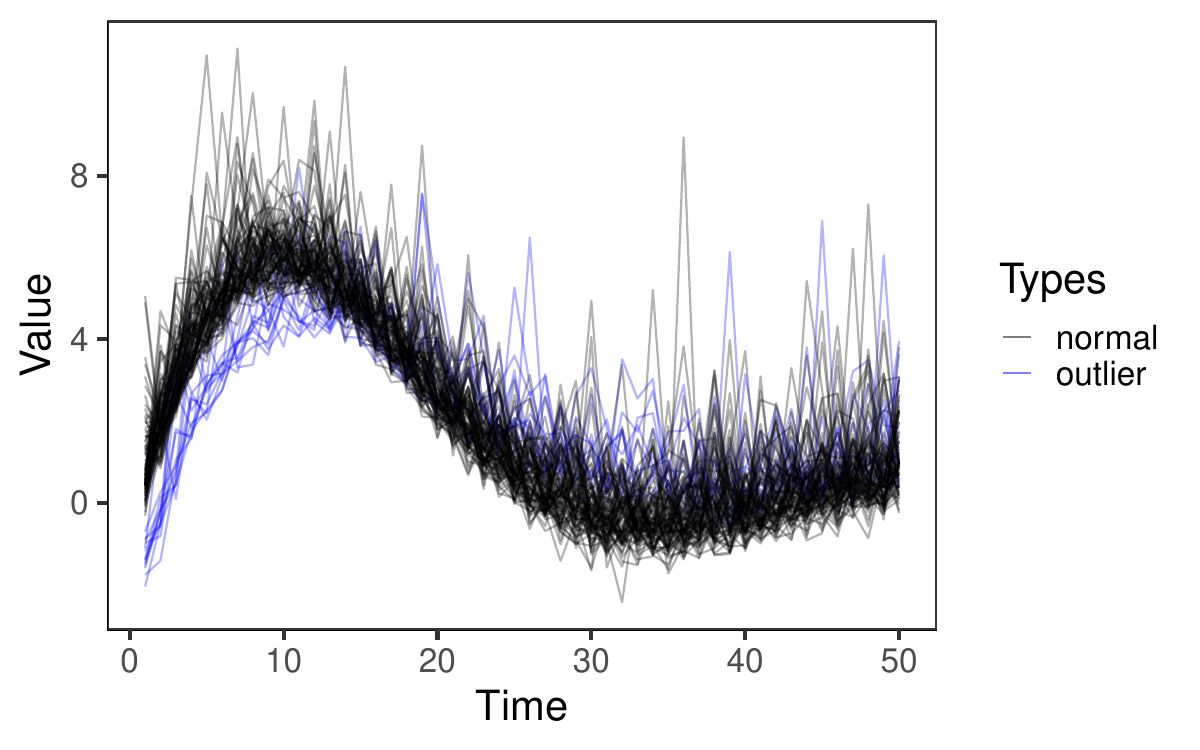}
  \caption{Simulation IV with $ST(0,1,5,5)$ noises: Clean samples with contaminated samples from the data generation process. The outlier percentage is 10\%.} \label{fig:sim4-raw_data_st}
\end{figure}

\begin{table}[htbp]
\small
\begin{tabular}{|c|c|c|c|c|c|c|}
\hline
    & $p$           & RB-FPCA                         & BFPCA   & FACE    & PACE    & \begin{tabular}[c]{@{}c@{}}RB-FPCA \\         outperforms others\end{tabular} \\ \hline
    & 0.05        & \textbf{39.137} & 50.766 & 43.282 & 44.375 & 100$\%$ \\ 
    & 0.10        & \textbf{39.472} & 50.803 & 43.589 & 44.517 & 100$\%$ \\ 
    \multirow{-3}{*}{\begin{tabular}[c]{@{}c@{}}Cov\\ Function\end{tabular}} & 0.15 & \textbf{39.531} & 51.588 & 43.410 & 44.596 & 100$\%$ \\ \hline
\end{tabular}
\caption{Simulation IV with $ST(0,1,5,5)$ noises: Estimations of the covariance function are evaluated by the distances between the estimates and the true covariance function with the $L_2$ norms.  \label{tab:sim4-cov_est_st}}
\end{table}

\begin{figure}[htbp]
    \centering
    \subfloat[]{\includegraphics[scale = 0.55]{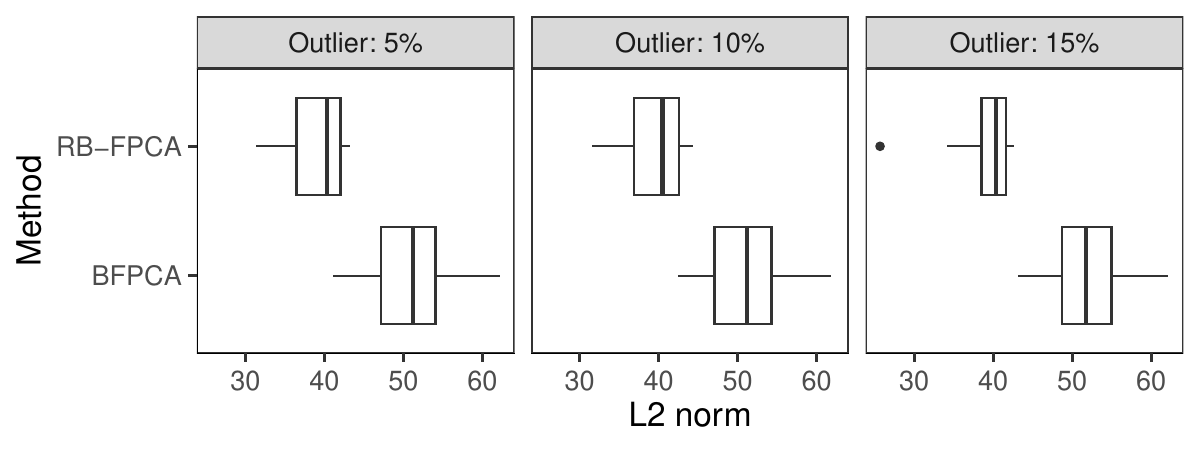}\label{fig:sim4-est_cov_1_st}} \\
    \subfloat[]{\includegraphics[scale = 0.55]{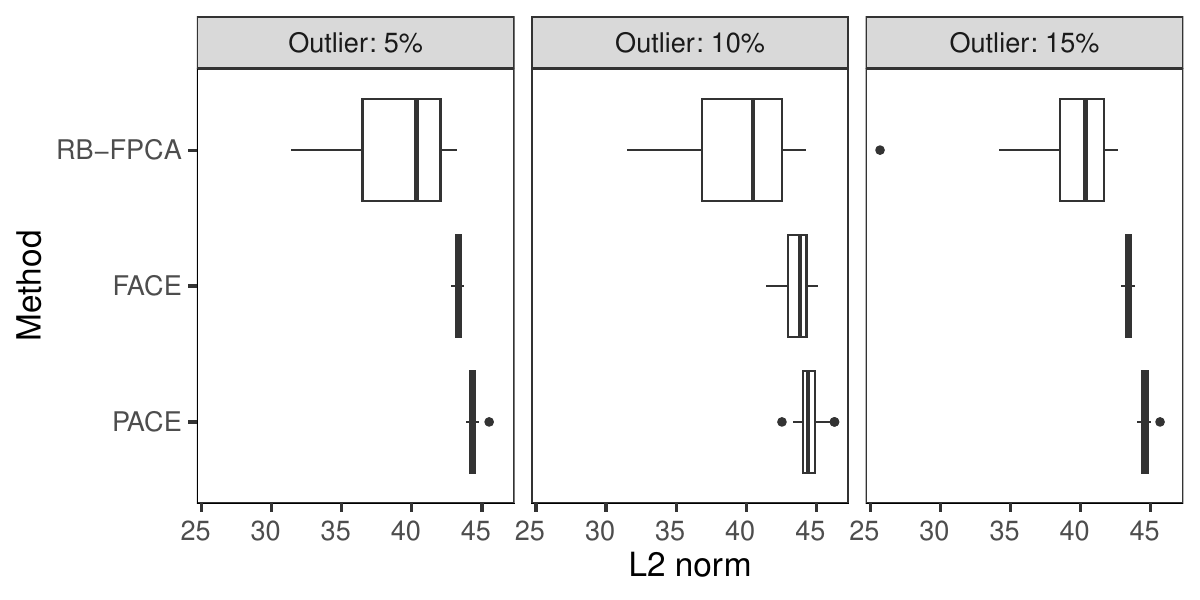}\label{fig:sim4-est_cov_2_st}}
    \caption{Simulation IV with $ST(0,1,5,5)$ noises: (a) Boxplots of distances between the estimates and the true covariance function with the $L_2$ norms. Comparison is between the proposed RB-FPCA method and the other Bayesian methods. (b) Boxplots of distances between the estimates and the true covariance function with the $L_2$ norms. Comparison is between the proposed RB-FPCA method and the two frequentist methods. \label{fig:sim4-est_cov_st}}
\end{figure}

\begin{table}[htbp]
\small
\begin{tabular}{|c|c|c|c|c|c|c|}
\hline
    & $p$      & PCs & RB-FPCA     & BFPCA     & FACE     & PACE  \\ \hline
    &          & PC1 & \textbf{1.042} & 1.691     & 1.661     & 1.688   \\ 
    &          & PC2 & \textbf{1.131} & 1.619     & 1.631     & 1.578   \\ 
    & \multirow{-3}{*}{0.05} & PC3   & \textbf{1.200}     & 1.423     & 1.587    & 1.573 \\ 
    &                        & PC1   & 1.555     & \textbf{1.421}     & 1.628    & 1.635 \\ 
    &                        & PC2   & \textbf{1.355} & 1.776     & 1.558     & 1.520 \\ 
    & \multirow{-3}{*}{0.10} & PC3   & \textbf{1.121}     & 1.695     & 1.503    & 1.595 \\ 
    &                        & PC1   & \textbf{0.990} & 1.647     & 1.576     & 1.578 \\ 
    &                        & PC2   & \textbf{1.054} & 1.665     & 1.545     & 1.588 \\ 
    \multirow{-9}{*}{PCs} & \multirow{-3}{*}{0.15} & PC3 & \textbf{1.205}    & 1.921     & 1.648    & 1.601 \\ \hline
\end{tabular}
\caption{Simulation IV with $ST(0,1,5,5)$ noises: Estimations of the first 3 principal components are compared with angles (in radians) between the truth and estimates. \label{tab:sim4-PCs_est_st}}
\end{table}

\clearpage

\section{Additional results for Simulation V}

In this simulation, we examine sparse functional data with outliers. We add independent noises to the sampling process through different noise distributions, such as Student-$t$ distribution, denoted as $t(5)$, skew Normal distribution, denoted as $SN(0,1,5)$, and skew Student-$t$ distribution, denoted as $ST(0,1,5,5)$. We included visualizations of $t(5)$, $SN(0,1,5)$ and $ST(0,1,5,5)$ noises and summarized the results from $t(5)$ and $ST(0,1,5,5)$ as follows.

\begin{figure}[htbp]
  \centering
  \subfloat[]{\includegraphics[scale = 0.28]{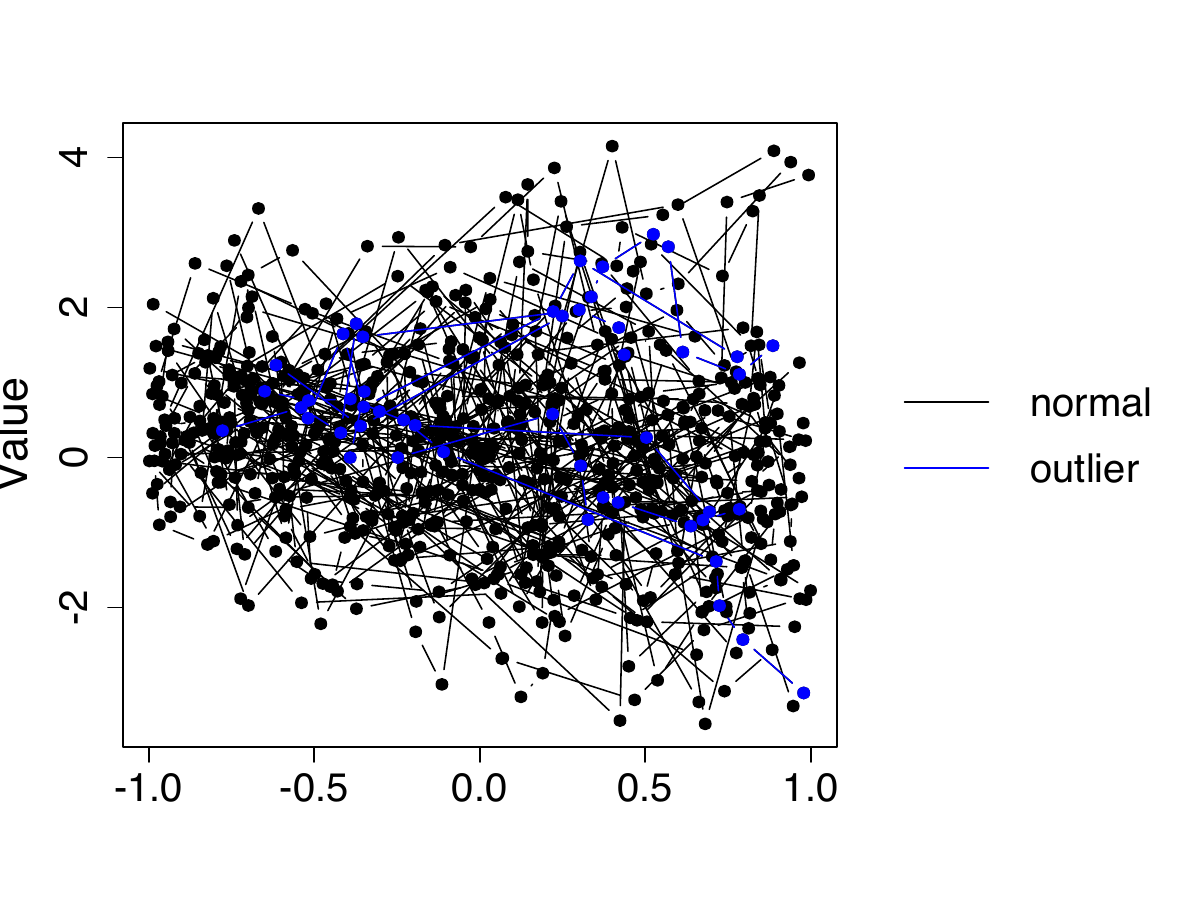}\label{fig:sim5-raw_data_new_t}} 
  \subfloat[]{\includegraphics[scale = 0.28]{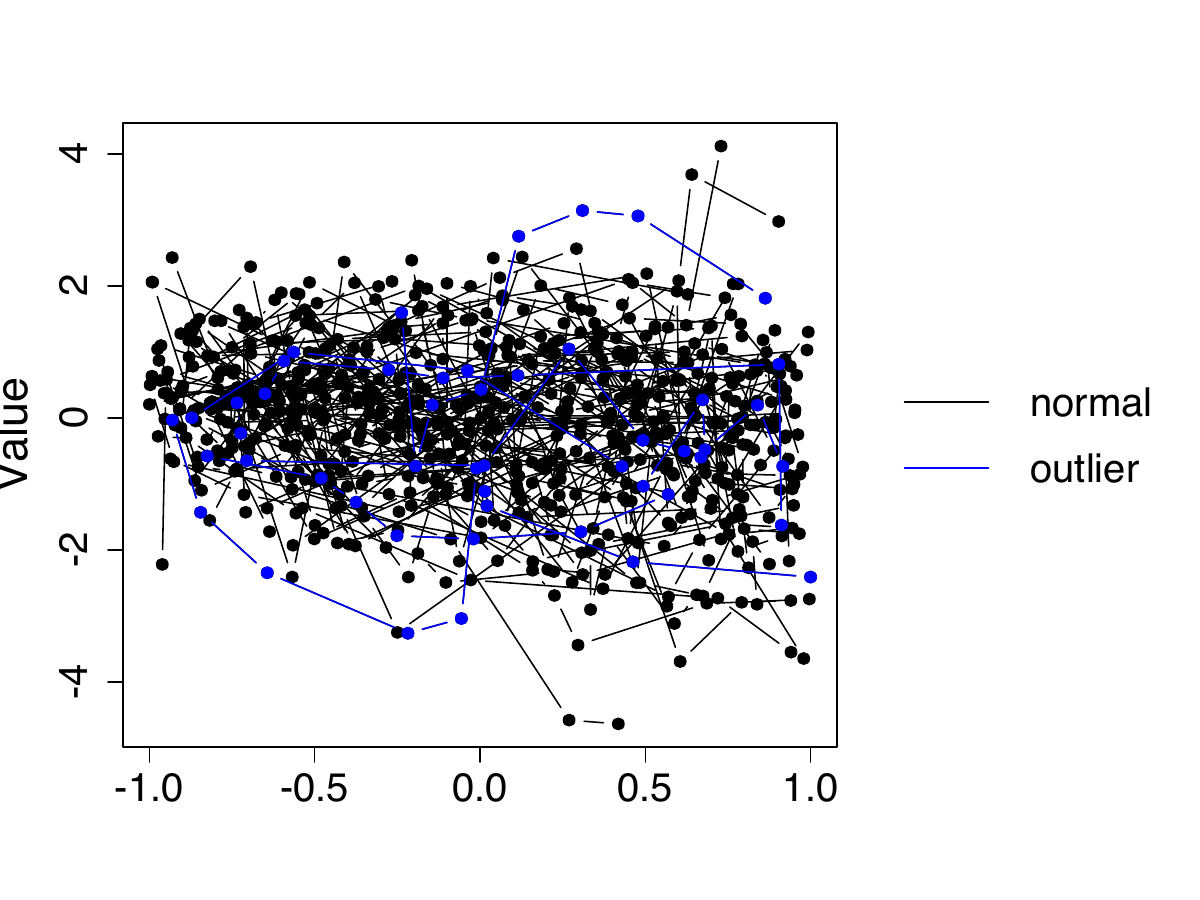}\label{fig:sim5-raw_data_new_sn}} \\
  \subfloat[]{\includegraphics[scale = 0.28]{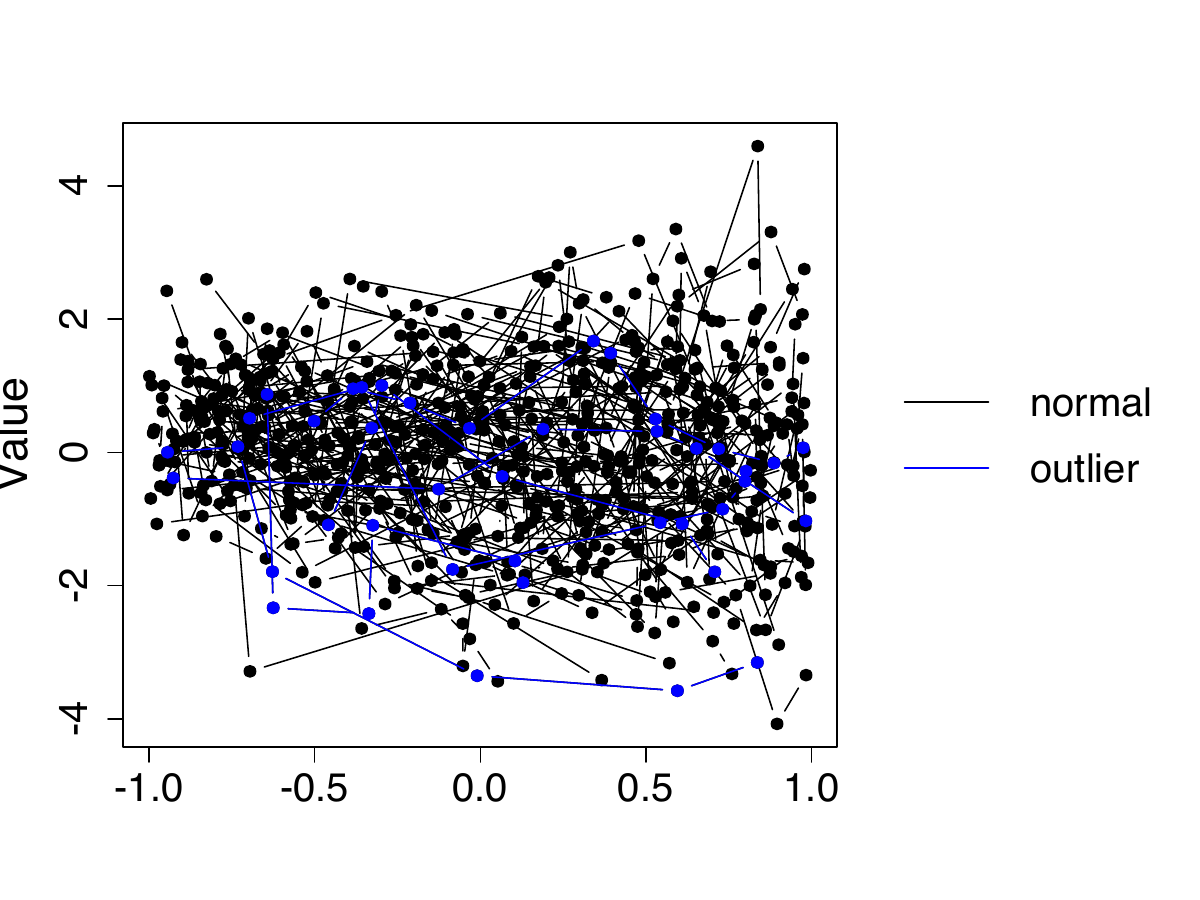}\label{fig:sim5-raw_data_new_st}}
  \caption{Simulation V: Clean samples with contaminated samples from the data generation process. The outlier percentage is 5\%. (a) $t(5)$ noises. (b) $SN(0,1,5)$ noises. (c) $ST(0,1,5,5)$ noises.} \label{fig:sim5-raw_data_new}
\end{figure}

\begin{figure}[htbp]
    \centering
    \subfloat[]{\includegraphics[width=0.44\textwidth]{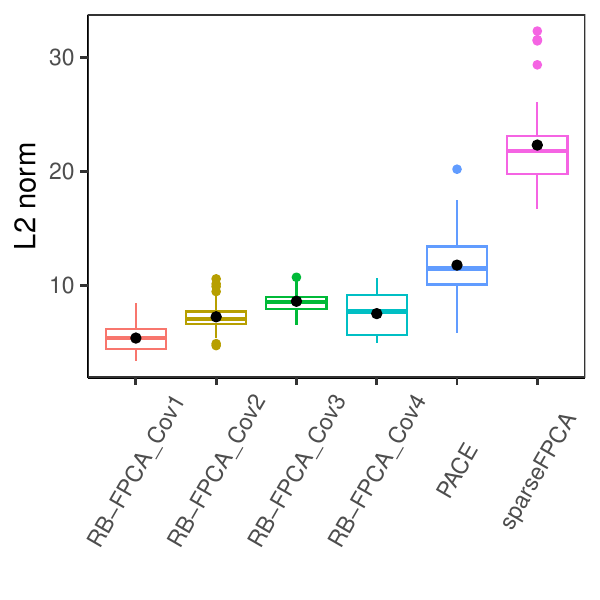}\label{fig:sim5_boxplot_a_t}}
    \subfloat[]{\includegraphics[width=0.44\textwidth]{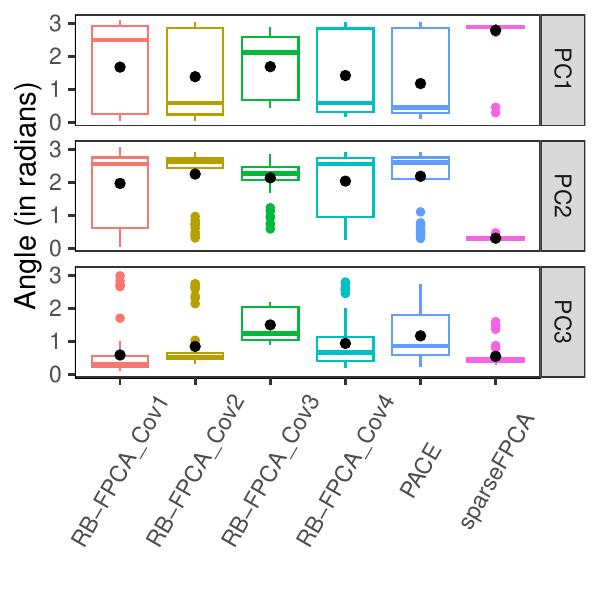}\label{fig:sim5_boxplot_b_t}} \\
    \subfloat[]{\includegraphics[width=0.44\textwidth]{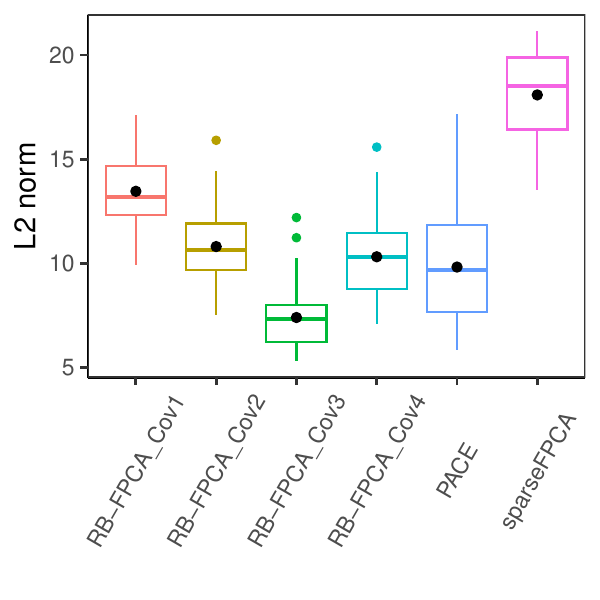}\label{fig:sim5_boxplot_c_t}} 
    \subfloat[]{\includegraphics[width=0.44\textwidth]{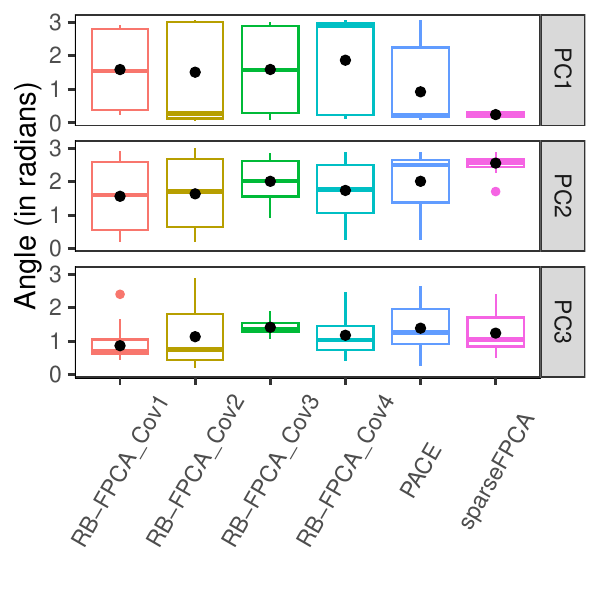}\label{fig:sim5_boxplot_d_t}}
    \caption{Simulation V with $t(5)$ noises: Comparison of RB-FPCA method with PACE and sparseFPCA methods. (a) and (c) shows boxplots of distances (in $L_2$ norms) between the estimates and the true correlation function. (b) and (d) shows boxplots of the angle (in radians) of the estimations for the first 3 principal components. The top row shows the results given the true covariance function is $\text{Cov}_{truth}(s,t) = \exp\{-3(t-s)^2\}$. The bottom row shows the results when the true covariance function is $\text{Cov}_{truth}(s,t) = \min\left(s+1, t+1\right)$. The black dot in each box corresponds to the value of the mean. \label{fig:sim5_boxplot_t}}
\end{figure}

\begin{figure}[htbp]
    \centering
    \subfloat[]{\includegraphics[width=0.44\textwidth]{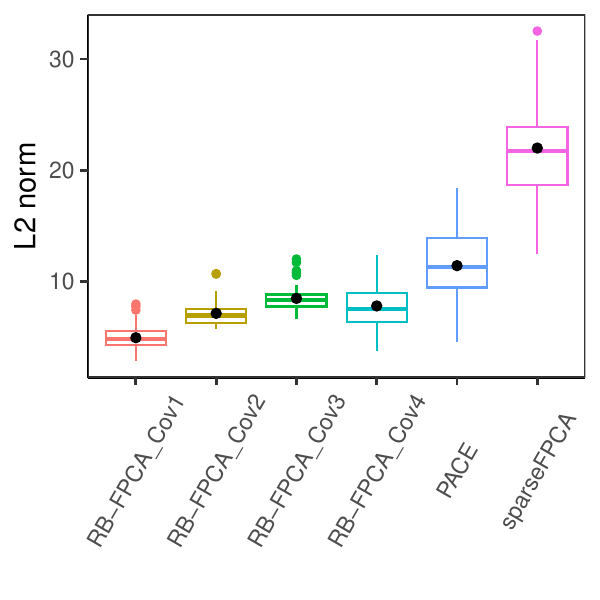}\label{fig:sim5_boxplot_a_st}}
    \subfloat[]{\includegraphics[width=0.44\textwidth]{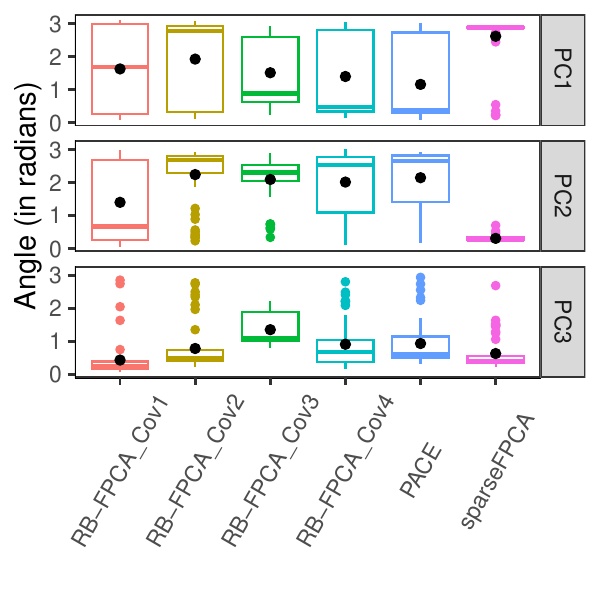}\label{fig:sim5_boxplot_b_st}} \\
    \subfloat[]{\includegraphics[width=0.44\textwidth]{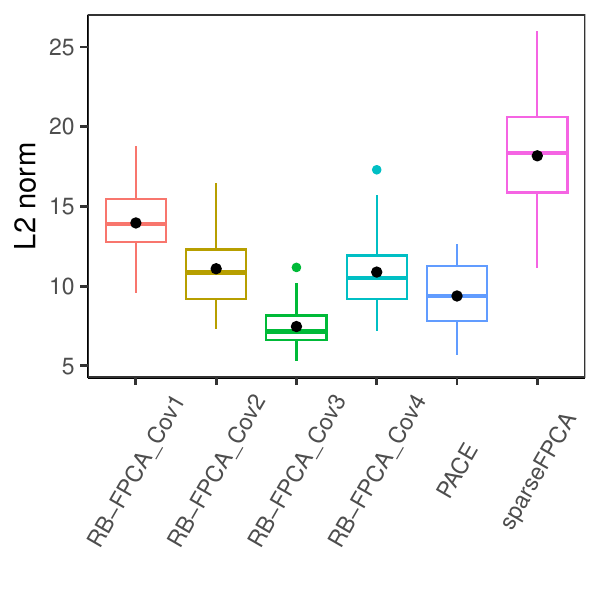}\label{fig:sim5_boxplot_c_st}} 
    \subfloat[]{\includegraphics[width=0.44\textwidth]{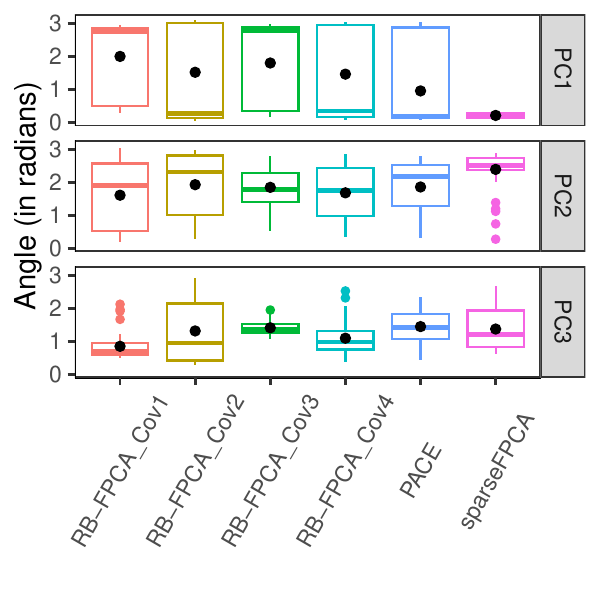}\label{fig:sim5_boxplot_d_st}}
    \caption{Simulation V with $ST(0,1,5,5)$ noises: Comparison of RB-FPCA method with PACE and sparseFPCA methods. (a) and (c) shows boxplots of distances (in $L_2$ norms) between the estimates and the true correlation function. (b) and (d) shows boxplots of the angle (in radians) of the estimations for the first 3 principal components. The top row shows the results given the true covariance function is $\text{Cov}_{truth}(s,t) = \exp\{-3(t-s)^2\}$. The bottom row shows the results when the true covariance function is $\text{Cov}_{truth}(s,t) = \min\left(s+1, t+1\right)$. The black dot in each box corresponds to the value of the mean. \label{fig:sim5_boxplot_st}}
\end{figure}